\documentclass[aps,prb,twocolumn,superscriptaddress,groupedaddress,10pt]{revtex4}
\usepackage{amsmath}
\usepackage{amssymb}
\usepackage{graphicx}
\usepackage{epsfig}
\usepackage{dcolumn}
\usepackage{bm}
\usepackage{bm}
\usepackage{blindtext}
\usepackage{natbib}

\usepackage{bbm}
\usepackage{multirow}

\def\be{\begin{equation}} \def\ee{\end{equation}}
\def\bea{\begin{eqnarray}} \def\eea{\end{eqnarray}}

\def\nn{\nonumber}

\newcommand{\ket}[1]{| #1 \rangle}
\newcommand{\bra}[1]{\langle #1 |}

\begin{document}
\title{Majorana surface modes of nodal topological pairings
in spin-$\frac{3}{2}$ semi-metals}

\author{Wang Yang}
\affiliation{Department of Physics, University of California,
San Diego, California 92093, USA}
\author{Tao Xiang}
\affiliation{Institute of Physics, Chinese Academy of Sciences,
P.O. Box 603, Beijing 100190, China}
\affiliation{Collaborative Innovation Center of Quantum Matter,
Beijing 100190, China}
\author{Congjun Wu}
\affiliation{Department of Physics, University of California,
San Diego, California 92093, USA}

\begin{abstract}
When solid state systems possess active orbital-band structures subject to
spin-orbit coupling, their multi-component electronic structures are often
described in terms of effective large-spin fermion models.
Their topological structures of superconductivity are beyond the
framework of spin singlet and triplet Cooper pairings
for spin-$\frac{1}{2}$ systems.
Examples include the half-Heusler compound series of RPtBi, where
R stands for a rare-earth element.
Their spin-orbit coupled electronic structures are described by the Luttinger-Kohn model with effective spin-$\frac{3}{2}$ fermions
and are characterized by band inversion.
Recent experiments provide evidence to unconventional superconductivity
in the YPtBi material with nodal spin-septet pairing.
We systematically study topological pairing structures in spin-$\frac{3}{2}$
systems with the cubic group symmetries and calculate the surface Majorana
spectra, which exhibit zero energy flat bands, or, cubic dispersion depending on the specific symmetry of the superconducting gap functions.
The signatures of these surface states in the quasi-particle interference
patterns of tunneling spectroscopy are studied, which can be tested in
future experiments.
\end{abstract}
\maketitle

\section{Introduction}
\label{sect:intro}

Topological superconductivity and paired superfluidity have been
attracting intense research interests in recent years \cite{Qi2011,
Volovik2003,Leggett1975}.
The nontrivial topology manifests itself in the Andreev-Majorana
zero modes on boundaries and topological defects like vortices \cite{Volovik1999,Read2000,Kitaev2001,Qi2009}.
Such Andreev-Majorana modes are of particular interests since they are
potentially useful for topological quantum computations
\cite{Ivanov2001,Alicea2011,Teo2010}.
Early topological classifications mostly focus
on the fully gapped superconducting systems\cite{Schnyder2008,
Kitaev2009,Ryu2010}, including the two-dimensional $p_x+ip_y$
superconductor \cite{Read2000},
and the three-dimensional $^3$He-B phase with the isotropic $p$-wave
triplet pairing \cite{Balian1963,Leggett1975,Chung2009}.
Recently, gapped topological superconductivity has
also been proposed for high T$_c$ cuprates in the very
underdoped regime \cite{Lu2014}.

Gapless, or, nodal, superconductors/pairing superfluids often exhibit unconventional pairing symmetries, such as the $d$-wave superconductors
of high $T_c$ cuprates\cite{Hu1994}, the three-dimensional $^3$He-A phase
with the $p_x+ip_y$ triplet pairing \cite{Volovik1992,Read2000}, the $p_z$-triplet pairing phase of electric dipolar fermions\cite{Wu2010},
and spin-orbit coupled $p$-wave pairing with total angular momentum
$J=1$ induced by magnetic dipolar interactions \cite{Li2012}.
For these examples, their gap functions exhibit
spherical or spin-orbit coupled harmonic symmetries.
In contrast, the doped magnetic Weyl semi-metals can support
monopole harmonic pairing\cite{Li2015}, which is
a class of topological superconducting states characterized by
non-trivial monopole structures.
Their pairing phases cannot be globally well-defined on Fermi surfaces.

The gapless superconducting systems also exhibit interesting topological
structures \cite{Chiu2016,Mizushima2016}, which are typically weaker
than those in the gapped cases in the sense that only suitably
oriented surfaces can support the zero energy Andreev-Majorana states.
These surfaces are with particular orientations such that a relative
sign change occurs between the gap functions along the incident
and reflected wavevectors.
There do not exist well-defined global topological numbers for a gapless system. 
However, a topological number can be defined for each momentum within the surface Brillouin zone, as the winding number of the effective one-dimensional system perpendicular to the surface with the fixed in-surface momentum. 
This topological number is related to the existence of zero energy Andreev-Majorana modes through the bulk-edge correspondence principle \cite{Schnyder2012,Schnyder2015}.

The non-centrosymmetric superconductors add to the diversity of gapless
superconductivity \cite{Schnyder2011,Brydon2011,Schnyder2012,Schnyder2015,
Smidman2016}.
In their normal state band structures, the spin degeneracy is lifted
by spin-orbit coupling, and pairing gap functions typically show
mixed-parity due to the breaking of inversion symmetry.
Depending on the pairing symmetry and the nodal structure, there
appear Majorana flat bands and zero energy arcs on the surfaces
with suitable orientations \cite{Sato2011,Schnyder2011,Brydon2011,Schnyder2015}.
The experimental signatures include zero-energy peaks in the
tunneling spectra \cite{Tanaka2000,Tanaka2010},
and certain patterns in the quasi-particle interference(QPI)
\cite{Hofmann2013}.
The instability of the Majorana flat bands has been studied with
respect to the spontaneous time-reversal (TR) symmetry breaking
effects arising from the Majorana fermion-superfluid phase
interaction \cite{Li2013}, or, the magnetic interactions \cite{Potter2014},
and also by the self-consistent mean field theory \cite{Timm2015}.

The superconducting pairing symmetries can be greatly enriched in
multi-component fermion/electron systems.
In ultra-cold fermion systems, many fermions carry
large-hyperfine-spin $S$ larger than $\frac{1}{2}$.
In solid state systems, electrons can be effectively multi-component
due to orbital degeneracy and spin-orbit splitting, such as
the effective spin-$\frac{3}{2}$ Luttinger-Kohn model for
the hole band of semi-conductors.
The spin of their Cooper pairing can take values from 0 to $2S$ beyond
the conventional singlet and triplet scenarios
\cite{Ho1999,Wu2003,Wu2006,Wu2010a,Yang2016}.
For example, the spin quintet pairing (the spin of Cooper pair $S=2$)
has been found to support the non-Abelian Cheshire charge in the presence
of half-quantum vortex loop \cite{Wu2010a}.
Recently, the $^3$He-B type isotropic topological pairing has been
generalized to multi-component fermion systems \cite{Yang2016}.
For the simplest case of spin-3/2 systems, both the $p$-wave triplet
and the $f$-wave septet pairings are non-trivial,
possessing topological index 4 and 2, respectively.
They support surface spectra with multiple linear and cubic Dirac-Majorana
cones\cite{Fang2015,Yang2016}.
The interaction-induced TR symmetry breaking effects are investigated
in Ref. [\onlinecite{Ghorashi2017}].
The topological nature of this class of pairings is most clearly seen
in the helicity basis, and thus it also applies to spin-orbit coupled
multi-orbital solid state band systems.

Recently, the half-Heusler compound YPtBi has attracted considerable
attention \cite{Liu2016,Kim2016}.
The band structure can be described by the effective spin-orbit coupled
spin-$\frac{3}{2}$  Kohn-Luttinger Hamiltonian with parity
symmetry breaking terms.
The chemical potential lies close to the $\Gamma$-point of
the $p$-like $\Gamma_8$ band \cite{Liu2016} as shown in the
angluar-resolved-photo-emission-spectroscopy.
Experiment evidence to unconventional superconductivity
has been found in the half-Heusler compound YPtBi \cite{Kim2016}.
A mixed parity pairing immune to pair breaking effect has been
proposed for YPtBi \cite{Brydon2016}, with a small fraction of
an $s$-wave singlet component superposed on the $p$-wave septet pairing.

Motivated by the advancements in non-centrosymmetric superconductors,
we systematically study the Majorana surface states of
topological superconductors
based on the spin-$\frac{3}{2}$ Luttinger-Kohn model
subject to the cubic symmetries.
For the $T_d$ point group symmetry, we show that the double degeneracy
along the $[0\,0\,1]$ direction and its equivalent ones are protected
by the little group $SD_{16}$, i.e., the semi-dihedral group of order $16$.
The pairing patterns in non-centrosymmetric systems are in general
of mixed-parity nature as discussed under concrete cubic
symmetry groups.
For the YPtBi, the proposed mixed $s$-wave singlet and $p$-wave
septet pairing exhibits line nodes on one of the spin split
Fermi surfaces as shown in Ref. [\onlinecite{Brydon2016}], and the six
nodal loops centering around $[0\,0\,1]$ and its equivalent
directions are topological.
We show that for the $[1\,1\,1]$ surface, the Majorana zero modes
appear in regions enclosed by the projections of the nodal loops to
the surface Brillouin zone, but disappear in the overlapping regions.
The QPI patterns are calculated on the $[1\,1\,1]$ surface due to the
scattering with a single impurity in the Born approximation.
For a non-magnetic impurity, the chiral symmetry forbids
the scatterings between Majorana islands with the same chiral
index, while for a magnetic impurity,
the scatterings between Majorana islands with opposite chiral
indices are forbidden.
The structures of the QPI patterns of a magnetic impurity
are richer than those of a non-magnetic impurity under
the $C_{3v}$ group, which is the symmetry group of the
$[1\,1\,1]$-surface.
Experiments on the QPI patterns will provide tests
to the proposed pairing symmetries of YPtBi.

The rest of this article is organized as follows.
In Sec. \ref{sec:model Ham}, the Luttinger-Kohn Hamiltonian and
the band inversion are reviewed.
The inversion breaking terms with the cubic symmetries are classified,
and the protected double degeneracy of the $T_d$ group along the
$[0\, 0\, 1]$ direction is proved.
In Sec. \ref{sec:noncentro_pairing}, the non-centrosymmetric Copper
pairings with the cubic symmetries are discussed.
The Majorana surface modes are solved in
Sec. \ref{sec:Majorana septet}.
The QPI patterns are calculated in  Sec. \ref{sec:qpi}.

\section{Non-centrosymmetric spin-$\frac{3}{2}$ systems with
the cubic symmetries}
\label{sec:model Ham}

In this section, we first discuss the Luttinger-Kohn Hamiltonian in
spin-orbit coupled systems, and classify the inversion breaking
terms according to the cubic point groups.
We then show that for the $T_d$ group, there exits a protected
double degeneracy along $[0\,0\,1]$ and its equivalent directions.
Finally the band structure properties of the YPtBi material are reviewed.

\subsection{The Luttinger-Kohn Hamiltonian and band inversion}
Although electrons carry spin-$\frac{1}{2}$, the effective
spin-$\frac{3}{2}$ systems are not rare in solid state materials
due to spin-orbit coupling.
Examples include the half-Heusler compounds, where spin-orbit coupling
recombines the outer-shell $s$- and $p$- orbitals into $s_{\frac{1}{2}}$
($\Gamma_6$), $p_{\frac{1}{2}}$ ($\Gamma_7$) and
$p_{\frac{3}{2}}$ ($\Gamma_8$) orbitals, with the subscripts denoting
the spin-orbit coupled total angular momentum.
The $p_{\frac{1}{2}}$-band is denoted the ``spin-split" band, which
is far from the Fermi energy, and will be neglected below.
The $s_{\frac{1}{2}}$ and $p_{\frac{3}{2}}$ bands are active,
and typically the $s_{\frac{1}{2}}$ band energy is higher.
The gap between them is tunable by varying the spin-orbit coupling strength,
which can be realized in experiments by substituting the heavy atom with
other atomic elements.
The gap vanishes at a critical spin-orbit coupling strength, and then
becomes negative, i.e, the energy of the $s_\frac{1}{2}$-band becomes
lower, which is termed as ``band inversion".
The $p_{\frac{3}{2}}$-band further splits due to spin-orbit coupling
according to the helicity quantum number, i.e., the spin projection
on the momentum direction.
The heavy hole band is of the helicity quantum numbers $\pm\frac{3}{2}$,
and the light hole one is of the helicity numbers $\pm\frac{1}{2}$.
The heavy and light hole bands touch at the $\Gamma$-point,
as protected by the cubic group symmetry.
All bands are doubly degenerate when TR and inversion symmetries are present.

\begin{figure}
\centering\epsfig{file=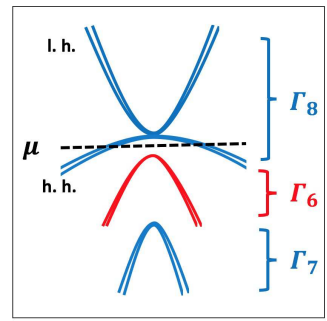,clip=0.6,
width=0.5 \linewidth,height=0.5 \linewidth,angle=0}
\caption{Schematic plots of the $p_{\frac{3}{2}}$- bands ($\Gamma_8$),
$p_{\frac{1}{2}}$- bands ($\Gamma_7$), and $s_{\frac{1}{2}}$- bands
($\Gamma_6$).
The abbreviations of ``h. h." and ``l. h." within the $p_{\frac{3}{2}}$-bands
represent the heavy and light hole bands, respectively.
The Fermi level crosses the heavy hole bands.
Each pair of bands exhibit spin splitting due to the presence of
the inversion symmetry breaking term $\frac{\delta}{k_f}
A(\vec{k})$ in Eq. (\ref{eq:H0}).
}
\label{fig:band}
\end{figure}

At the critical spin-orbit coupling strength, where the $s_\frac{1}{2}$-
and $p_\frac{3}{2}$-bands touch at the $\Gamma$-point,
the dispersions of $s_{\frac{1}{2}}$- and light hole bands become linear,
while the heavy hole band remains parabolic.
After the band inversion, the curvature of the dispersion of
$s_{\frac{1}{2}}$-band becomes negative, while that of the
light hole actually is positive.
A schematic plot of the band structure after band inversion
is shown in Fig. \ref{fig:band}.

The process of band inversion can be understood by a $k\cdot p$
analysis as follows.
Consider systems with the full spin-orbit coupled $SO(3)$ symmetry
for simplicity.
The $k\cdot p$ basis for $s_{\frac{1}{2}}$ and $p_{\frac{3}{2}}$-bands
are chosen as
\bea
&\ket{s;\uparrow}, ~~ \ket{s;\downarrow},
\label{eq:kp_basis_s}
\eea
and
\bea
&&\ket{p_x+ip_y;\uparrow}, \nn \\ &&\frac{1}{\sqrt{3}}\left(\ket{p_x+ip_y;\downarrow}+\sqrt{2}\ket{p_z;\uparrow}
\right), \nn\\
&&\frac{1}{\sqrt{3}}\left(\ket{-p_x+ip_y;\uparrow}+\sqrt{2}\ket{p_z;\downarrow}
\right), \nn \\ &&\ket{-p_x+ip_y;\downarrow},
\label{eq:kp_basis_p}
\eea
which are eigen-basis of $S_z=L_z+\frac{1}{2}\sigma_z$ with eigenvalues
in a descending order.
The three basis at the $\Gamma$-point with positive eigenvalues of $S_z$
are $s_{\frac{1}{2}}$, $\ket{p_x+ip_y\uparrow}$, $\frac{1}{\sqrt{3}}(\ket{p_x+ip_y\downarrow}+\sqrt{2}\ket{p_z\uparrow})$,
and the other ones with negative $S_z$ can be obtained from them by
the TR operation.
Apart from an overall constant the Hamiltonian at the $\Gamma$-point
in the positive $S_z$ sector is given by
\bea
H^{+}(\Gamma)=\left(\begin{array}{ccc}
m&&\\
&-m&\\
&&-m
\end{array}\right),
\eea
in which $m$ is half of the band gap.
To obtain the Hamiltonian away from the $\Gamma$-point, it is sufficient
to consider the $z$-direction due to the rotation invariance.
The little group along this direction is the $U(1)$ group $e^{-iS_z\phi}$,
hence, hybridizations only occur between states with the same $S_z$.
The $k\cdot p$ Hamiltonian in the positive $S_z$ sector up to
the linear order in $k_z$ is
\bea
H^{+}(k_z)=\left(\begin{array}{ccc}
m&& \lambda k_z\\
&-m&\\
\lambda^{*} k_z&&-m
\end{array}\right),
\label{eq:Ham_posiSz}
\eea
in which $m>0$ and $<0$ correspond to before and after
band inversion, respectively.
The Hamiltonian $H^{-}$ in the negative $S_z$ sector can be obtained
by applying the TR operation to Eq. (\ref{eq:Ham_posiSz}).
The $k\cdot p$ Hamiltonian in the basis of Eqs. (\ref{eq:kp_basis_s},
\ref{eq:kp_basis_p}) along a general direction of $\vec k$
can be constructed by performing the rotation operation
$U=e^{-iS_z \phi_k}e^{-iS_y \theta_k}$
on $H^{+}(k_z)+ H^{-}(k_z)$ where $\theta_k$ and $\phi_k$ are the
polar and azimuthal angles of $\vec k$, respectively.
At the critical point $m=0$, the $s_{\frac{1}{2}}$- and
light hole band dispersions become linear with
the velocity $\pm \frac{|\lambda|}{\hbar}$.

Consider the case with band inversion as illustrated in Fig. \ref{fig:band},
where the Fermi energy lies close to the $\Gamma$-point of the
$p_{\frac{3}{2}}$-bands.
Only the $p_{\frac{3}{2}}$-bands are taken into account with
the $k\cdot p$ basis chosen as the four states in Eq. (\ref{eq:kp_basis_p}),
and the band structure is captured by the Luttinger-Kohn Hamiltonian
\bea
H_L(\vec{k})&=&(\lambda_1+\frac{5}{2} \lambda_2)k^2
-2\lambda_2 (\vec{k}\cdot \vec{S})^2
\nn \\
&+& \lambda_3  \sum_{i\neq j} k_ik_j S_i S_j,
\label{eq:Luttinger}
\eea
in which $\vec{S}=(S_x\,\,S_y\,\,S_z)$ are the spin-$\frac{3}{2}$ operators.
The $\lambda_3$ term breaks the full spin-orbit coupled $SO(3)$ rotational symmetry,
but is allowed for the cubic symmetry group.
When $\lambda_3=0$, the mass of helicity $\pm \frac{3}{2}$ bands is $\frac{\hbar^2}{2(\lambda_1-2\lambda_2)}$,
and that of the helicity $\pm \frac{1}{2}$ bands is $\frac{\hbar^2}{2(\lambda_1+2\lambda_2)}$.
For the band inverted case, we need $-2\lambda_2<\lambda_1<2\lambda_2$
to ensure the opposite signs of the light and heavy hole masses.

\subsection{Non-centrosymmetric spin-orbit couplings with the cubic symmetries}
We classify all the TR and cubic symmetry allowed $k\cdot p$
terms up to the quadratic order in $k$.
The Hamiltonian Eq. (\ref{eq:Luttinger}) includes all the inversion
invariant terms up to the $k^2$-apart from an overall constant.
Inversion breaking terms are allowed for the three cubic groups
$O, T_d, T$.
The corresponding band Hamiltonian becomes
\bea
H_0(\vec{k})=H_L(\vec{k})+\frac{\delta}{k_f} A(\vec{k}),
\label{eq:H0}
\eea
in which $A(\vec{k})=-A(-\vec{k})$ breaks inversion symmetry,
$k_f$ is the Fermi wave vector,
and $\delta$ parameterizes the inversion breaking strength.
To the lowest order in momentum, the TR invariant $A(\vec{k})$ takes the form
\bea
T_d&:& k_i\cdot (S_{i+1}S_iS_{i+1}-S_{i+2}S_iS_{i+2}),
\label{eq:A_Td}
\nn \\
O&: &  k_i\cdot S_i + a_1 k_i\cdot S_i^3,
\label{eq:A_O}
\nn \\
T&: & k_i\cdot S_i+ b_1 k_i\cdot S_i^3+ b_2 k_i\cdot (S_{i+1}S_iS_{i+1}
\nn \\
&& -S_{i+2}S_iS_{i+2}),
\label{eq:A_T}
\eea
in which $a_1,\,b_1,\,b_2$ are numerical factors, and the indices
$i$, $i+1$ are defined cyclicly for $x,y,z$ and the summation
over $i$ is assumed.
Detailed discussions are included in Appendix \ref{sec:Cubic groups}.

\subsection{Protected degeneracy along the $[0\, 0\, 1]$
direction with the $T_d$ symmetry}
\label{sect:degenercy}

The band structure of Eq. (\ref{eq:H0}) does not exhibit double degeneracy
in general due to the breaking of inversion symmetry.
Interestingly, for the $T_d$ group, there is a protected non-Kramers
degeneracy along $[0\, 0\, 1]$ and its equivalent directions
explained as follows.

\begin{figure}
\centering\epsfig{file=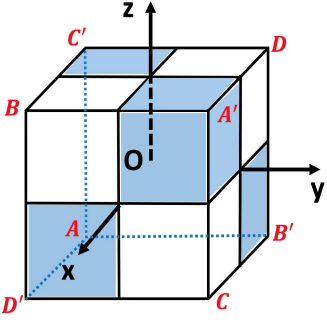,clip=0.6,
width=0.5\linewidth,height=0.5\linewidth,angle=0}
\caption{The decorated cube with the $T_d$ symmetry.
}
\label{fig:TdCube}
\end{figure}

The illustration of the $T_d$ symmetry in terms of a decorated cube
is shown in Fig. \ref{fig:TdCube}.
Its little group along the $[0\, 0\, 1]$ direction contains
two mirror reflections along the diagonal directions $y=\pm x$
denoted as $x^\prime$ and $y^\prime$, respectively.
The corresponding operations are denoted as $M_{x^{'}}$ and $M_{y^{'}}$,
respectively.
Since $M_{x^{'}}M_{y^{'}}=R(\hat{z},\pi)$, where $R(\hat n, \phi)$ denotes
the rotation around the $\hat n$ axis at the angle of $\phi$,
the little group $L_0$ of $T_d$ along the $[0\, 0\, 1]$ direction
is
\bea
L_0=\{\mathbbm{1},M_{x^{'}},M_{y^{'}},R(\hat{z},\pi)\},
\eea
which is isomorphic to the dihedral group $D_2$.
In half-odd integer spin representations, the $2\pi$-rotation
equals $-1$, and hence $L_0$ is doubled as
\bea
L_1&=&\{\mathbbm{1},M_{x^{'}},M_{y^{'}},R(\hat{z},\pi), \nn \\
&& \bar{\mathbbm{1}},\bar{\mathbbm{1}}M_{x^{'}},\bar{\mathbbm{1}}M_{y^{'}},
\bar{\mathbbm{1}}R(\hat{z},\pi)\},
\eea
in which $\bar{\mathbbm{1}}$ denotes the $2\pi$-rotation.
As discussed in Appendix \ref{sec:little group}, $L_1$ is in fact
isomorphic to the quaternion group $Q_8$.
Different from $D_2$, $Q_8$ is non-Abelian, and has four
1D irreducible representations and one 2D irreducible representation.
The half-odd integer spin representations do not contain the 1D
representation of $Q_8$ shown as follows:
\bea
M_{x^{'}}M_{y^{'}}=-M_{y^{'}}M_{x^{'}}
\eea
for half-odd integer spins, since $(M_{x^{'}}M_{y^{'}})^2=
R^2(\hat {z}, \pi)=\bar{\mathbbm{1}}$.
This anti-commutativity protects the double degeneracy along the
$[0\, 0\, 1]$ direction.

There is another mechanism of the degeneracy protection based on an
anti-unitary symmetry $\mathcal{S}$.
It is constructed based on the $T_d$ group and the TR operation
$\mathcal{T}$ as
\bea
\mathcal{S}=R(\hat{z},\frac{\pi}{2})\mathcal{I}\cdot\mathcal{T}
\label{eq:S}
\eea
where $\mathcal{I}$ is the inversion operator which flips the
momentum direction and acts as the identity operator in spin
space, and $T$ is the Kramers TR operation satisfying $\mathcal{T}^2=-1$.
$\mathcal{S}$ leaves the $[0\,0\,1]$-direction invariant.
It is easy to verify that $\mathcal{S}$'s quartic power
is $-1$, {\it i.e.},
\bea
\mathcal{S}^4=-1.
\eea
Nevertheless, unlike the Kramers operation, $S^2=-R(z,\pi)$
which remains an operator instead of a constant.
Still $\mathcal{S}$ ensures the double degeneracy of electron states
with $\vec k\parallel \hat z$ \cite{yangliu2016}, which can be
proved by contradiction.
If there was no degeneracy, each Bloch wave state $|\psi_k\rangle$
must be a simultaneous eigenstate of both the Hamiltonian and the
operator $\mathcal{S}$, then
$
\mathcal{S}|\psi\rangle =\lambda |\psi\rangle,
$
where $\lambda$ is a complex number of unit norm.
This implies that $\mathcal{S}^2 |\psi\rangle=|\lambda|^2 |\psi\rangle$
due to the anti-unitarity of $\mathcal{S}$, and then
$\mathcal{S}^4|\psi\rangle=|\psi\rangle$ which is in contradiction with
the property $\mathcal{S}^4=-1$.

Including the anti-unitary operation $S$, the little group along
the $[0\, 0\, 1]$ direction is extended from the double group $L_1$
to $SD_{16}$, the semi-dihedral group of order $16$
Detailed discussions about $SD_{16}$ are included
in Appendix \ref{sec:little group}.

Both degeneracy protection mechanisms are general beyond the $k\cdot p$
approximation.
The degeneracy is held for any band Hamiltonian with the $T_d$
and TR symmetries realized by half-integer fermions.
It even applies to the Bogoliubov excitation spectra in the
superconducting states which maintain these symmetrys.

\subsection{The YPtBi material}
\label{subsec:The_YPtBi}

We briefly review the YPtBi material and its band structure.
It is an half-Heusler compound with band inversion.
The active atomic orbitals are the $6s$ and $6p$ orbitals from the Bi atom.
The Pt and Bi atoms form a zinc-blende sublattice, and the Y atoms fill
in the lattice such that the Pt and Y atoms form another
zinc-blende sublattice \cite{Chadov2010,Lin2010,Butch2011}.
The system has the TR and $T_d$ point group symmetries, but is not
inversion symmetric.

The charge carrier density of YPtBi is very low around $2\times10^{18}
cm^{-3}$ as revealed by the Shubnikov-de Hass (SdH)
oscillation experiments \cite{Kim2016}.
The corresponding Fermi energy is on the order of $10^2K$ with the
Fermi wave vector one order smaller than the Brillouin zone boundary,
such that the $k\cdot p$ description around the $\Gamma$-point is applicable.
The $k\cdot p$ Hamiltonian $H_0(\vec{k})$ of effective spin-$\frac{3}{2}$
particles has been proposed for YPtBi \cite{Brydon2016} as a combination
of the Luttinger-Kohn Hamiltonian in Eq. (\ref{eq:Luttinger})
and an inversion breaking term given corresponding to
the $T_d$ group in Eq. (\ref{eq:A_Td}).
The inversion breaking term breaks the double degeneracy
except along $[0\, 0\, 1]$ directions, and leads to the
spin-split Fermi surfaces with the energy splitting
on the order of $10K$.
The distortion of the Fermi surfaces away from the perfect sphere
induced by the $\lambda_3$-term in Eq. (\ref{eq:Luttinger})
is shown to be a small effect as revealed by the SdH oscillation
experiments \cite{Kim2016}, which will be neglected in later calculations.


\section{Non-centrosymmetric spin-$\frac{3}{2}$ Cooper pairings}
\label{sec:noncentro_pairing}
In this section, we first discuss the non-centrosymmetric
Cooper pairings under cubic group symmetries.
Then the possible pairing symmetries of YPtBi are briefly reviewed.

\subsection{Non-centrosymmetric Cooper pairings with cubic symmetries}
\label{subsec:lattice analogues}

We discuss the Cooper pairings in non-centrosymmetric sytsems with
the TR and cubic group symmetries, which can be viewed as
analogues of the $^3$He-$B$ pairing in the lattice.
The generalization of the $^3$He-$B$ pairing to the large spin and
high partial-wave channels in continuum has been studied in
Ref. [\onlinecite{Yang2016}].

The Bogoliubov-de Gennes (B-deG) Hamiltonian of a spin-$\frac{3}{2}$
superconductor is
\bea
H_{\text{B-deG}}= \sum_{\vec{k}}^{'} (c^{\dagger}(\vec{k}),
c(-\vec{k})^T ) H(\vec{k}) \left(\begin{array}{c}
c(\vec{k})\\
c^{\dagger}(-\vec{k})
\end{array}\right),
\eea
in which $\sum_{\vec{k}}^{'}$ denotes summing over half of momentum
space, and  $c({\vec{k}})=(c_{3/2}(\vec{k}), c_{1/2}(\vec{k}),c_{-1/2}(\vec{k}),c_{-3/2}(\vec{k}))^T$.
The matrix kernel $H(\vec{k})$ is represented as
\bea
H(\vec{k})=\left( \begin{array}{cc}
H_0(\vec{k})-\mu & \Delta(\vec{k})  \\
\Delta^{\dagger} (\vec{k}) & -(H_0(-\vec{k})-\mu)^{T}
\end{array}\right),
\label{eq:Hamiltonian}
\eea
in which $H_0(\vec{k})$ is the band structure given by Eq. (\ref{eq:H0}),
and $\mu$ is the chemical potential measured from the $\Gamma$
point of the $p_{3/2}$-bands.
The pairing term $\Delta(\vec{k})$ is given as
\bea
\Delta(\vec{k})=K(\vec{k}) R,
\label{eq:pairing}
\eea
in which $K(\vec{k})$ denotes the pairing kernel, and $R$ is the charge conjugation matrix defined as $R_{\alpha\beta}=
(-)^{\alpha+\frac{1}{2}}\delta_{\alpha, -\beta}$
with $\alpha$ spin indices \cite{Wu2006}.
For non-centrosymmetric systems, the breaking of the inversion symmetry
mixes pairings with different parities.
The pairing kernel $K(\vec{k})$ has been proposed to
take the form\cite{Frigeri2004}
\bea
K(\vec{k})=\Delta_s+\frac{\Delta_p}{k_f} A(\vec{k}),
\label{eq:K_k}
\eea
in which $A(\vec{k})$ is given by Eq. (\ref{eq:A_T}) for
the cubic groups $O,T_d,T$; $\Delta_s$ and $\Delta_p$
parameterize the strength of the $s$ and $p$-wave components,
respectively.
Such pairing avoids the pair breaking effect induced by the
inversion breaking term $\frac{\delta}{k_f} A(\vec{k})$
in the band structure \cite{Frigeri2004}, and thus is conceivably
to be energetically favorable.
The pairing only takes place between electrons from
the same spin-split Fermi surface.
Nevertheless, we would like to emphasize that the actual superconducting gap symmetry of the YPtBi material is still undetermined, and this is only one possibility that needs to be tested in future experiments.

\subsection{The superconducting properties of the YPtBi material}

\begin{figure}
\centering\epsfig{file=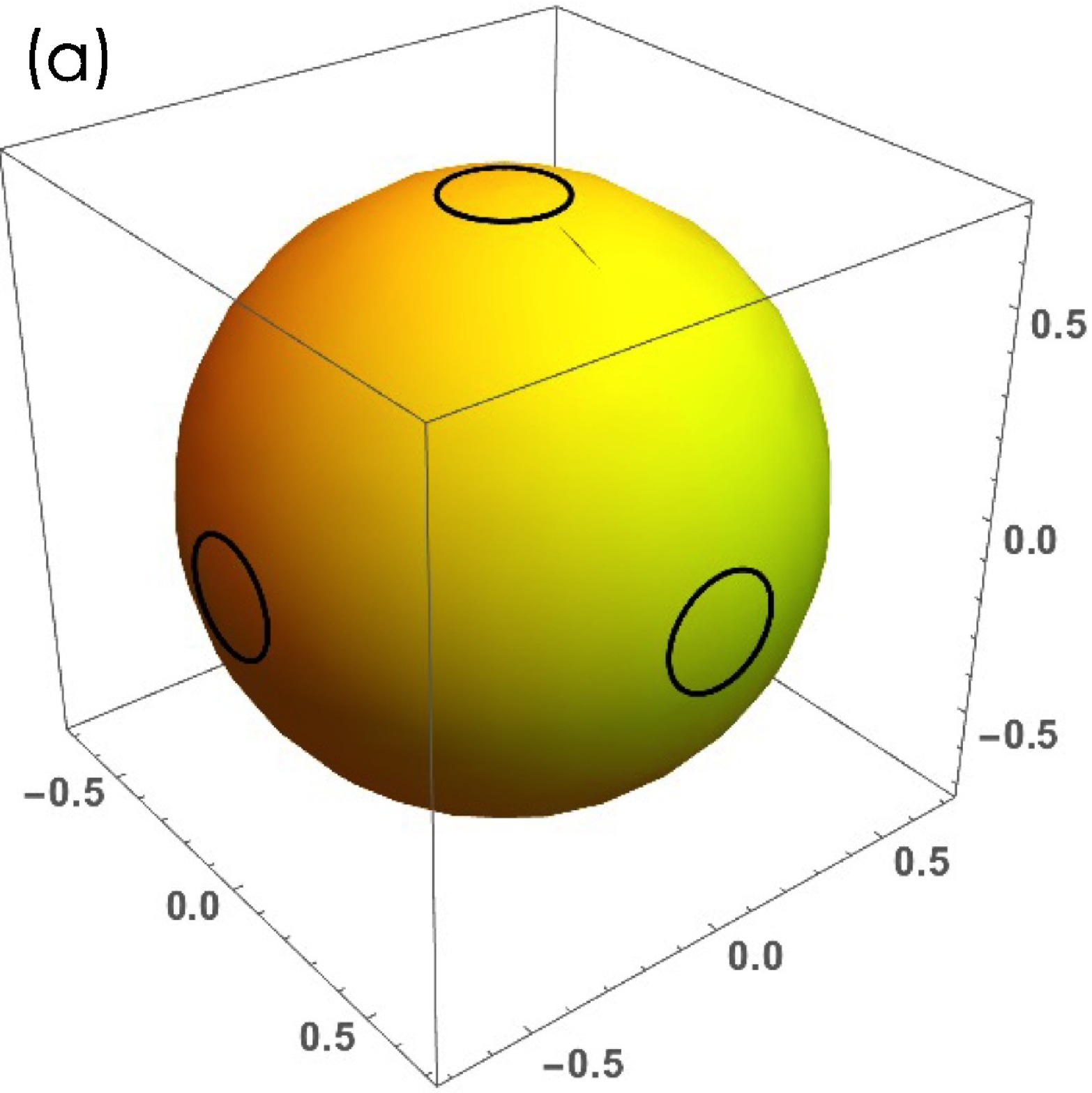,clip=0.45,
width=0.45\linewidth,height=0.45\linewidth,angle=0}
\centering\epsfig{file=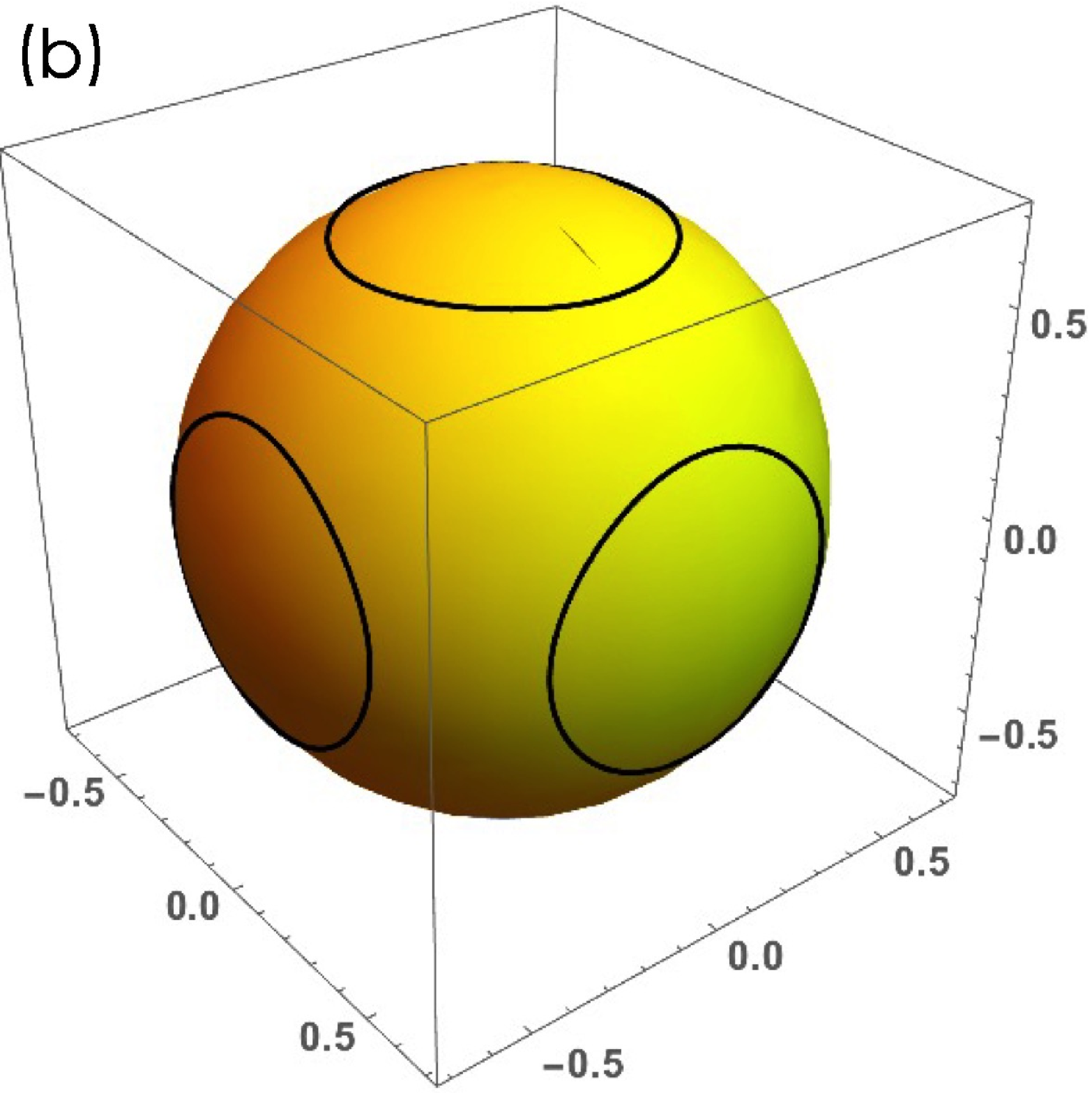,clip=0.45,
width=0.45\linewidth,height=0.45\linewidth,angle=0}
\centering\epsfig{file=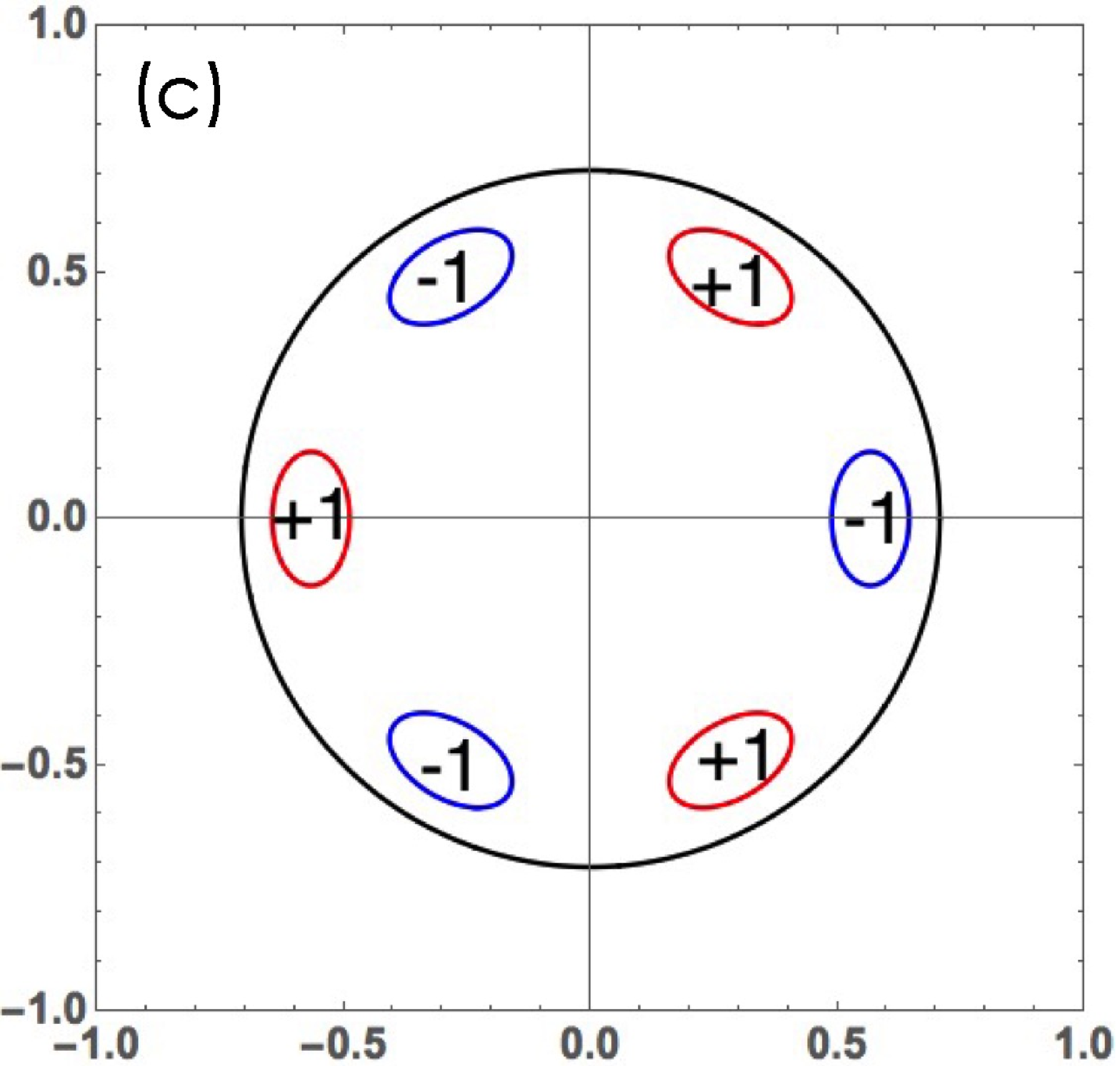,clip=0.45,
width=0.45\linewidth,height=0.45\linewidth,angle=0}
\centering\epsfig{file=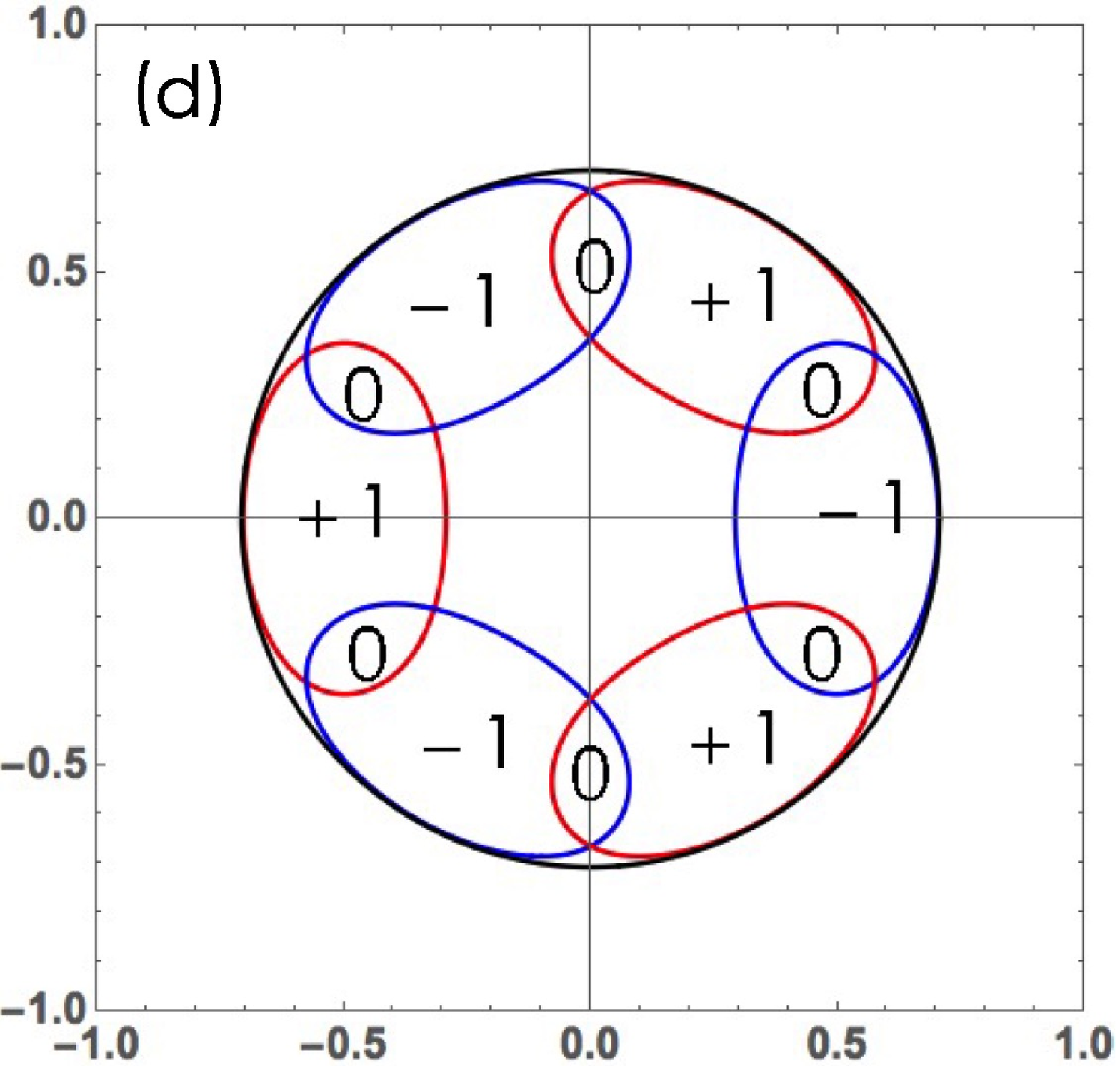,clip=0.45,
width=0.45\linewidth,height=0.45\linewidth,angle=0}
\caption{Nodal loops of the superconducting gap function on the larger
spin-split Fermi surface for $a)$ $\Delta_s/\Delta_p=0.3$ and
$b)$ $\Delta_s/\Delta_p=0.7$,
and projections of the nodal loops to the $(1\,1\,1)$-surface for $c)$ $\Delta_s/\Delta_p=0.3$ and $d)$ $\Delta_s/\Delta_p=0.7$.
In (c) the topological numbers are $+1$ and $-1$ in the regions
enclosed by the red loops and the blue loops, respectively,
and are zero outside the loops.
In (d) the situation is similar except that the topological
numbers are zero in the overlapping regions of the loops.
The three axes in $a)$, $b)$ and the two axes in $c)$, $d)$
are momenta in the bulk Brillouin zone and surface Brillouin
zone, respectively, in which the momenta are measured in
the unit of $\sqrt{2}k_f$.
The Luttinger parameters are taken as $\lambda_1=0$ and
$\lambda_2=|\mu|/(2k_f^2)$.
}
\label{fig:nodal FS}
\end{figure}

In this part we give a brief review to the superconducting properties
of the YPtBi material, particularly its topological nodal line
structure in the gap function.
Its transition temperature is $T_c=0.78K$.
Its London penetration depth exhibits a linear temperature
dependence \cite{Kim2016}, which is a strong evidence for a
gap function with nodal lines.

Let us first consider the band eigenstates of $H_0$ defined in
Eq. (\ref{eq:H0}) with $A$ defined as
\bea
A(\vec{k})=\sum_{i=x,y,z} k_i (S_{i+1}S_iS_{i+1}-S_{i+2}S_iS_{i+2}),
\eea
which maintains the TR and $T_d$ symmetries.
At $|\delta|\ll |\mu|$, the main effect of the inversion symmetry
breaking term is within the heavy hole band and the light hole
one to split the double degeneracy.
The mixing between heavy and light hole bands is small and
will be neglected.
Since the Fermi surface cuts the heavy hole band, we project the $4\times 4$
matrix kernel $A(\vec k)$ into the heavy hole bands.
Then $A(\vec k)$ becomes a $2\times 2$ matrix $\vec{\Lambda}(\vec{k})
\cdot \vec{\tau}$  where $\vec{\tau}$ is the Pauli matrices under
the basis with helicities $\pm \frac{3}{2}$, and
\bea
\Lambda_x(\vec k)&=&-\frac{9k}{8} \sin 2\theta_k \cos 2\phi_k, \nn \\
\Lambda_y(\vec k)&=&\frac{9k}{16} (3+\cos 2\theta_k)\sin\theta_k
\sin 2\phi_k, \nn \\
\Lambda_z(\vec k)&=&0,
\eea
with $\theta_k$ and $\phi_k$ the polar and azimuthal angles of $\vec{k}$,
respectively.
Then the heavy hole energies exhibit the splitting of $\delta/k_f |
\Lambda(\vec k)|$ which becomes zero along the $[0\, 0\, 1]$
and its equivalent directions.

It has been proposed for YPtBi that its pairing symmetry is likely
of the mixed $s$-wave singlet and $p$-wave septet \cite{Brydon2016}.
The B-deG Hamiltonian is given by Eq. (\ref{eq:Hamiltonian}),
which maintains the TR and $T_d$ symmetries.
The Bogoliubov quasi-particle spectrum exhibits line nodes centering
around $[0\, 0\, 1]$ directions when $\Delta_s/\Delta_p
\neq 0$ \cite{Brydon2016}.
Based on a similar basis as above, when projected into the two
spin-split Fermi surfaces belonging to the heavy holes, the
diagonalized gap functions becomes $\Delta_s \pm  \Delta_p
|\vec{\Lambda}(\vec{k})|$.
Hence, the gap function becomes nodal on one of the spin-split
Fermi surfaces at $|\vec{\Lambda}(\vec{k})|=\Delta_s/\Delta_p$,
and the node lines form closed loops centering around $[0 \,0\,1]$
and its equivalent directions.
The plots of the nodal loops on the corresponding Fermi surface
are shown in Fig. \ref{fig:nodal FS} ($a$,$b$) for two representative
ratios between $\Delta_s$ and $\Delta_p$, for which the nodal
loops do not cross, or, cross, respectively.

The nodal loops centering around $[0\, 0\, 1]$ and its equivalent
directions are topologically non-trivial \cite{Kim2016,Schnyder2012}.
An integer-valued index can be assigned to each nodal loop as the
topological number of a closed loop in momentum space linked with
the nodal one on the Fermi surface \cite{Schnyder2012}.
For a non-trivial nodal loop, the gap functions
located inside and outside the nodal loop on the Fermi surface
are with opposite signs.
Now consider a 2D surface and the associated surface Brillouin zone,
the projection of the nodal loop in the 2D Brillouin zone becomes
a 2D planar loop.
For each momentum in the surface Brilliouin zone, it represents
an effective 1D system perpendicular to the surface.
Hence, a topological number can be defined for each 2D surface momentum
except that on the nodes \cite{Schnyder2012}.
The gap functions of the incident wave and the reflection wave
perpendicular to the surface change sign if their common
2D momentum is inside the planar projection loop.
Hence, for the surface states, their topological numbers are
non-trivial for those inside the 2D projection loop but trivial
for those outside.
But for those in the overlapping regions between two projection loops,
they become trivial again.
The schematic plot for the case of the $[1\,1\,1]$-surface is taken
as an example for illustration, as shown in Fig. \ref{fig:nodal FS}
($c$, $d$) for two representative ratios between $\Delta_s$ and $\Delta_p$.
Detailed discussions of the topological properties of the
nodal loops are included in Appendix \ref{sec:topo}.

\section{Majorana surface states}
\label{sec:Majorana septet}

In this section, we first derive the equation solving the Majorana
surface states for the spin-$\frac{3}{2}$ systems.
The corresponding calculations were performed for the fully gapped
isotropic $p$-wave triplet and $f$-wave septet for the continuum model
before \cite{Yang2016}.
Here we consider the gapless mixed parity pairing state explained
in the last section.
For simplicity, we set $\lambda_3=0$ in the band Hamiltonian,
and consider the limit of $\Delta_s,\Delta_p<< \delta<< |\mu|$.
This limit is justified in YPtBi, since $\Delta_s,\Delta_p \sim 1K$,
$\delta \sim 10K$, and $\mu \sim 10^2K$.
The equation is then applied to obtain the Majorana surface modes
for systems with TR and $T_d$ symmetries and also for those with the
$T$, or, $O$ symmetry.

\subsection{The methodology}
Consider a surface with the normal direction along
$\hat{n}=(\sin \theta_n \cos \phi_n,\,
\sin \theta_n \sin \phi_n, \cos\theta_n)$.
To simplify calculations, we rotate $\hat n$ into the $z$-axis.
The bulk of the system lies at $z<0$, and the side of
$z>0$ is vacuum.
The boundary condition is that the wavefunction vanishes at $z=0$
and exponentially decays to zero when $z\rightarrow -\infty$.
Within the rotated coordinates, the Luttinger Hamiltonian, the inversion
breaking term in band structure, and the pairing Hamiltonian
are expressed as follows,
\bea
H_L(-i \vec \nabla)&=&-(\lambda_1+\frac{5}{2} \lambda_2) \nabla^2
- 2\lambda_2 \left(R_n (-i \vec \nabla)\cdot \vec{S}\right)^2 \nn \\
&-& \mu, \nn \\
H_A(-i \vec \nabla )&=& \frac{\delta}{k_f} A\left(R_n (-i \vec \nabla)
\right), \nn \\
\Delta(-i \vec \nabla)&=& K\left(R_n (-i\vec \nabla)\right) R,
\eea
where $R_n=R(\hat{z}, \phi_n) R(\hat{y},\theta_n)$ is the rotation
operation transforming $\hat{z}$ to $\hat{n}$,
$R_n(-i\vec{\nabla})$ represents the rotation of the vector
operator $-\vec i\nabla$, i.e,
\bea
R_{n,a}(-i\vec {\nabla})= -i R_{n,ab} \nabla_b,
\eea
where $a=x,y,z$, and $R_{n,ab}$ is the $3\times 3$ rotation matrix.
The momenta $k_x$ and $k_y$ parallel to the surface plane remain
conserved.

We use the trial plane wavefunction as
\bea
\Psi(\vec{r}) =\sum_l C_l \Phi_l e^{i \vec{k}_l \cdot \vec{r}},
\label{eq:wavefunction}
\eea
where $\Phi_l$ is an eight-dimensional column vector with four spin
components of both particle and hole degrees of freedom.
Plug Eq. (\ref{eq:wavefunction}) into the eigen-equation, we obtain
\bea
\left( \begin{array}{cc}
H_0(R_a \vec{k}_l)-\mu& \Delta(R_a \vec{k}_l)  \\
\Delta^{\dagger} (R_a \vec{k}_l) & -(H_0(-R_a\vec{k}_l)-\mu)^{T}
\end{array}\right)
\Phi_l=E_s \Phi_l,\nn \\
\label{eq:boundary1}
\eea
where $\vec{k}_l=(k_x\,k_y\,k_{zl})$, and $E_s$ is the surface state
energy.
The boundary condition requires that $\mbox{Im} k_{zl}<0$,
and there are eight solutions of $k_{zl}$ satisfying this
condition, i.e., $1\le l \le 8$:
Four come from the sector of helicity eigenvalues $\pm\frac{3}{2}$
and the other four from helicity eigenvalues $\pm \frac{1}{2}$ .
Furthermore, the boundary conditions require that
\bea
\sum_{l=1}^8 C_l \Phi_l=0.
\eea
In order to have non-trivial solutions of $C_l$, the determinant
composed of the eight column vectors needs to be zero, {\it i.e.},
\bea
\det\Big( \{\Phi_l\}_{1\leq l \leq 8} \Big)=0,
\eea
which determines the surface state eigen-energies.
The explicit forms of the eight column vectors $\Phi_l$'s
in the limit of $\Delta_s,\Delta_p<< \delta<< |\mu|$
are derived in Appendix \ref{sec:solve Majorana}.

\subsection{Majorana zero modes under the $T_d$ symmetry}
\label{subsec:Majorana Td}

\begin{figure}
\centering\epsfig{file=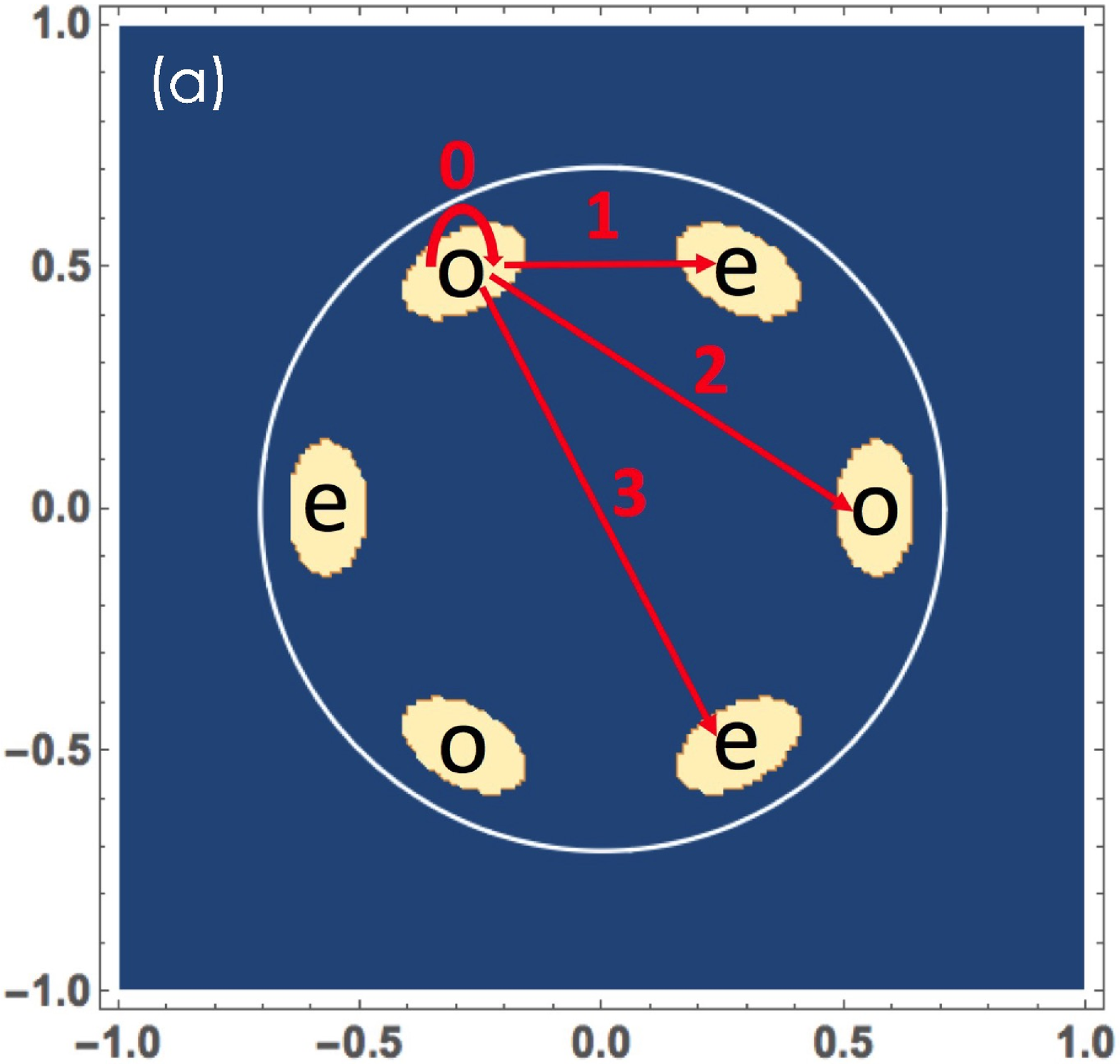,clip=0.45,
width=0.45\linewidth,height=0.45\linewidth,angle=0}
\centering\epsfig{file=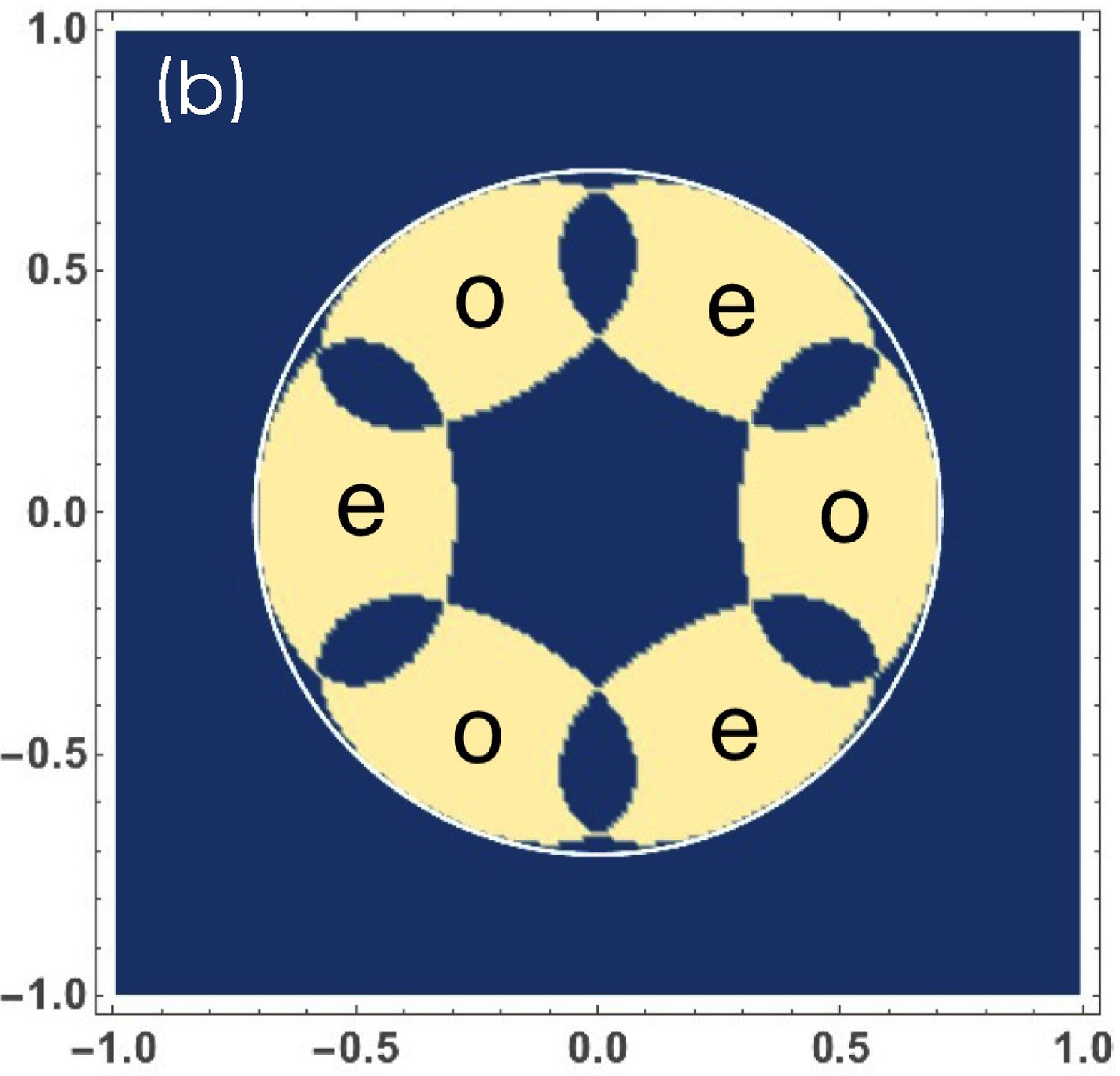,clip=0.45,
width=0.45\linewidth,height=0.45\linewidth,angle=0}
\caption{Distribution of Majorana zero modes in the surface Brillouin zone in the $(1\,1\,1)$-surface for $a)$ $\Delta_s/\Delta_p=0.3$ and $b)$ $\Delta_s/\Delta_p=0.7$.
The white circle denotes the boundary of the projection of the Fermi surface.
Majorana zero modes exist in the bright regions.
The chiral index is marked for each island of the Majorana flat band,
where ``{\it e}" and ``{\it o}" represent ``even" and ``odd" with
the chiral eigenvalue $\pm 1$, respectively.
The horizontal and vertical axes are momenta in the surface Brillouin
zone measured in the unit of $\sqrt{2} k_f$.
The Luttinger parameters are taken as $\lambda_1=0$ and $\lambda_2=|\mu|/(2k_f^2)$.
The numerical computations are carried out for a $200\times 200$
lattice in momentum space.
}
\label{fig:surface spectrum}
\end{figure}

In this part, we solve for the zero energy Majorana surface modes.
Let us consider the $(0\,0\,1)$, and $(1\, 1\, 0)$-surfaces.
From the gap nodal loop configurations shown in Fig. \ref{fig:nodal FS},
the nodal loop projections fully overlap with each other, such that
all the momentum-dependent topological numbers are trivial.
Hence we will only consider the $(1\,1\,1)$-surface.
Through out the calculations the Luttinger parameters are
taken as $\lambda_1=0$, $\lambda_2=|\mu|/(2k_f^2)$ and $\lambda_3=0$.

The results of Majorana spectra on the $(1\,1\,1)$-surface are presented in
Fig. \ref{fig:surface spectrum}.
Fig. \ref{fig:surface spectrum} ($a$) shows the case that the projections
of the gap nodal loops do not overlap, and the surface zero Majorana
modes appear inside the projection loops.
In Fig. \ref{fig:surface spectrum} ($b$), the projections overlap,
and the zero Majorana modes in the overlap region disappear.
The former corresponds to a smaller value of $\Delta_s/\Delta_p=0.3$,
and the latter is with a larger one $\Delta_s/\Delta_p=0.7$.
Since each non-trivial momentum dependent topological index equals
either 1 or -1 as shown in Fig. \ref{fig:nodal FS}, these surface
zero Majorana modes are non-degenerate.
The surface spectra solved from matching boundary conditions are
consistent with the previous analysis based on bulk topological number.
This can be seen as a verification of the bulk-edge correspondence
principle in the spin-$3/2$ situation.

We then analyze the symmetry properties of the Majorana surface states.
Based on the $T_d$ and the TR symmetry, the symmetry subgroup for
the $(1\,1\,1)$-surface is $C_{3v}\times \{1,\mathcal{T}\}$,
where $\mathcal{T}$ is the TR operation defined as $\mathcal{T}c^{\dagger}(\vec{k})\mathcal{T}^{-1}
=c^{\dagger}(\vec{k})R\cdot K$, with $K$ the complex
conjugate operation.
The surface spectra in Fig. \ref{fig:surface spectrum} exhibit
the $C_{3v}$-symmetry.
The TR operation reverses the momentum direction, and the spectra
are also invariant under the TR operation.
Consider the particle-hole operation $P_H$ defined as
$P_H c^{\dagger}_\alpha (\vec{k}) P_H^{-1}=c_\alpha(\vec{k})K$.
$P_H$ anti-commutes with B-deG Hamiltonian Eq. (\ref{eq:Hamiltonian})
and transforms a state to another state with the opposite energy.
Since Majorana surface modes are at zero energy, they are
particle-hole symmetric.

We can also define a chiral operator as
\bea
C_{ch}=i\mathcal{T} P_H.
\label{eq:chiral}
\eea
It is Hermitian in half-odd integer spin spaces:
$C_{ch}^{\dagger}=-i P_H^{\dagger} \mathcal{T}^{\dagger}=C_{ch}$,
since $\mathcal{T}^{\dagger}=\mathcal{T}^{-1}=-\mathcal{T}$
(as $\mathcal{T}^2=-1$),
$P_H^{\dagger}=P_H$, and $[\mathcal{T},P_H]=0$.
Both TR and particle-hole operations reverse the sign of the momentum,
hence, $C_{ch}$ maintains momentum invariant.
This implies that each Majorana zero mode solved above can be chosen
as a chiral eigen-state with a chiral index of $\pm 1$.
Since $C_{ch}$ commutes with the $C_{3v}$ group, the islands related
by the $C_{3v}$ group carry the same chiral index.
While the two islands related by TR operation carry opposite
chiral indices, because $C_{ch}$ anti-commutes with the TR operator.

\subsection{The $T$ and $O$ groups}

\begin{figure}
\centering\epsfig{file=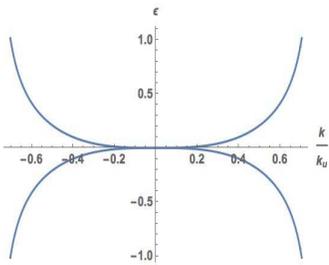,clip=0.4,
width=0.5\linewidth,height=0.4\linewidth,angle=0}
\caption{Surface spectrum of $p$-wave triplet pairing along radial direction.
The horizontal axis is the momentum measured in the unit of $k_u=\sqrt{2}k_f$,
and vertical axis is the surface energy measured in  the unit of $\Delta_p$.
}
\label{fig:tripletspectra}
\end{figure}

In this part we briefly present the surface spectra calculation
of the $p$-wave triplet pairing for point groups $O$ and $T$.
While the situation of the $p$-wave triplet pairing with band inversion
and the Fermi energy tuned to cross the heavy hole bands has been
sketched based on physical intuitions \cite{Fang2015}, here, we
perform a detailed calculation.
The band structure and the Cooper pairing are given by Eq. (\ref{eq:H0})
and Eq. (\ref{eq:pairing}), respectively, in which $A(\vec{k})$ takes
the corresponding form for $O$ and $T$ groups in Eq. (\ref{eq:A_T}).
Here we consider the $p$-wave triplet dominant pairing, i.e., $A(\vec{k})=k_i\cdot S_i$. 
Unlike the case of $p$-wave septet pairing under the $T_d$ symmetry, the $p$-wave triplet pairing under the $T$ and $O$ point groups is $SO(3)$ invariant under the combined orbital and spin rotations. 
Hence, the quasi-particle spectrum is fully gapped on the Fermi surface.
For small enough parameters $a_1,\,b_1,\,b_2$ in Eq. \ref{eq:A_T}
and the ratio $\Delta_s/\Delta_p$ in Eq. (\ref{eq:K_k}), they can
be set to zero without affecting topological properties since
the pairing is fully gapped.
In this case, the B-deG Hamiltonians for the $T$ and $O$ point
group symmetries are the same.
The system belongs to the $D$III class due to the TR and particle-hole
symmetries.
The bulk topological number is $3$ as a sum of those from the helicity
$\pm\frac{3}{2}$ bands \cite{Yang2016}.
Fig. 5 shows the surface spectrum along the radial direction in the polar coordinate within the surface 
Brillouin zone. 
The full spectrum can be obtained by performing a rotation around the normal direction.
The overall surface spectrum exhibit the cubic Dirac dispersions
\cite{Yang2016,Fang2015}, which is consistent with
the bulk topological number.

\section{Quasi-particle interference patterns}
\label{sec:qpi}
The scanning tunneling spectroscopy measures the local density of
states (LDOS) on the sample surface.
In the presence of an impurity, the LDOS exhibits interference pattern
due to the scattering of electrons by impurities.
The quasi-particle interference (QPI) patterns provide information
to the pairing symmetry of high-$T_c$ superconductors \cite{Wang2003},
orbital ordering in $2D$ materials \cite{Lee2009}, surface states
of topological insulators \cite{Lee2009a} and semi-metals \cite{Kourtis2016}.
The QPI patterns have also been discussed for various spin-$1/2$
non-centrosymmetric  superconductors \cite{Hofmann2013}.

In this section, we use the wavefuntions of the surface states solved from Sec. \ref{subsec:Majorana Td} to compute the QPI patterns of a single impurity
on the surface of the spin-$\frac{3}{2}$ topological superconductor with
the TR and $T_d$ symmetries.
Experiments on QPI patterns provide test to the proposed
pairing form in Eq. (\ref{eq:K_k}).

\subsection{Spin-resolved local density of states}
The impurity can be either magnetic or non-magnetic, and the tunneling
spectroscopy can be either spin-resolved or non-spin-resolved.
In consideration of these, the LDOS tensor $\rho^{\mu\nu}$ is defined
with $\mu,\nu=0,1,2,3,4$, whose $(\mu, \nu)$ element is the spin-resolved
LDOS for spin in $\mu$-direction with impurity spin polarized in
$\nu$-direction.
The cases of $\mu=0$ and $\nu=0$ correspond to non-spin-resolved
and non-magnetic impurity, respectively.
To facilitate the symmetry analysis, we rotate the coordinate frame
to $\hat{x}^{'}=\frac{1}{\sqrt{6}}(1,1,-2)$, $\hat{y}^{'}=\frac{1}{\sqrt{2}}(-1,1,0)$,
$\hat{z}^{'}=\frac{1}{\sqrt{3}}(1,1,1)$,
The little group of $T_d$ along the $(1\,1\,1)$-direction is $C_{3v}$.
The three-fold rotation axis in $C_{3v}$ is the $\hat{z}^{'}$-axis,
and the three vertical reflection planes are the $x^{'}z^{'}$-plane
and the $\pm \frac{2\pi}{3}$ rotations of the $x^{'}z^{'}$-plane
around $z^{'}$-axis.
To simply notation, below we still use $x,y,z$ to represent
$x^\prime$, $y^\prime$ and $z^\prime$, i.e., to suppress
the $'$ symbol, and use the Greek index $\nu=1,2,3$
to represent them, respectively.

Let us comment on the modeling of the impurity potential in calculations
below.
For simplicity, the non-magnetic impurity potential $V_{\text{imp}}$ is
used as an example, which can be straightforwardly extended to the magnetic
impurity potential.
Typically, it is taken a short range $\delta$-potential in real space.
However, due to the open boundary condition, the surface state wavefunctions
vanish on the surface.
If the impurity is located exactly on the surface, then it will not
cause any scattering.
This artifact can be cured by using a more realistic model to take into
account the finite range of the impurity potential.
Hence, we assume the following form of $V_{\text{imp}}$,
\bea
V_{\text{imp}}(\vec{r})= N_a V_0 e^{-z/a_0} \delta(x)\delta(y),
\label{eq:Himp}
\eea
in which $V_0$ parameterizes the potential strength and $N_a$ is a normalization factor.
Although the detailed distribution of QPI patterns depends on the explicit form of $V_{\text{imp}}$,
the overall characteristic features should not be sensitive on the choice of $V_{\text{imp}}$.


For an impurity whose spin is polarized in $\nu$-direction
with potential $V_{\text{imp}}(\vec{r})$, or, a non-magnetic impurity
for $\nu=0$, the retarded Green's function $\mathcal{G}^{\nu}_R(\omega,\vec{r})$
of the frequency $\omega$ at the position $\vec{r}$ is defined by
\bea
\mathcal{G}^{\nu}_R(\omega,\vec{r})= \bra{\vec{r}} \frac{1}{\omega- (H+H^{\nu}_{\text{imp}})+i\epsilon } \ket{\vec{r}},
\label{eq:retarded}
\eea
in which $\ket{\vec{r}}$ represents the coordinate eigen-state
located at $\vec r$.
In principle, $\mathcal{G}_R^\nu$ exhibits a matrix structure with
respect to all the spin, Nambu, coordinate, and frequency indices.
For $\mathcal{G}^{\nu}_R(\omega,\vec{r})$, it takes the diagonal
elements in terms of coordinate and frequency, and leave the spin
and Nambu indices general.
$H$ is the matrix-kernel of the B-deG Hamiltonian without impurity
in Eq. (\ref{eq:Hamiltonian}) expressed in the coordinate
representation.
$H^\nu_{\text{imp}}$ is the impurity potential in the Nambu representation,
defined as
\bea
H^{\nu}_{\text{imp}}= V_{\text{imp}}(\vec{r}) \Sigma^{\nu},
\eea
where $\Sigma^\nu=\tau_3\otimes S^\nu$ for $\nu=0,1,3$,
and $\Sigma^2=\tau_0\otimes S^2$.
$\tau_3$ is the Pauli matrix and $\tau_0$ is the
identity matrix acting in Nambu space, and $S^{0}$
is the identity matrix acting in spin space.

We define the 3D local density of states (LDOS) tensor as
\bea
\rho^{\mu\nu}(\omega,\vec{r})=-\frac{1}{2\pi} \text{Im}
\text{Tr}\Big((1+\tau_3) \Sigma^{\mu} ~\mathcal{G}^{\nu}_R (\omega,\vec{r})
\Big ),
\label{eq:def_LDOS}
\eea
in which $1+\tau_3$ is the abbreviation of $(I_2+\tau_3)\otimes I_4$.
$\rho^{\mu\nu}(\omega, \vec r)$ refers to the density distribution
in the presence of the magnetic or non-magnetic impurity for $\mu=0$,
and the spin-density distribution at polarization $\mu$ at $\mu\neq 0$.
Then the surface LDOS tensor $\rho_{sf}^{\mu\nu}(\omega,\vec{r_\parallel})$
for the position $\vec{r}_{\parallel}$ on the surface is defined by
\bea
\rho_{sf}^{\mu\nu}(\omega,\vec{r}_{\parallel})=
\int d\vec{r} \mathcal{F}(\vec{r}-\vec{r}_c)
\rho^{\mu\nu}(\omega,\vec{r}),
\label{eq:rho_F}
\eea
where $\vec{r}_c=(\vec{r}_{\parallel},0)$ is the 3D coordinate,
and $\mathcal{F} (\vec{r}-\vec r_c)$ is an envelop function
describing the sensitivity of the STM tip to the local density of
state distribution along the $z$-axis.
Here, we do not use the $\delta$-function along the $z$-direction
either, due to the open boundary condition used for the
calculation of surface states.
Instead, a Gaussian distribution envelop function is used
\bea
\mathcal{F} (\vec{r})=N_b e^{-(\frac{z}{b_0})^2} \delta(x)\delta(y),
\label{eq:envelop}
\eea
in which $N_b$ is the normalization factor.
The characteristic features of the LODS should not be sensitive
to the detailed form of the function $\mathcal{F}$.

Subtracting the background contribution in the absence of the impurity  from  $\rho_{sf}^{\mu\nu}(\omega,\vec{r}_{\parallel})$,
we extract the impurity contribution to the LDOS defined as
\begin{flalign}
\Delta \rho_{sf}^{\mu\nu}(\omega,\vec{r}_{\parallel})&=
-\frac{1}{2\pi i}\int_{-\infty}^{0} dz \mathcal{F}(z)
\text{Tr} \Big[(1+\tau_3) \Sigma^{\mu} \nn \\ &\Big(\Delta\mathcal{G}_R^{\nu}(\omega,\vec{r}_{\parallel},z)
-\Delta\mathcal{G}^{\nu*}_R(\omega,\vec{r}_{\parallel},z)\Big)
\Big ],
\end{flalign}
where
\bea
\Delta\mathcal{G}_R^{\nu}(\omega,\vec{r})
=\mathcal{G}_R^{\nu}(\omega,\vec{r})-\mathcal{G}_{R}^{(0)}(\omega,\vec{r}),
\eea
with
\bea
\mathcal{G}_{R}^{(0)}(\omega,\vec{r})
=\bra{\vec{r}} \frac{1}{\omega- H+i\epsilon } \ket{\vec{r}}.
\eea

The QPI pattern in momentum space $\Delta \rho_{sf}^{\alpha\beta}
(\omega,\vec{q})$ is defined to be the Fourier transform
of $\Delta\rho_{sf}^{\alpha\beta}(\omega,\vec{r}_{\parallel})$
with respect to $\vec{r}_{\parallel}$, as
\bea
\Delta \rho_{sf}^{\mu\nu} (\omega,\vec{q})&=&
-\frac{1}{2\pi i}\int_{-\infty}^{0} dz \mathcal{F}(z)
\big(\Lambda^{\mu\nu}(\omega,\vec{q},z) \nn \\
&-&\Lambda^{\mu\nu*}(\omega,-\vec{q},z)\big),
\label{eq:rho_Lambda}
\eea
in which $\Lambda^{\mu\nu}(\omega,\vec{q},z)$ is the Fourier
transform of $\text{Tr} [(1+\tau_3)\Sigma^{\mu} \Delta\mathcal{G}_R^{\nu}(\omega,\vec{r}_{\parallel},z)]$.

\vspace{2mm}
\subsection{The $T$-matrix formalism and Born approximation}

The retarded Green's function $\mathcal{G}_{R} (\omega,\vec{r}_{\parallel},z)$
can be evaluated using the T-matrix formalism.
Define the operator $\mathcal{G}_R(\omega)$ as
\bea
\mathcal{G}_R(\omega)=\frac{1}{\omega-H-H_{\text{imp}} +i\epsilon},
\eea
and that in the absence of impurity $\mathcal{G}^{(0)}_R(\omega)$  as
\bea
\mathcal{G}_R^{(0)}(\omega)=\frac{1}{\omega-H +i\epsilon}.
\eea
$\mathcal{G}_R(\omega)$ can be solved through
\bea
\mathcal{G}_R(\omega)=\mathcal{G}^{(0)}_R(\omega)
+\mathcal{G}^{(0)}_R(\omega) T(\omega) \mathcal{G}^{(0)}_R(\omega),
\label{eq:G_Tmatrix}
\eea
in which the $T$-matrix operator $T(\omega)$ satisfies the equation
\bea
T(\omega)=H_{\text{imp}}+H_{\text{imp}} \mathcal{G}^{(0)}_R(\omega) T(\omega).
\eea
The Born approximation will be performed to solve the $T$-matrix.

Now we outline the procedure of calculating $\Delta \rho^{\mu\nu}_{sf}$
based on  Eq. \ref{eq:rho_Lambda}.
First, $\Delta \mathcal{G}^{\nu}_R(\omega,\vec{q},z)$ can be evaluated
by inserting the complete basis to both the left and right hand sides of
$T(\omega)$ in Eq. (\ref{eq:G_Tmatrix}).
A general eigenstate of the Hamiltonian $H$ is of the form
$\frac{1}{L} e^{i \vec{k}_{\parallel}\cdot \vec{r}_{\parallel}} \Psi_{\vec{k}_{\parallel},\alpha}(z)$,
where $L$ stands for the average inter-impurity distance,
and the index $\alpha$ labels the states
with fixed in-surface momentum $\vec{k}_{\parallel}$,
which can be either scattering state or surface state.
$\Psi_{\vec{k}_{\parallel},\alpha}(z)$ is the $8$-component
normalized wavefunction in $z$-direction.
After carrying out the Fourier transform, we obtain
\bea
\Lambda^{\mu\nu}(\omega,\vec{q},z)&&=
\frac{N_a}{L^2}\sum_{\vec{k}_{\parallel}\alpha,\vec{k}^{'}_{\parallel}\beta} \delta_{\vec{k}_{\parallel}-\vec{k}^{'}_{\parallel},\vec{q}}
\frac{1}{\omega-E_{\vec{k}_{\parallel},\alpha}+i\epsilon}
\nn \\
&\times&\bra {\Psi_{\vec{k}_{\parallel},\alpha}} H_{\text{imp,z}}^{\nu} \ket{\Psi_{\vec{k}^{'}_{\parallel},\beta}}  \frac{1}{\omega-E_{\vec{k}^{'}_{\parallel},\beta}+i\epsilon}
\nn \\
&\times&
\text{Tr} \Big [ (1+\tau_3)\Sigma^{\mu}  \Psi_{\vec{k}_{\parallel},\alpha} (z)
\Psi^{\dagger}_{\vec{k}^{'}_{\parallel},\beta} (z)  \Big ],
\label{eq:Lambda}
\eea
in which $E_{\vec{k}_{\parallel},\alpha}$ is the energy of the wavefunction $\Psi_{\vec{k}_{\parallel},\alpha}(z)$,
$H^{\nu}_{\text{imp,z}}(z)$ is the impurity potential in $z$-direction
defined as $H^{\nu}_{\text{imp,z}}(z)=\Sigma^{\nu} V_0 e^{-z/a_0}$,
and $\bra {\Psi_{\vec{k}_{\parallel},\alpha}} H_{\text{imp,z}}^{\nu} \ket{\Psi_{\vec{k}^{'}_{\parallel},\beta}}$
represents $\int_{-\infty}^{0} dz\,V_0 e^{-z/a_0}  \Psi^{\dagger}_{\vec{k}_{\parallel},\alpha} (z) \Sigma^{\nu}\Psi_{\vec{k}^{'}_{\parallel},\beta} (z)$.
In Eq. \ref{eq:Lambda}, the expression of $\Psi_{\vec{k}_{\parallel},\alpha} (z)
 \Psi^{\dagger}_{\vec{k}^{'}_{\parallel},\beta} (z)$ represents a
$8\times 8$ matrix structure in the combined spin and Nambu space.

We consider the QPI patterns for at the frequency less than
the gap energy.
Since the density of states is singular at zero energy arising from the surface flat bands, only the Majorana zero modes are kept in the summation over states.
With this approximation, Eq. (\ref{eq:Lambda}) becomes
\bea
&&\Lambda^{\mu\nu}(\omega,\vec{q},z)=
\frac{N_a}{N}\sum_{\vec{k}_{\parallel},\vec{k}^{'}_{\parallel}} \delta_{\vec{k}_{\parallel}-\vec{k}^{'}_{\parallel},\vec{q}}
(\frac{1}{\omega+i\epsilon} )^2 \nn \\
&\times&\bra {\Psi^{M}_{\vec{k}_{\parallel}}} H_{\text{imp,z}}^{\nu} \ket{\Psi^{M}_{\vec{k}^{'}_{\parallel}}}
\text{Tr}
\Big [ (1+\tau_3)\Sigma^{\mu}
\Psi^{M}_{\vec{k}_{\parallel}}(z)
\Psi^{M\dagger}_{\vec{k}^{'}_{\parallel}} (z)  \Big ], ~~~~~~
\label{eq:Lambda_Majorana}
\eea
in which  $\Psi^{M}_{\vec{k}_{\parallel}}(z)$ represents the wavefunction
of the Majorana state with the in-plane momentum $\vec{k}_{\parallel}$.

The surface Majorana modes exhibit flat-band structure at zero energy.
In the case of the dilute limit, in which the single impurity scattering
can be justified, the scattering matrix element scales as $V_0/L^2$,
thus the scattering occurs nearly at zero energy.
Below, we set $\omega$ at the order of $\epsilon$, which can be viewed
as the inverse lifetime of the Majorana states.

Before presenting the detailed QPI patterns, there is a general property
of $\Delta\rho^{\mu\nu}(\omega,\vec q)$.
Since both the density and spin-density distributions are real fields,
their Fourier transforms satisfy
\bea
\Delta\rho^{\mu\nu}(\omega,\vec q)=\Delta\rho^{\mu\nu,*}(\omega,-\vec q).
\eea
In other words, $\mbox{Re}\Delta\rho^{\mu\nu}(\omega,\vec q)$ and
$\mbox{Im}\Delta\rho^{\mu\nu} (\omega,\vec q)$ are even and odd
with respect to $\vec q$.
This symmetry has been clearly shown in all of Fig. \ref{fig:Delta03},
Fig. \ref{fig:Delta07}, Fig. \ref{fig:spin_QPI_Re},
and Fig. \ref{fig:spin_QPI_Im}, which present the Fourier
transforms of $\Delta\rho_{sf}^{\mu\nu}(\omega,\vec{q})$ at
$\omega =\epsilon$ in $(1\,1\,1)$-surface.

\subsection{The QPI pattern for a non-magnetic impurity}
\label{eq:nonspin_QPI}

\begin{figure}
\centering\epsfig{file=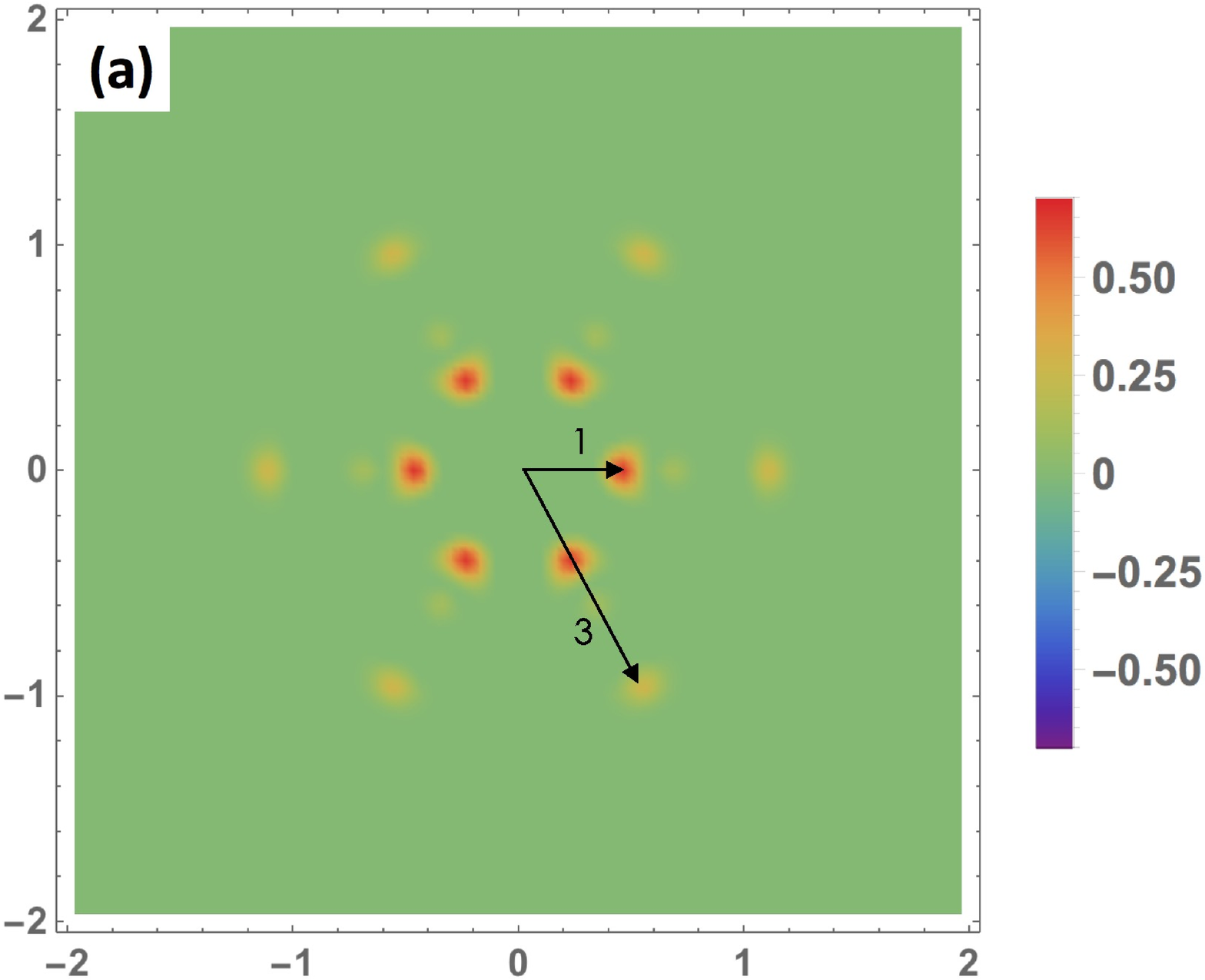,clip=0.45,
width=0.47\linewidth,height=0.4\linewidth,angle=0}
\centering\epsfig{file=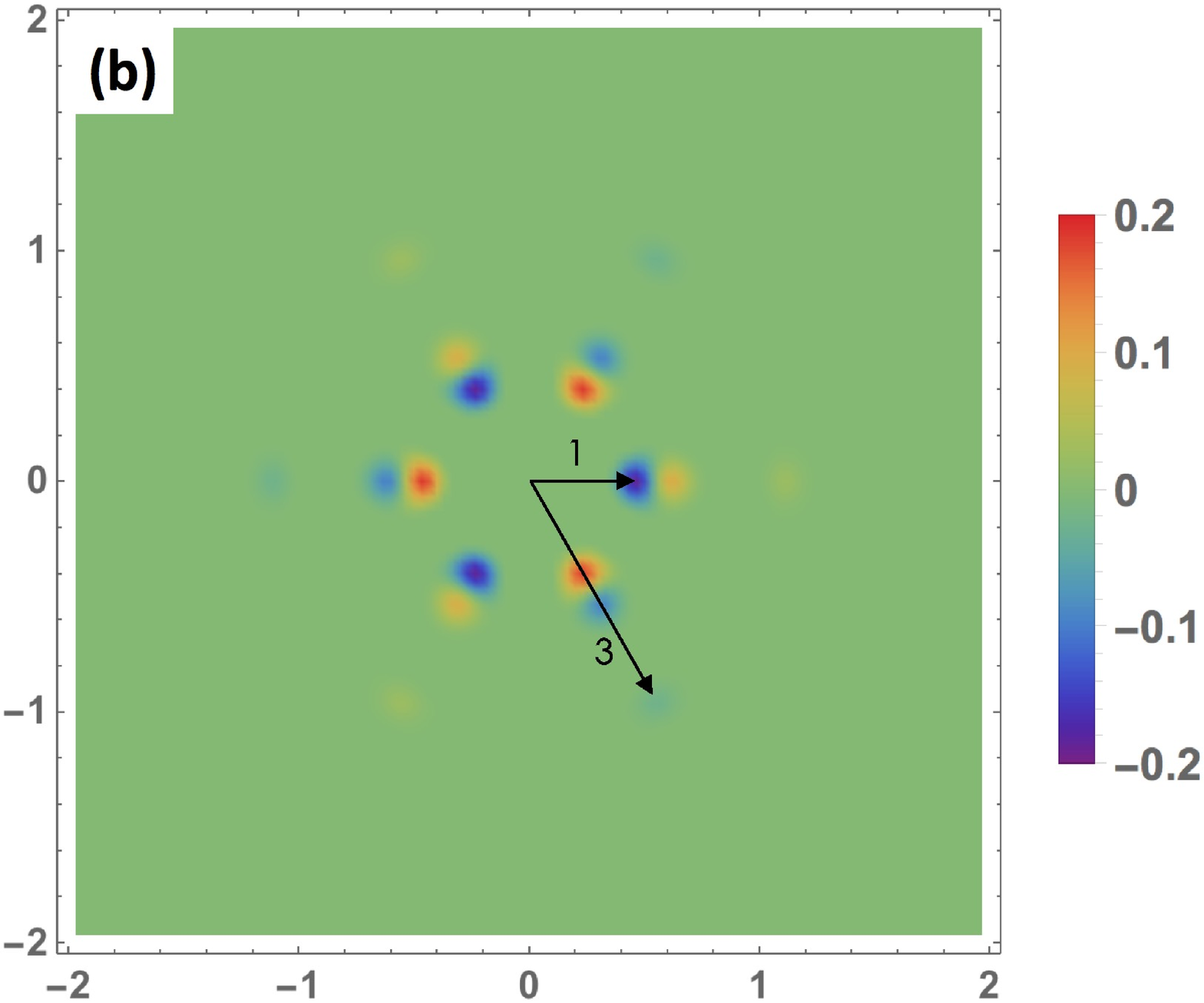,clip=0.45,
width=0.47\linewidth,height=0.4\linewidth,angle=0}
\caption{$a)$ $\text{Re}\Delta\rho^{00}_{sf}(\epsilon,\vec q)$
and $b)$ $\text{Im}\Delta\rho_{sf}^{00} (\epsilon,\vec{q})$ in the
$(1\,1\,1)$-surface for $\Delta_s/\Delta_p=0.3$ under the
Born approximation.
The background contribution in the absence of impurity
is subtracted.
The numerical computations are carried out for a $60\times 60$
lattice in momentum space.
The tip resolution $b_0$ in Eq. (\ref{eq:envelop}) is set
to be $b_0=1/(\sqrt{2}k_f)$, and the impurity range $a_0$ in
Eq. (\ref{eq:Himp}) is taken as $a_0=1/(\sqrt{2}k_f)$,
both of which are at the order of Fermi wavelength.
The impurity potential strength is taken as $V_0= g\Delta_0$ where $g$ is a scaling factor satisfying $g\ll 1$ to justify the Born approximation.
Other parameters are $\lambda_1=0,\lambda_2=|\mu|/(2k_f^2)$,
and $\lambda_3=0$.
$\Delta_0=0.02|\mu|$, $N_a=\frac{1}{2k_f^2}$,
 $N_b=\sqrt{2}k_f$.
$\epsilon=2\times 10^{-5} |\mu|$, which is the
inverse of the Majorana life time.
The color bar is in the unit of $g\frac{\sqrt{2} k_f}{800\pi^2|\mu|}$.
}
\label{fig:Delta03}
\end{figure}

\begin{figure}
\centering\epsfig{file=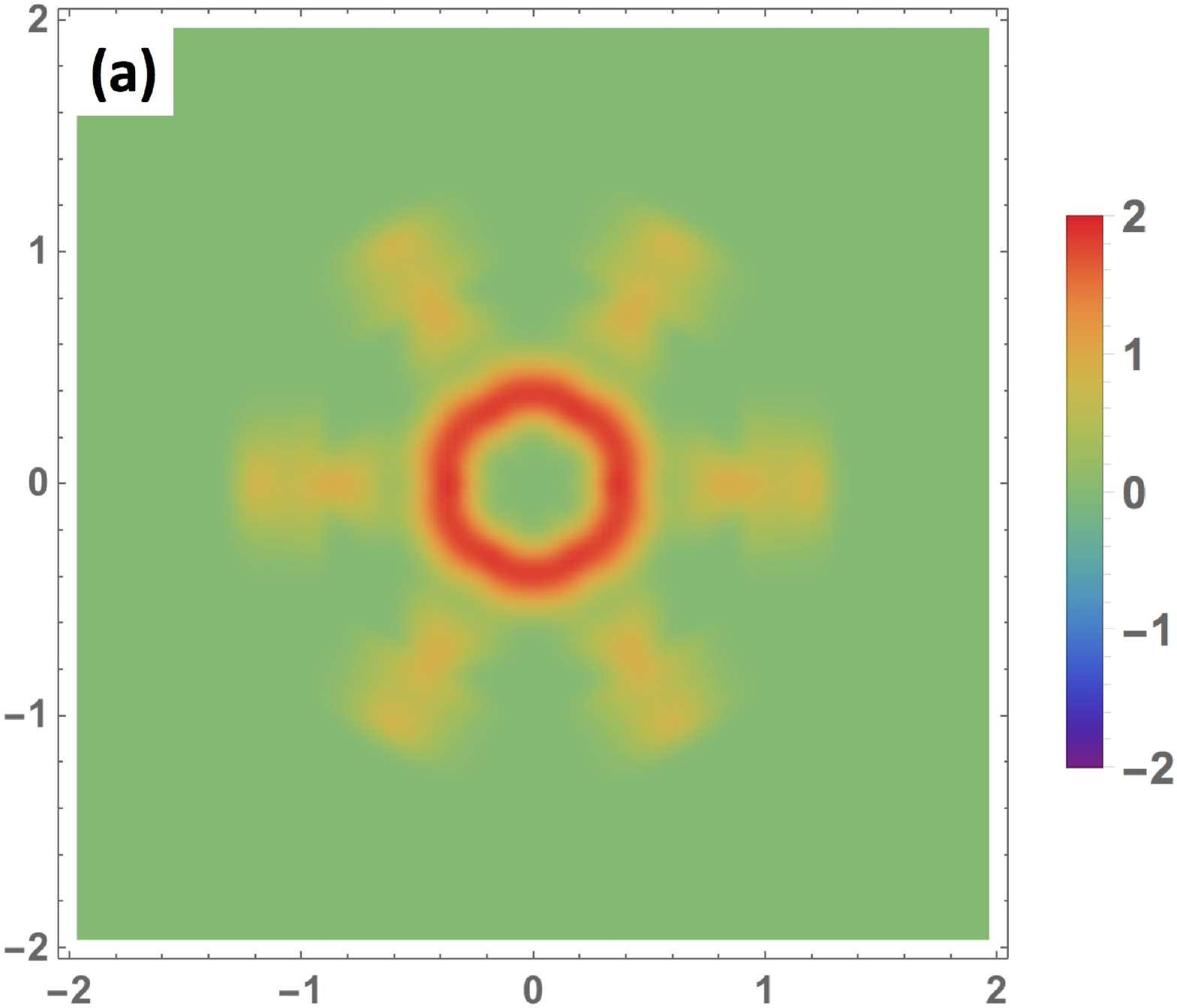,clip=0.45,
width=0.47\linewidth,height=0.4\linewidth,angle=0}
\centering\epsfig{file=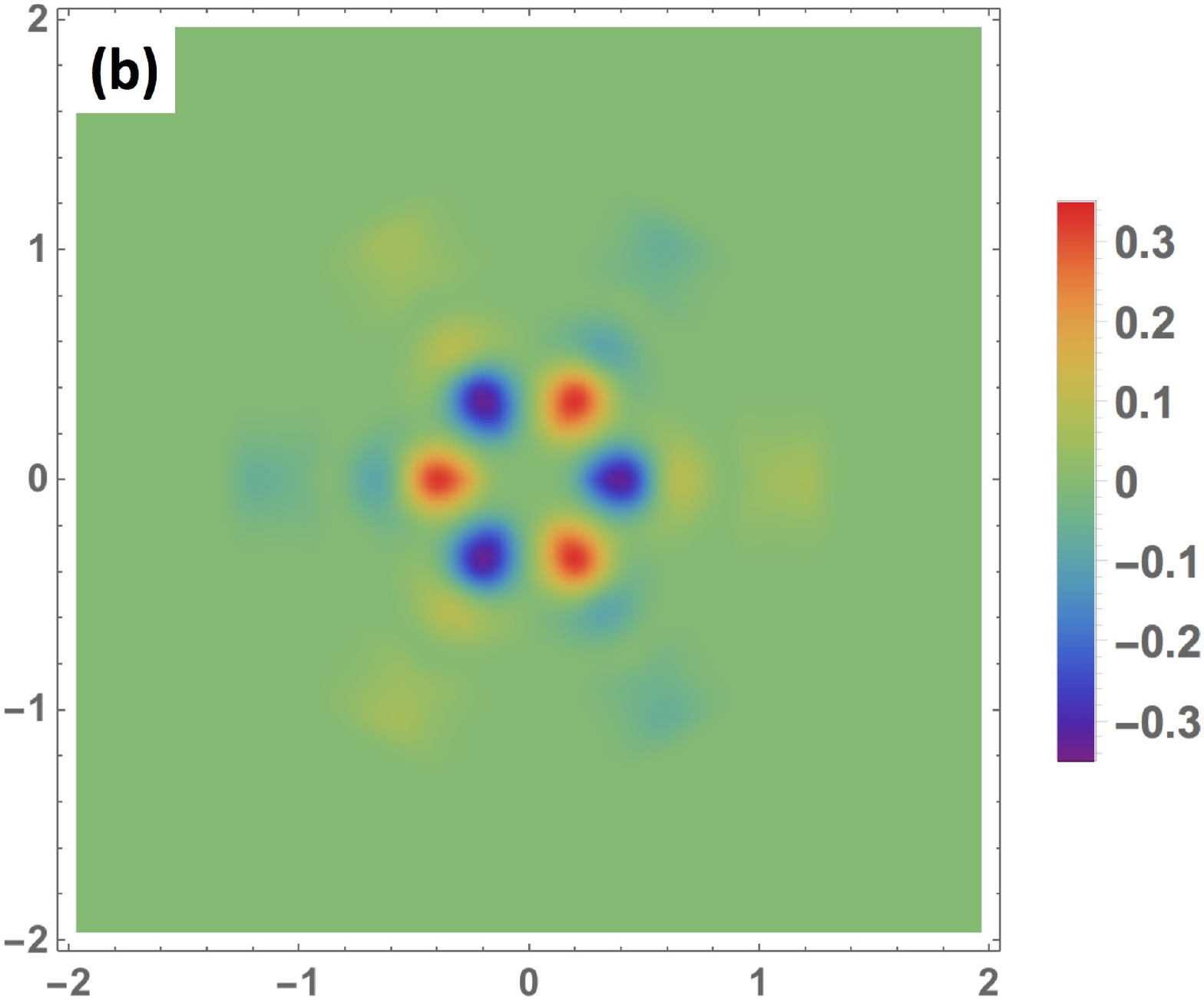,clip=0.45,
width=0.47\linewidth,height=0.4\linewidth,angle=0}
\caption{
$ a) \text{Re}\Delta\rho^{00}_{sf}(\epsilon,\vec q)$
and $b)$ $\text{Im}\Delta\rho_{sf}^{00} (\epsilon,\vec{q})$
in the $(1\,1\,1)$-surface with the parameter $\Delta_s/\Delta_p=0.7$.
The other parameters are as the same as in Fig. \ref{fig:Delta03}.
}
\label{fig:Delta07}
\end{figure}

In this section, the QPI patterns for non-magnetic impurities are
presented.
$\Delta \rho_{sf}^{\mu0}(\omega,\vec{q})$ vanishes when $\mu\neq 0$
due to TR symmetry, hence, only the results of $\Delta \rho_{sf}^{00}
(\omega,\vec{q})$ are displayed.

Fig. \ref{fig:Delta03} and Fig. \ref{fig:Delta07} present $\Delta\rho_{sf}^{00}(\omega,\vec{q})$ at two representative pairing ratios $\Delta_s/\Delta_p$.
Both figures exhibit the $C_{3v}$ symmetry with three vertical
reflection planes.
For a non-magnetic impurity, the Hamiltonian remains odd under the
chiral operation defined in Eq. (\ref{eq:chiral}), which imposes
strong restrictions on the QPI patterns.
It can only couple the Majorana zero modes with opposite
chiral indices as shown in Fig. \ref{fig:surface spectrum}
($a$) and ($b$).
In Fig. \ref{fig:surface spectrum} ($a$), four representative scattering
wavevectors between Majorana islands are drawn in red arrowed lines,
among which $``0"$ represents the intra-island scattering, and $``1"$,
$``2"$, and $``3"$ represent inter-island scatterings.
The scatterings $``1"$ and $``3"$ are between islands with opposite chiral
indices, hence are allowed in the Born approximation, as shown in
Fig. \ref{fig:Delta03}.
In contrast, the scatterings of $``0"$ and $``2"$ connect islands with the
same chiral index, and hence are forbidden.
For example, the QPI spectra vanish near $q=0$, which is the consequence
of the absence of intra-island scatterings.

While both scatterings $``1"$ and $``3"$ appear in the QPI patterns, the QPI
spectral magnitudes of $``3"$ are much weaker than that of $``1"$, which
is a consequence of TR symmetry.
The impurity matrix element vanishes between two Majorana states
with in-plane momenta $\vec k_{2d}$ and $\vec k_{2d}^\prime$
with $\vec k_{2d}^\prime=-\vec k_{2d}$, forming a Kramers pair
with $T^2=-1$.
Nevertheless, the TR symmetry does not completely forbid the scattering
from $\vec k_{2d}$ to $\vec k_{2d}^\prime$ in the neighbourhood
of $-\vec k_{2d}$, although this kind of scatterings are weakened.
Hence, unlike the case of chiral symmetry which completely forbids
scatterings between islands with the same chiral index,
the TR symmetry only reduces while not strictly forbids the scatterings
between TR related Majorana islands.
For the case of a large pairing ratio $\Delta_s/\Delta_p$ as shown
in Fig. \ref{fig:Delta07}, the phase space for non-TR related Majorana
states scatterings for inter-island scattering ``3"
is much larger than the case shown in Fig. \ref{fig:Delta03}.

\subsection{The QPI for a magnetic impurity}
\label{sec:QPI_magnetic}

\begin{figure}
\centering\epsfig{file=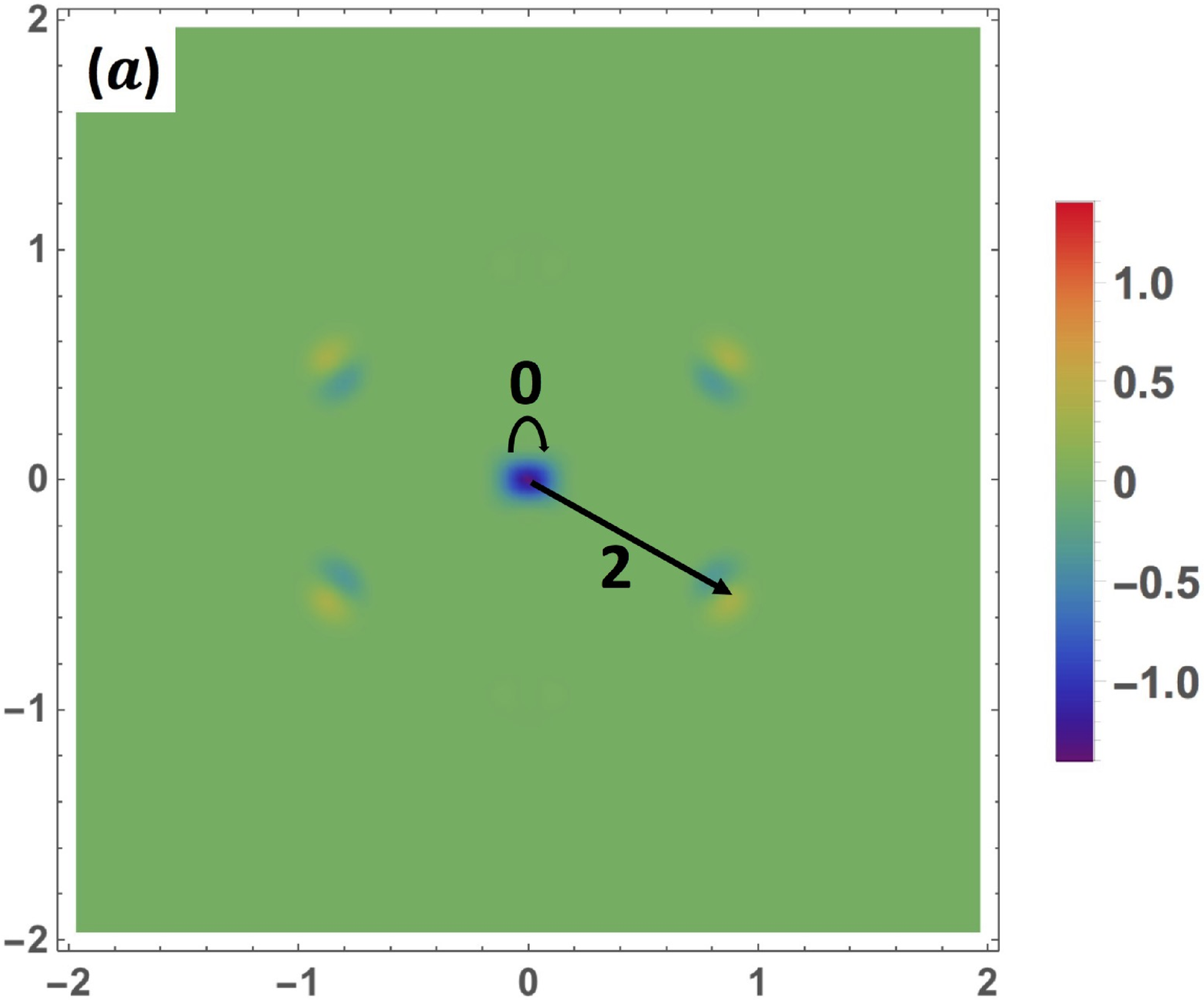,clip=0.45,
width=0.32\linewidth,height=0.26\linewidth,angle=0}
\centering\epsfig{file=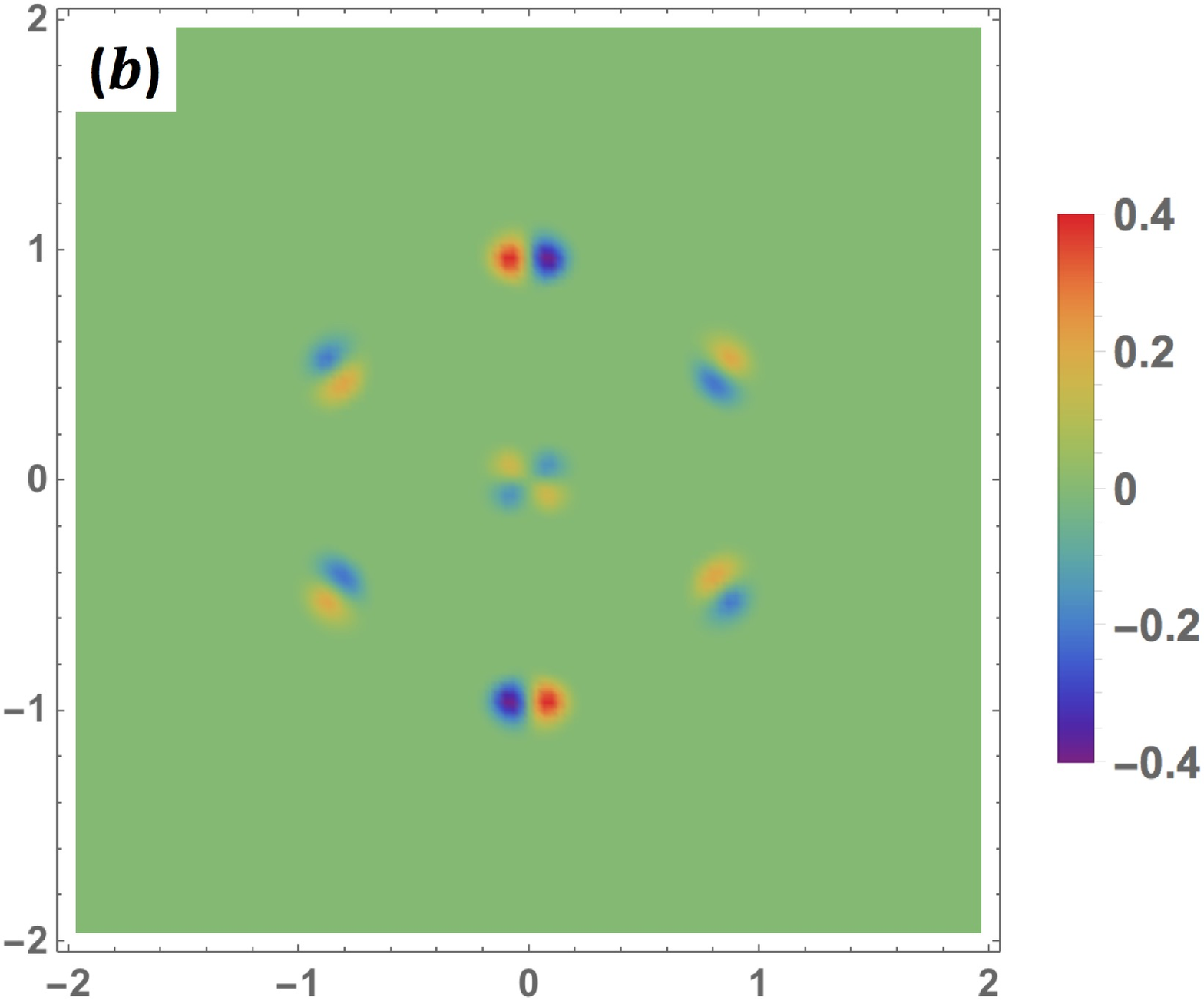,clip=0.45,
width=0.32\linewidth,height=0.26\linewidth,angle=0}
\centering\epsfig{file=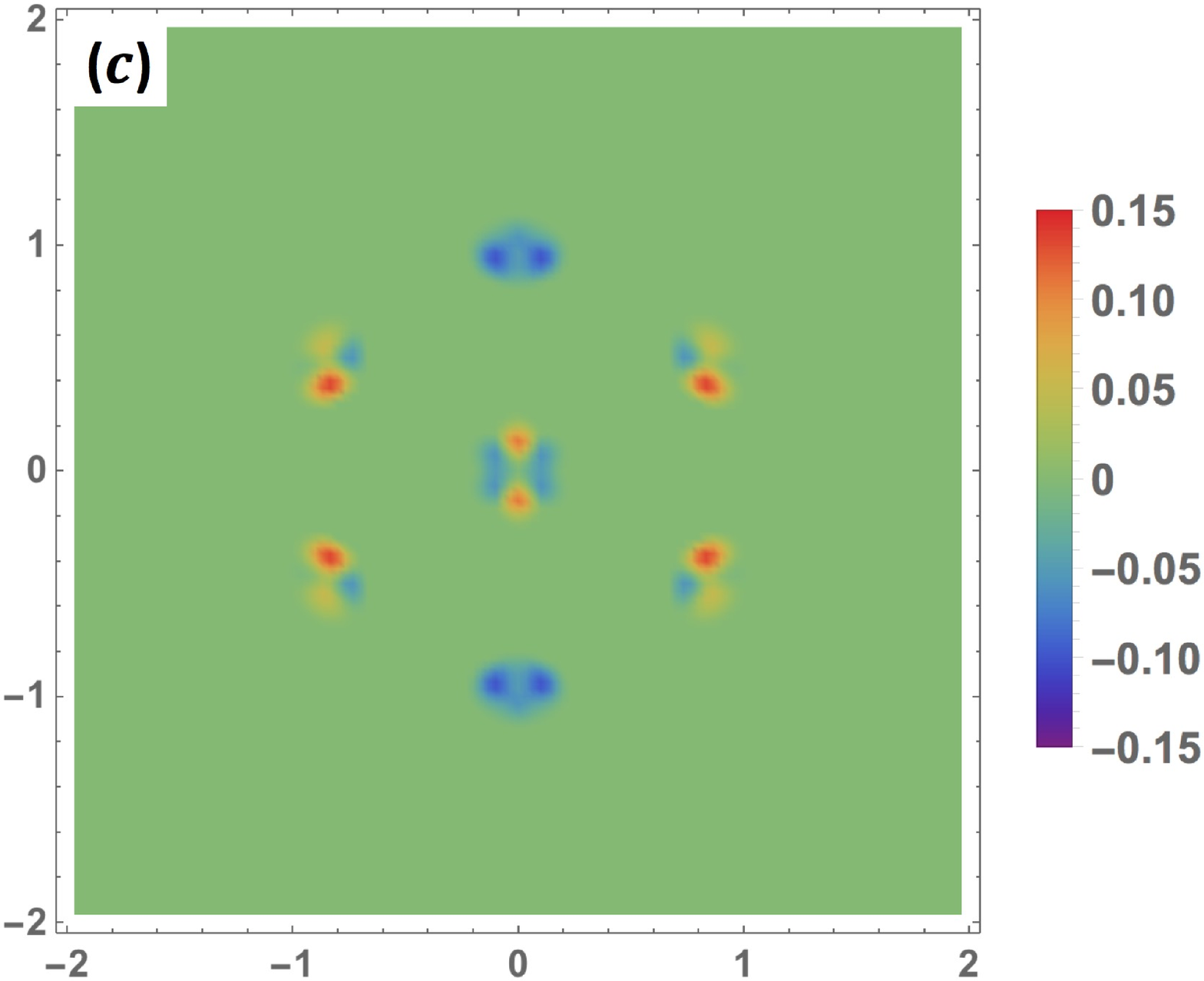,clip=0.45,
width=0.32\linewidth,height=0.26\linewidth,angle=0}
\centering\epsfig{file=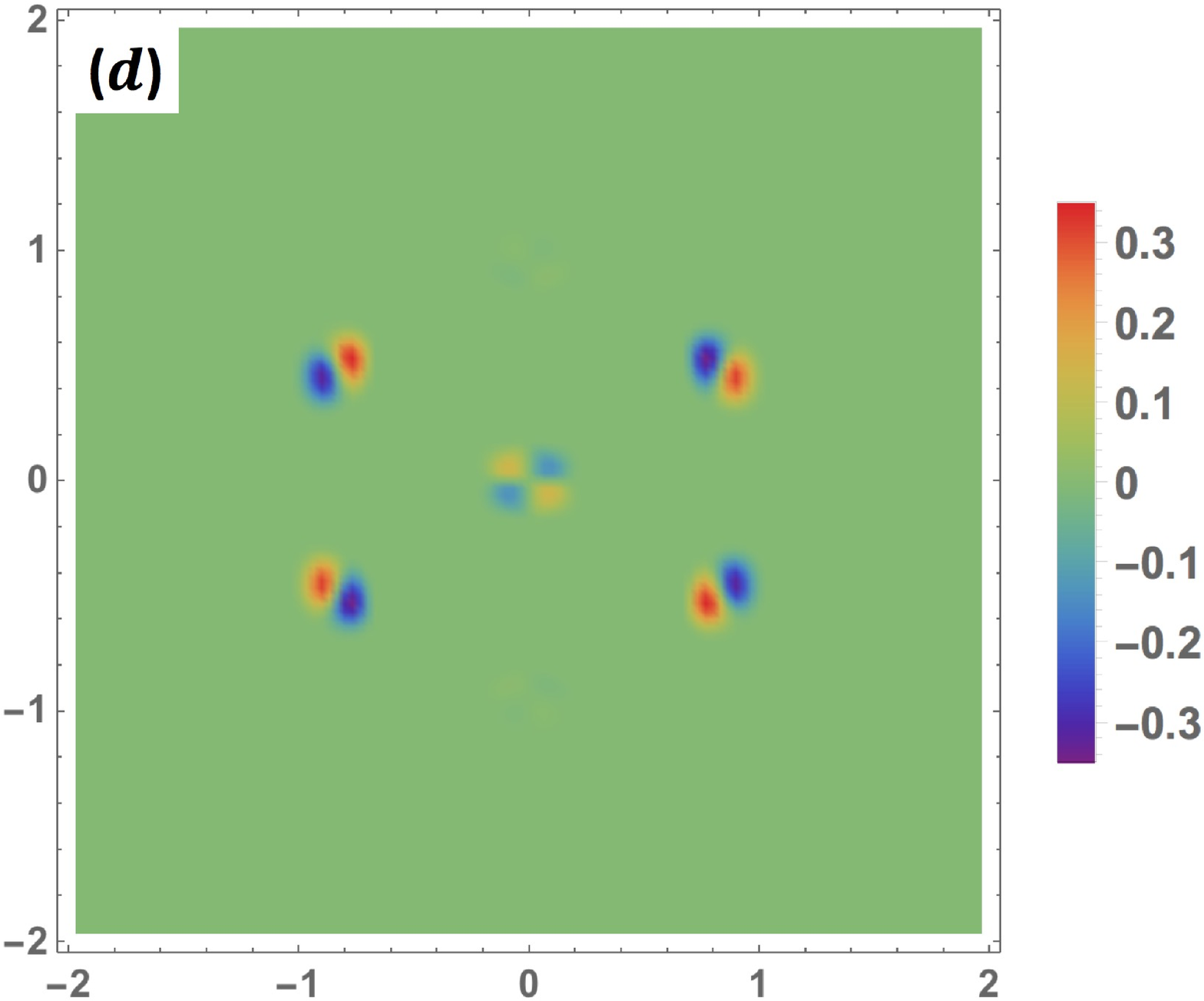,clip=0.45,
width=0.32\linewidth,height=0.26\linewidth,angle=0}
\centering\epsfig{file=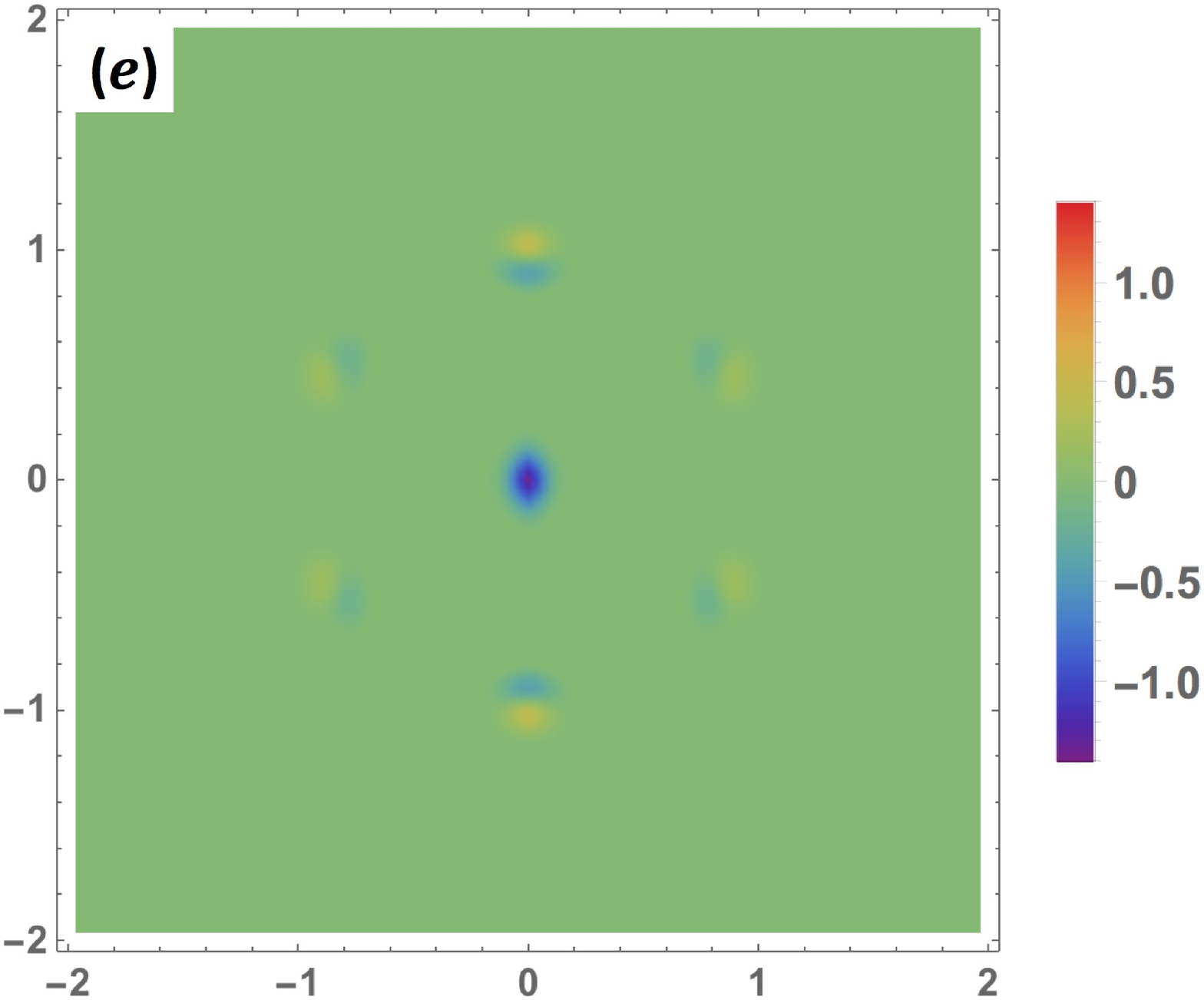,clip=0.45,
width=0.32\linewidth,height=0.26\linewidth,angle=0}
\centering\epsfig{file=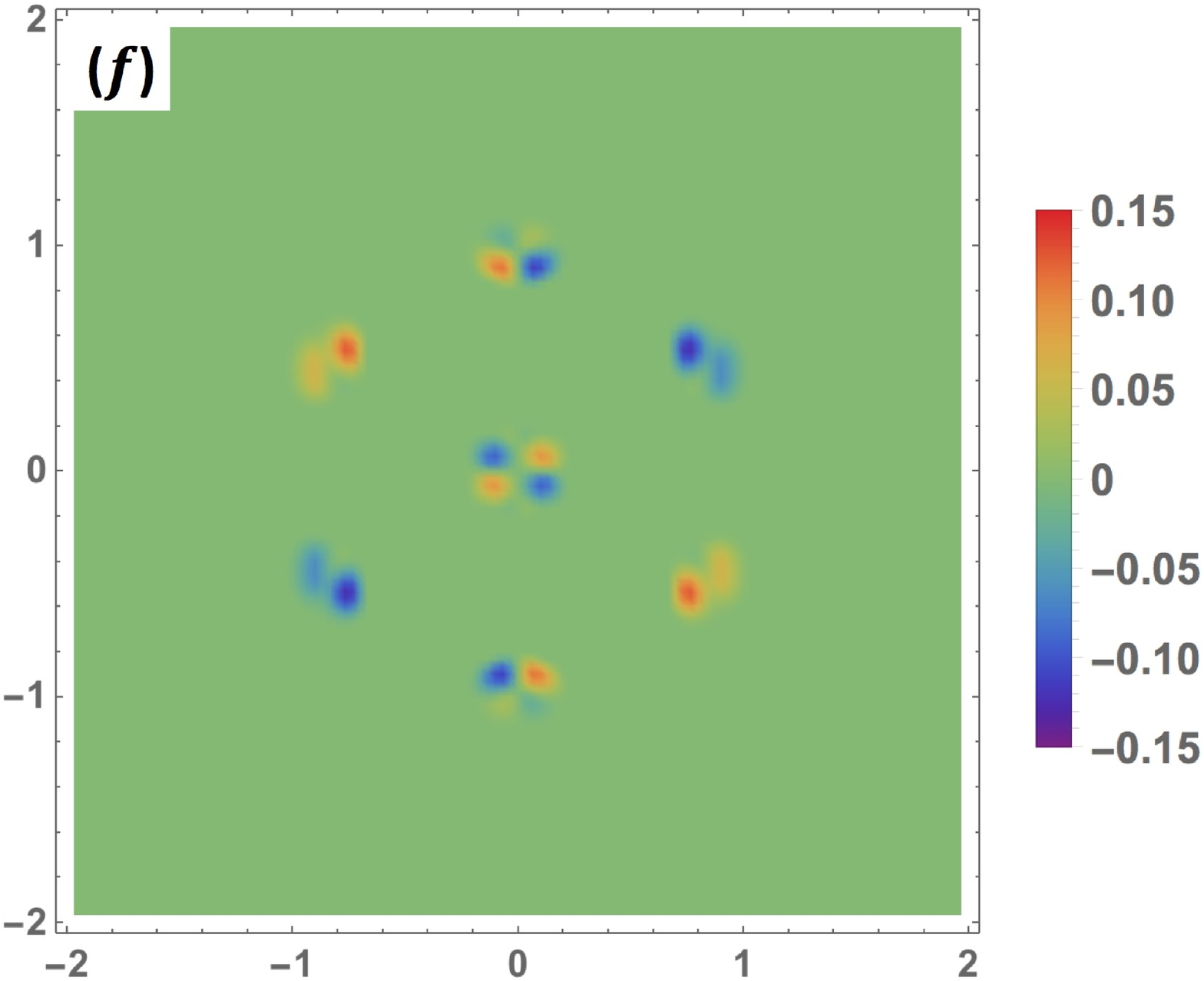,clip=0.45,
width=0.32\linewidth,height=0.26\linewidth,angle=0}
\centering\epsfig{file=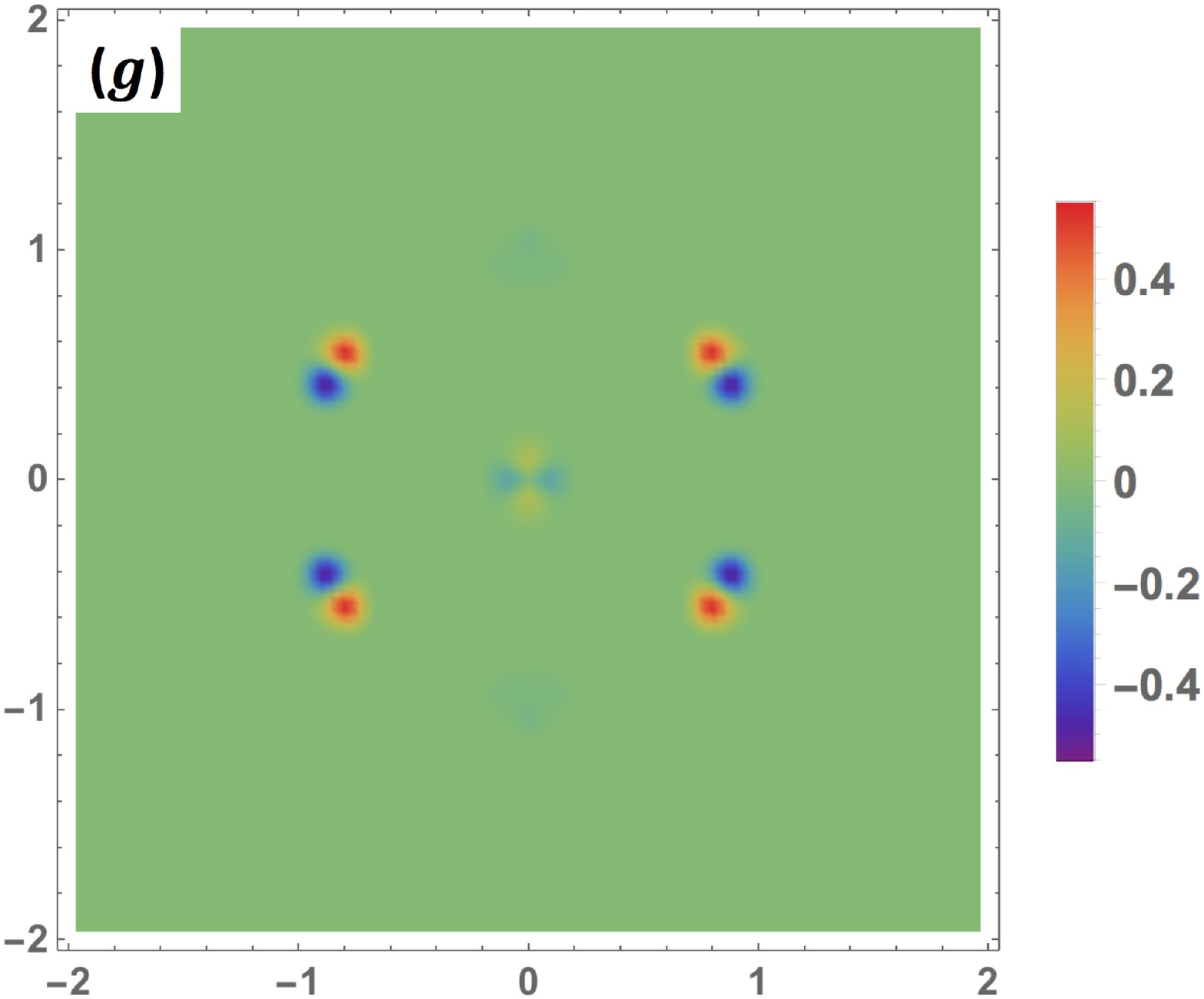,clip=0.45,
width=0.32\linewidth,height=0.26\linewidth,angle=0}
\centering\epsfig{file=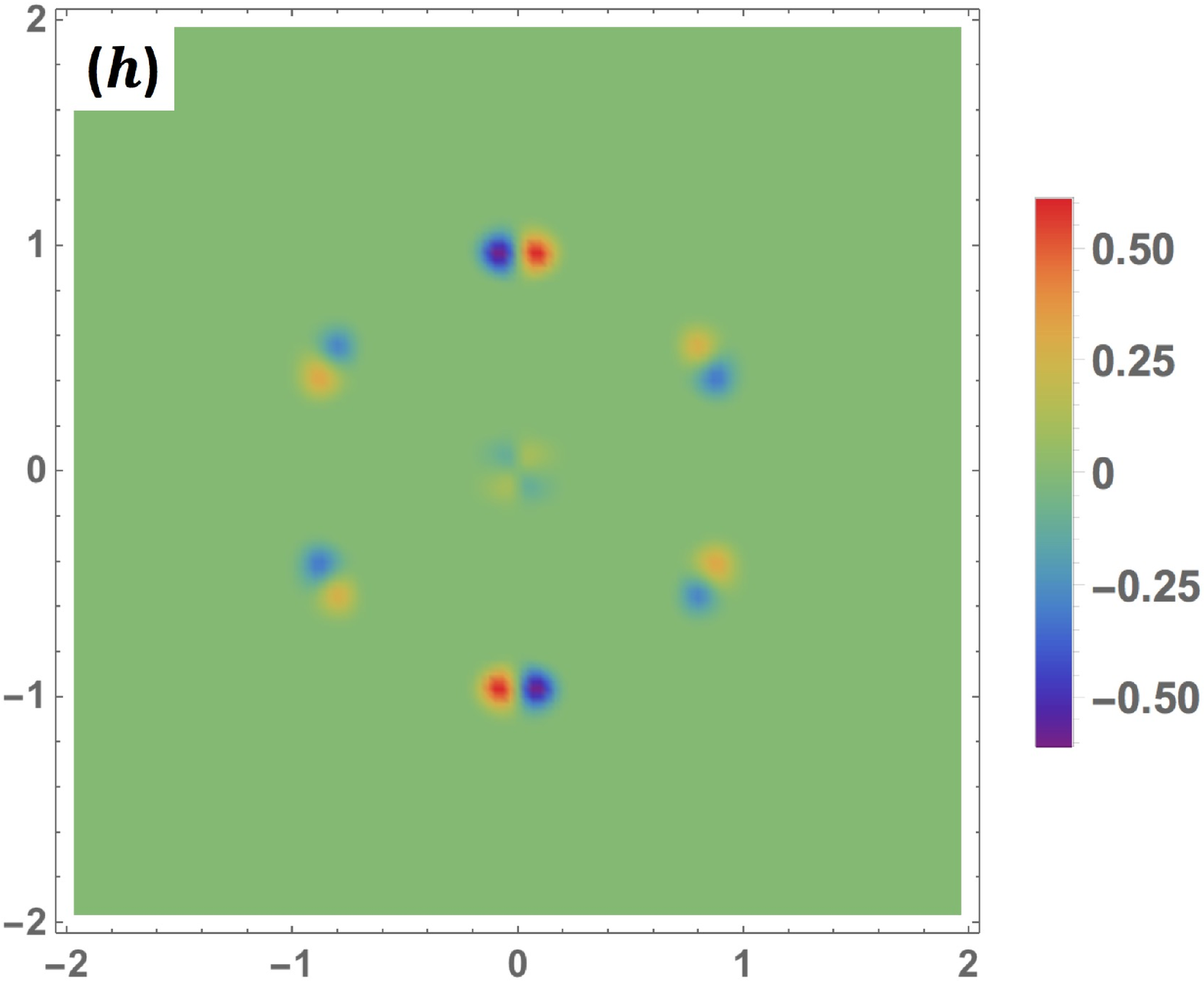,clip=0.45,
width=0.32\linewidth,height=0.26\linewidth,angle=0}
\centering\epsfig{file=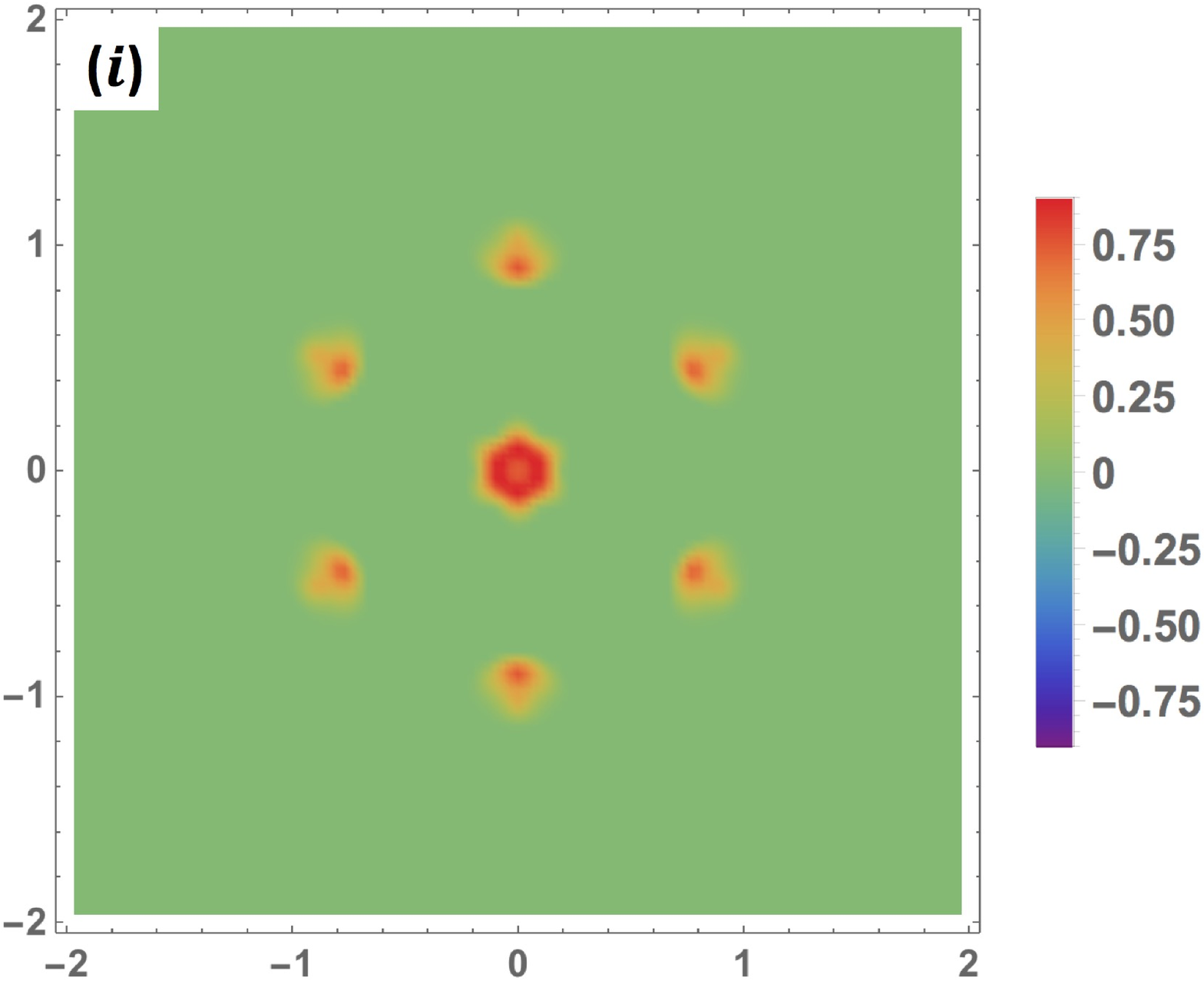,clip=0.45,
width=0.32\linewidth,height=0.26\linewidth,angle=0}
\caption{$\text{Re}\Delta \rho_{sf}^{ij}(\epsilon,\vec{q})$ with
 $ij$ equal to $a)11$, $b) 12$, $c) 13$, $d) 21$, $e) 22$, $f) 23$,
$g) 31$, $h) 32$, $i) 33$, for $\Delta_s/\Delta_p=0.3$ in
the $(1\,1\,1)$-surface.
The background contribution in the absence of impurity is subtracted.
The numerical computations are carried out for a
$60\times 60$ lattice in momentum space.
The parameters are taken the same as
Fig. \ref{fig:Delta03}.
}
\label{fig:spin_QPI_Re}
\end{figure}

\begin{figure}
\centering\epsfig{file=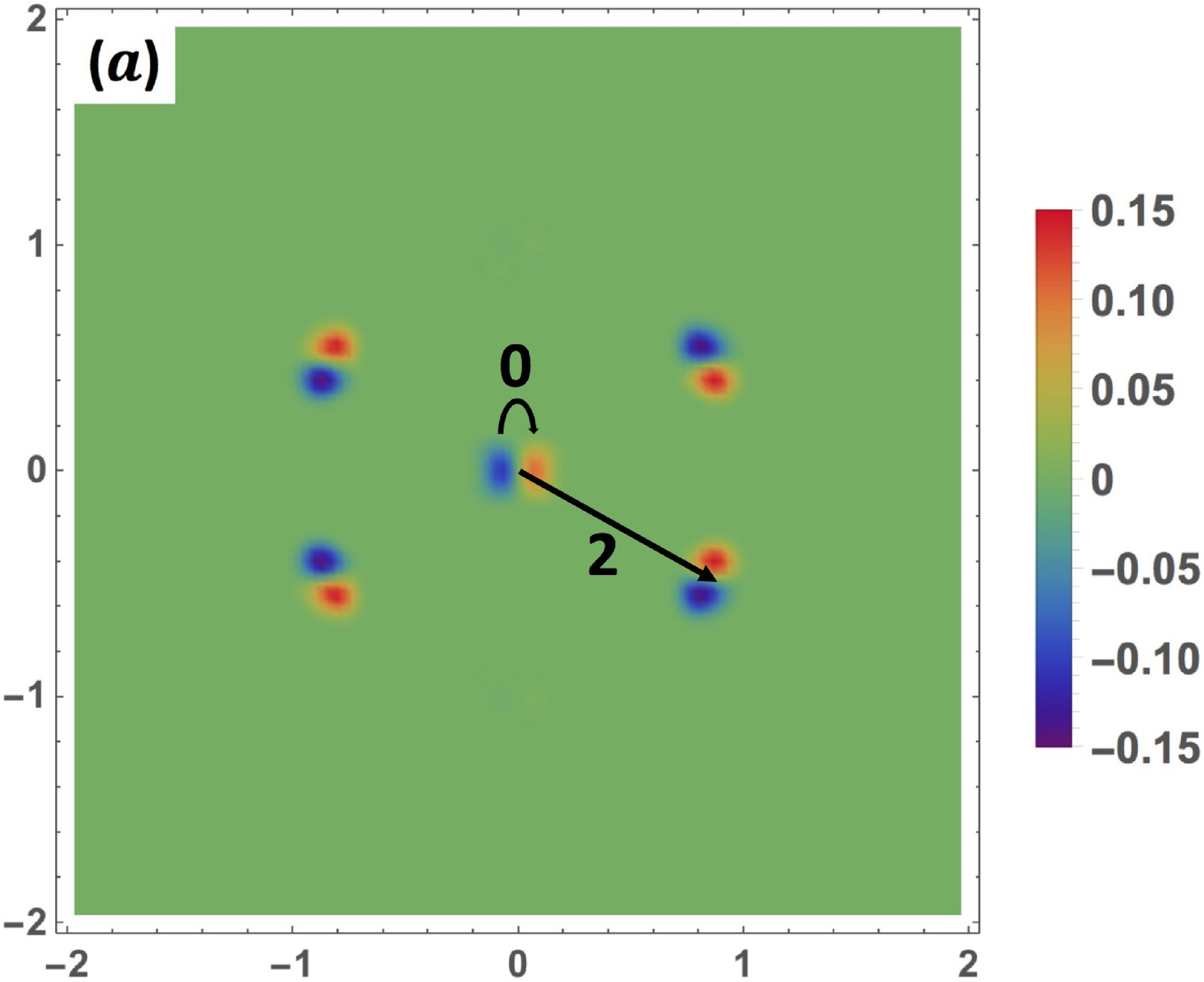,clip=0.45,
width=0.32\linewidth,height=0.26\linewidth,angle=0}
\centering\epsfig{file=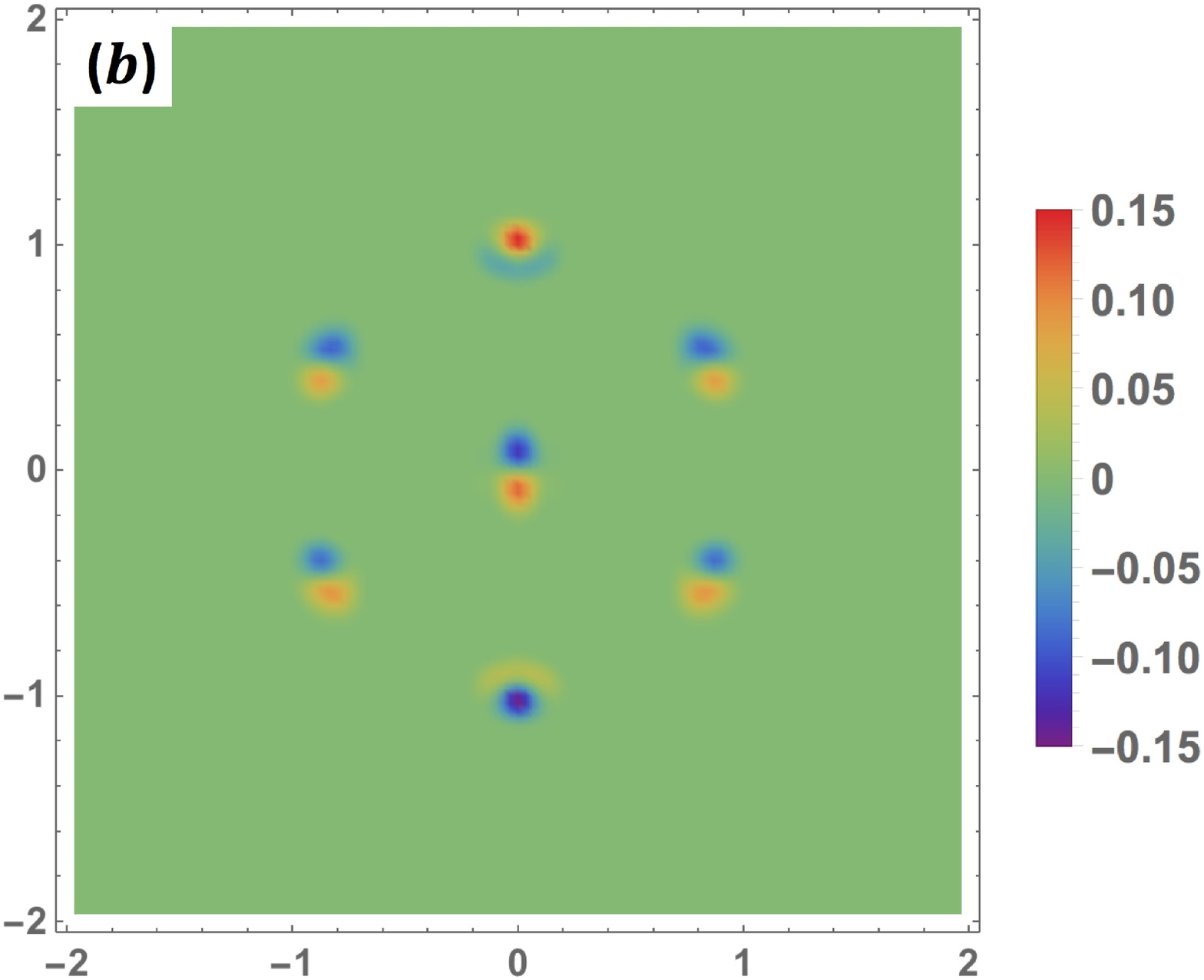,clip=0.45,
width=0.32\linewidth,height=0.26\linewidth,angle=0}
\centering\epsfig{file=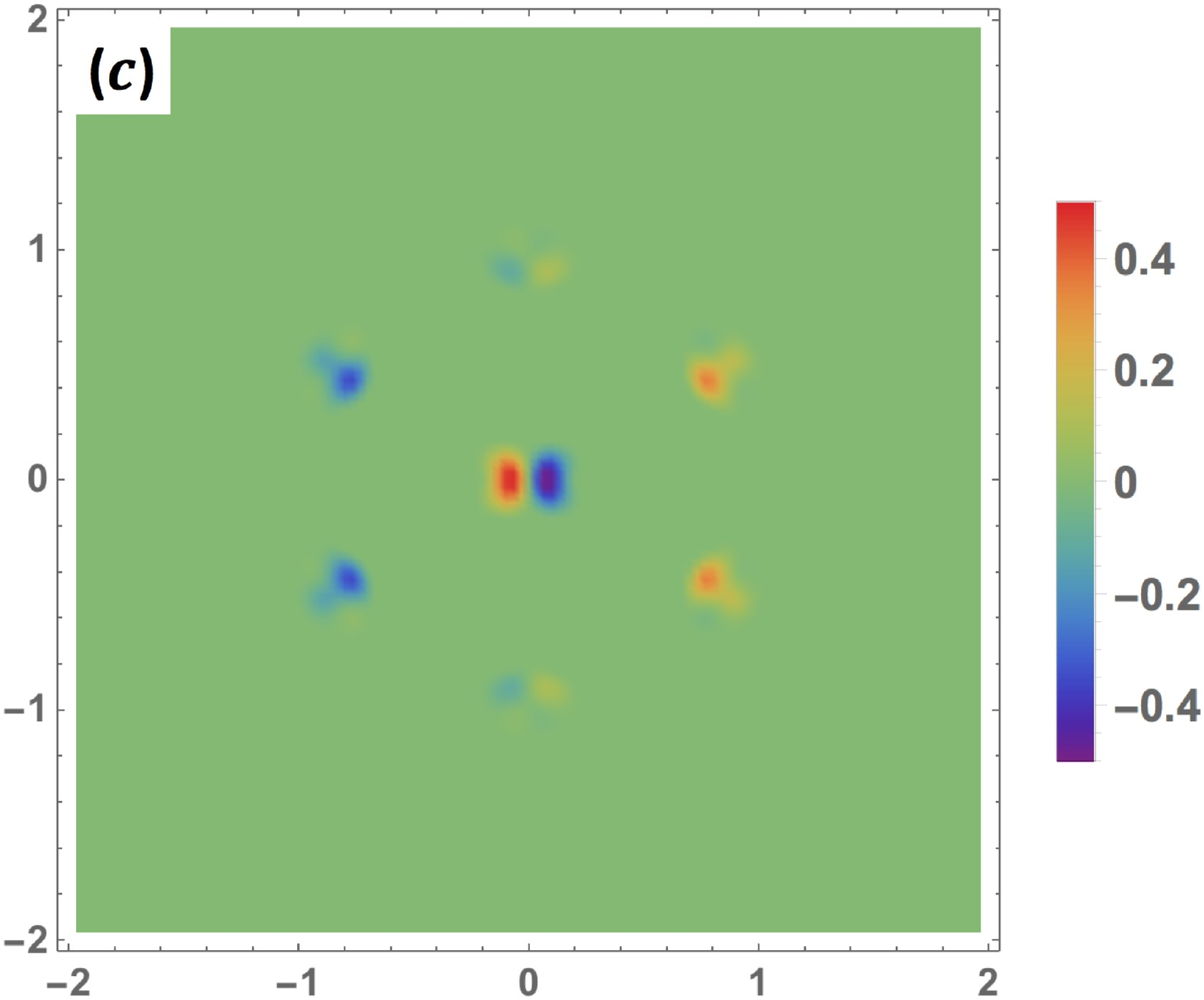,clip=0.45,
width=0.32\linewidth,height=0.26\linewidth,angle=0}
\centering\epsfig{file=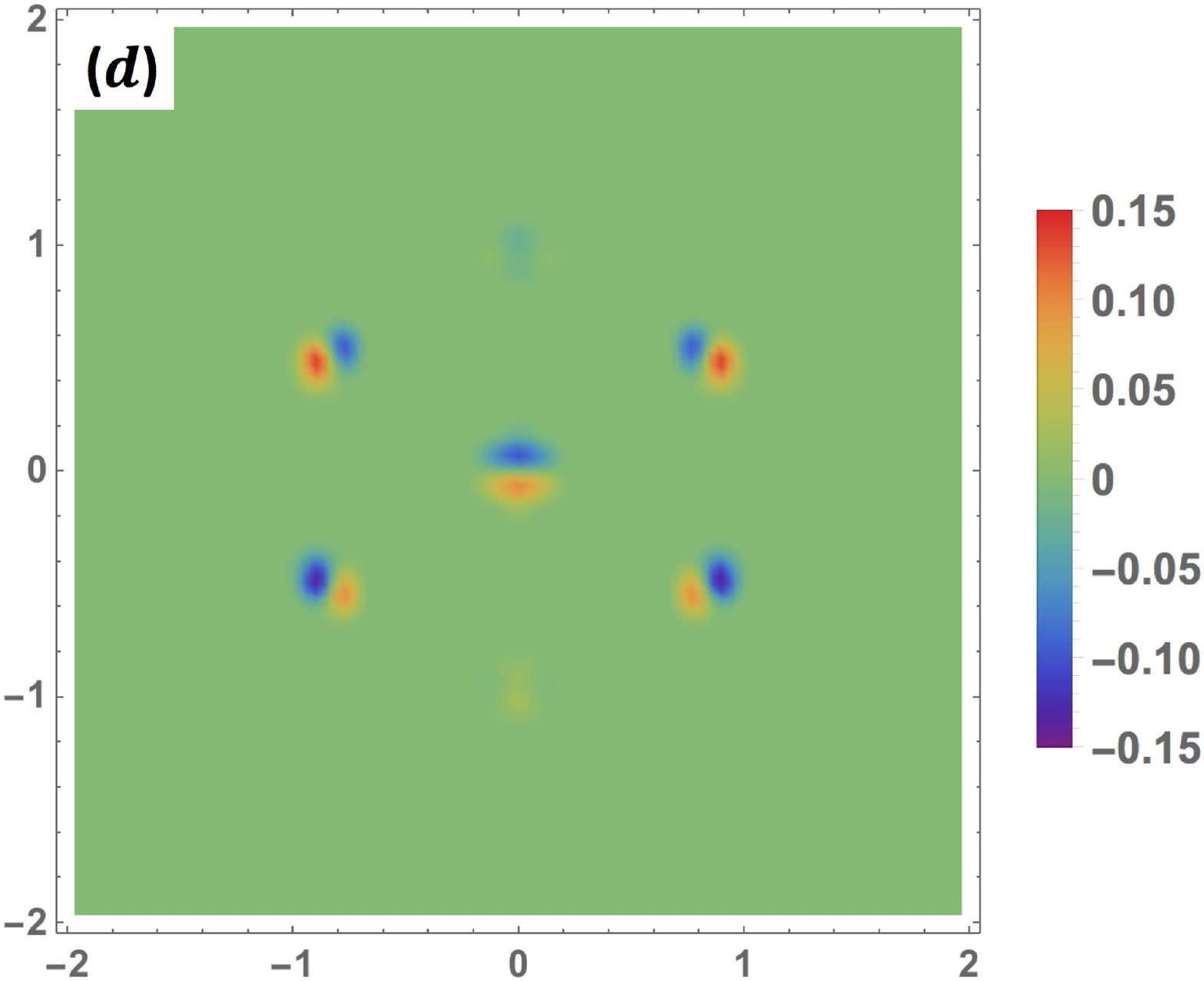,clip=0.45,
width=0.32\linewidth,height=0.26\linewidth,angle=0}
\centering\epsfig{file=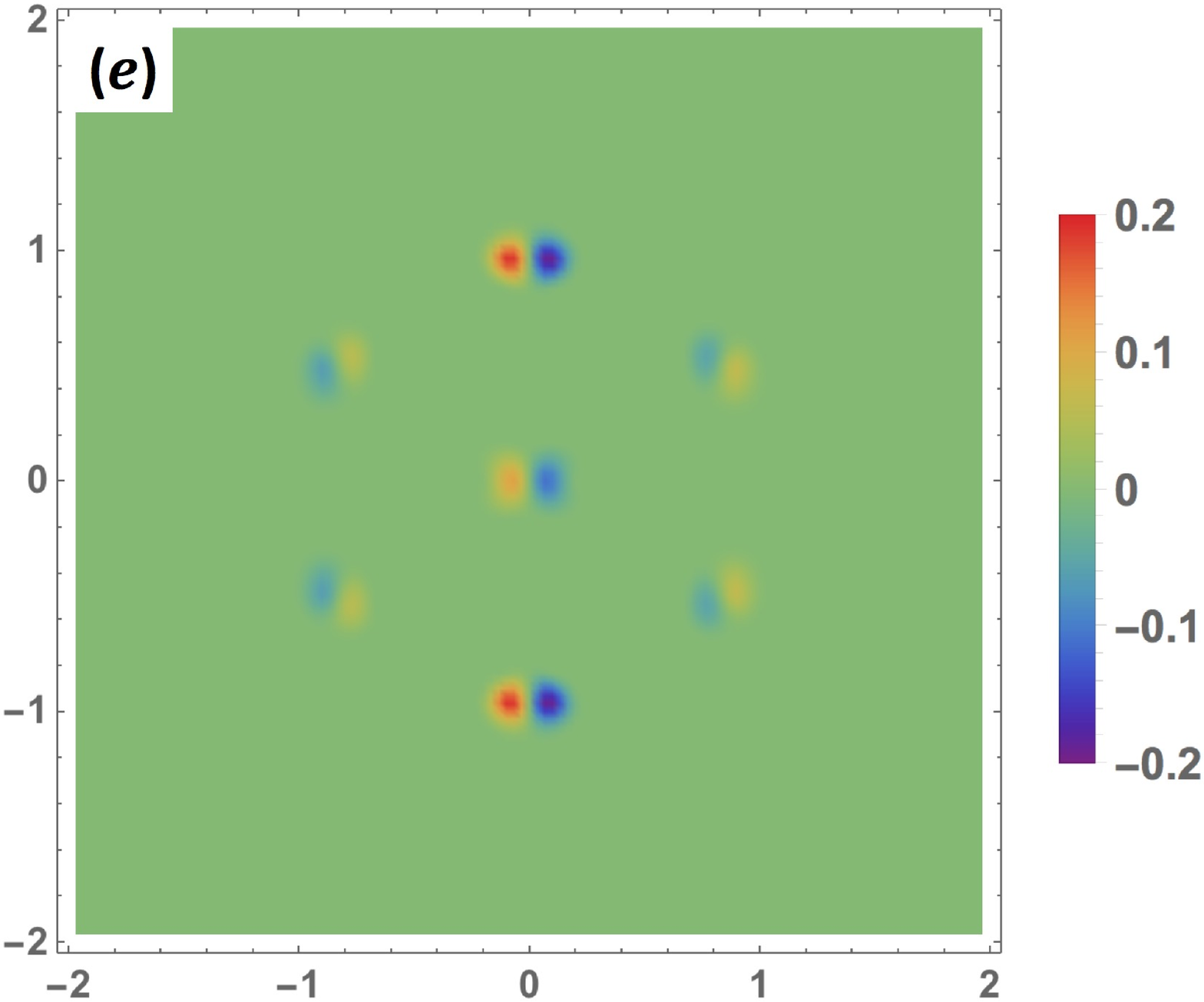,clip=0.45,
width=0.32\linewidth,height=0.26\linewidth,angle=0}
\centering\epsfig{file=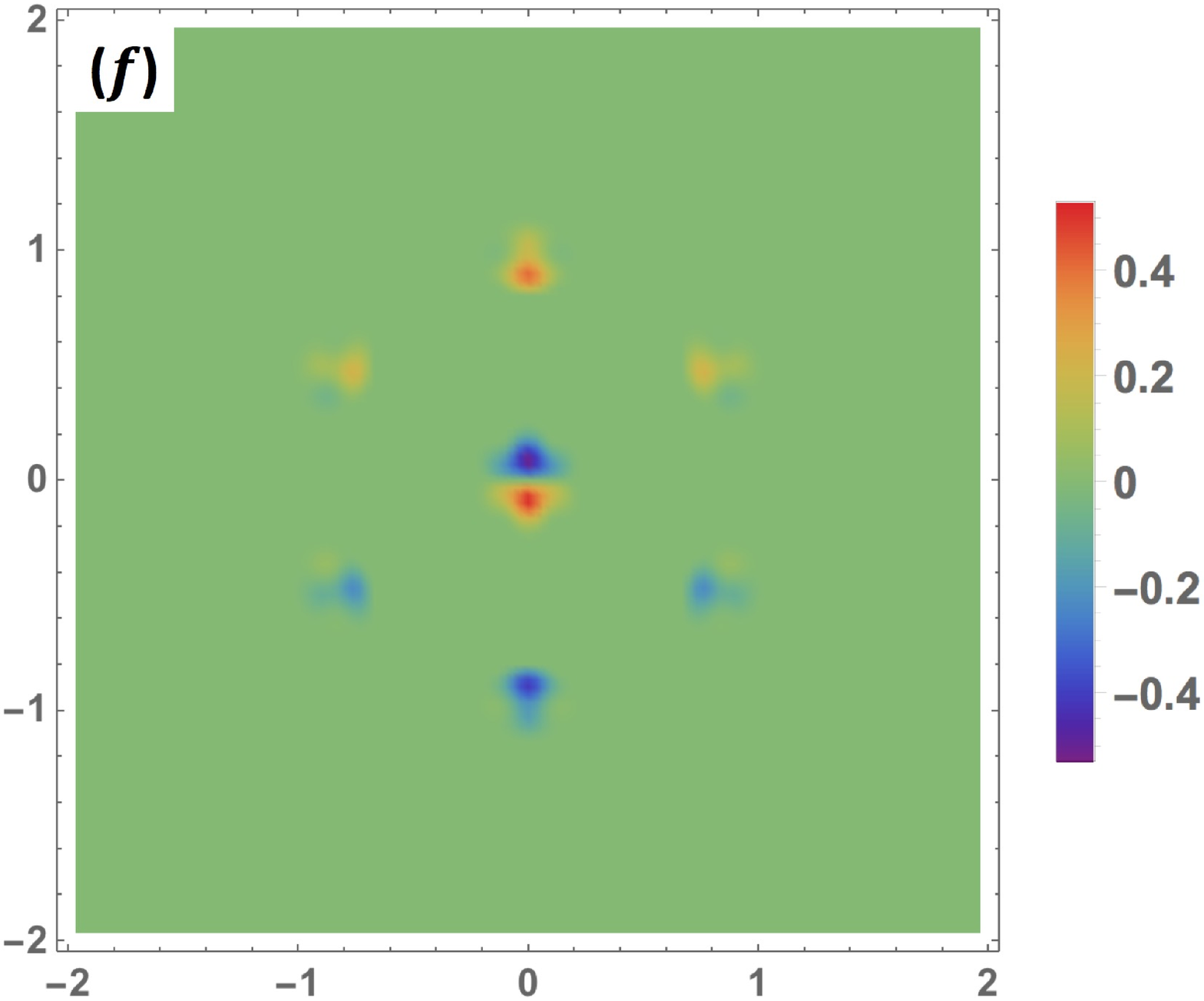,clip=0.45,
width=0.32\linewidth,height=0.26\linewidth,angle=0}
\centering\epsfig{file=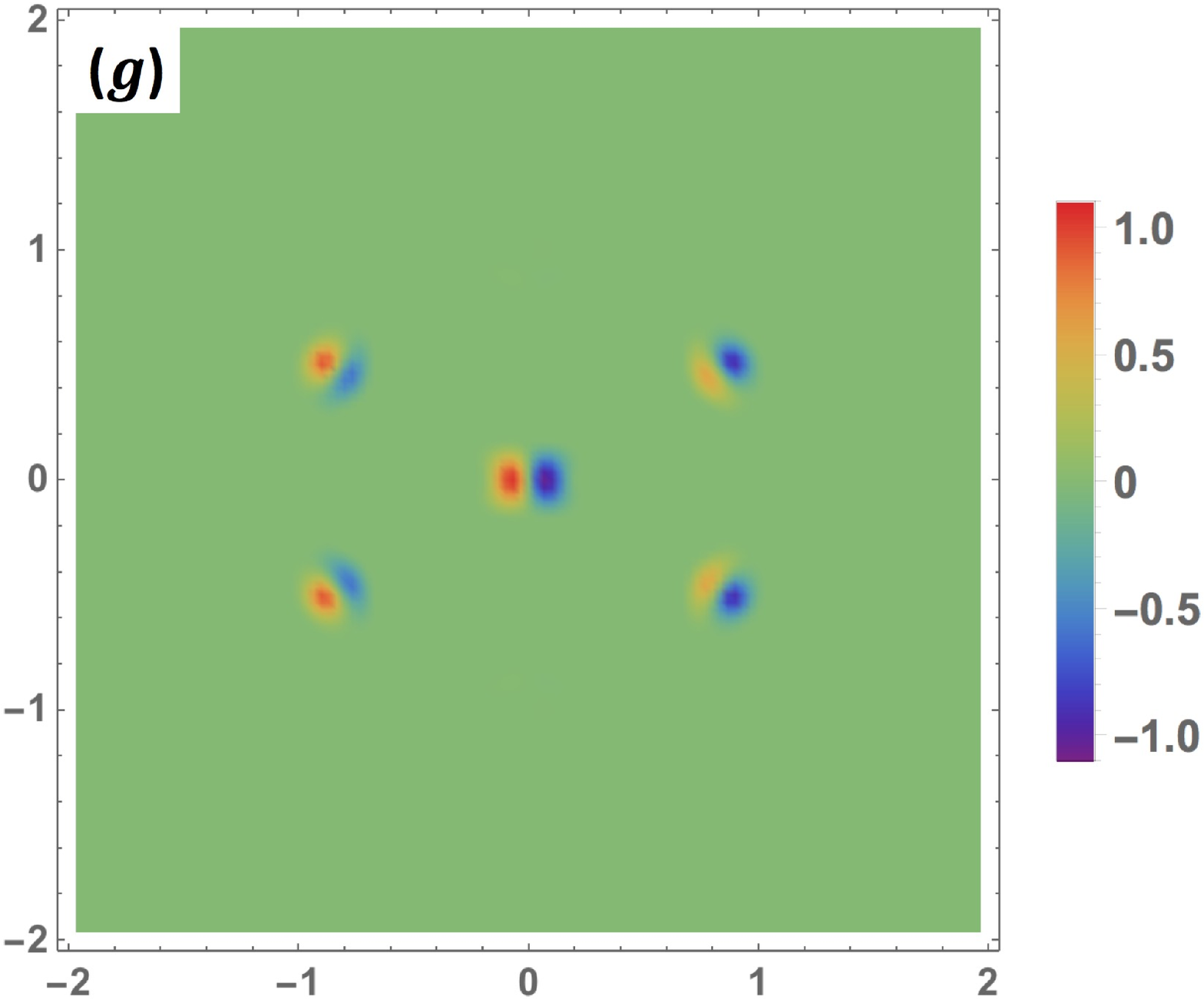,clip=0.45,
width=0.32\linewidth,height=0.26\linewidth,angle=0}
\centering\epsfig{file=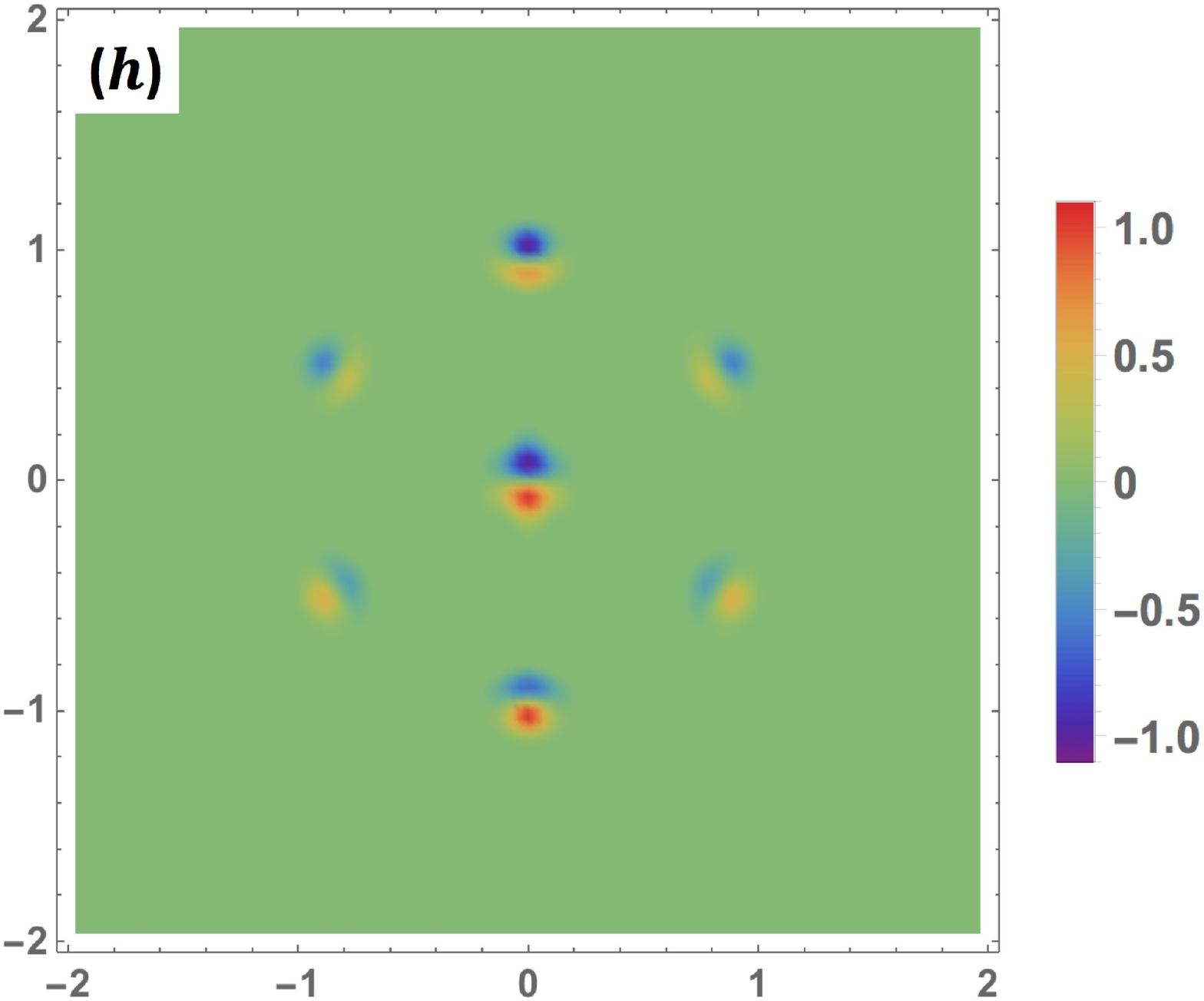,clip=0.45,
width=0.32\linewidth,height=0.26\linewidth,angle=0}
\centering\epsfig{file=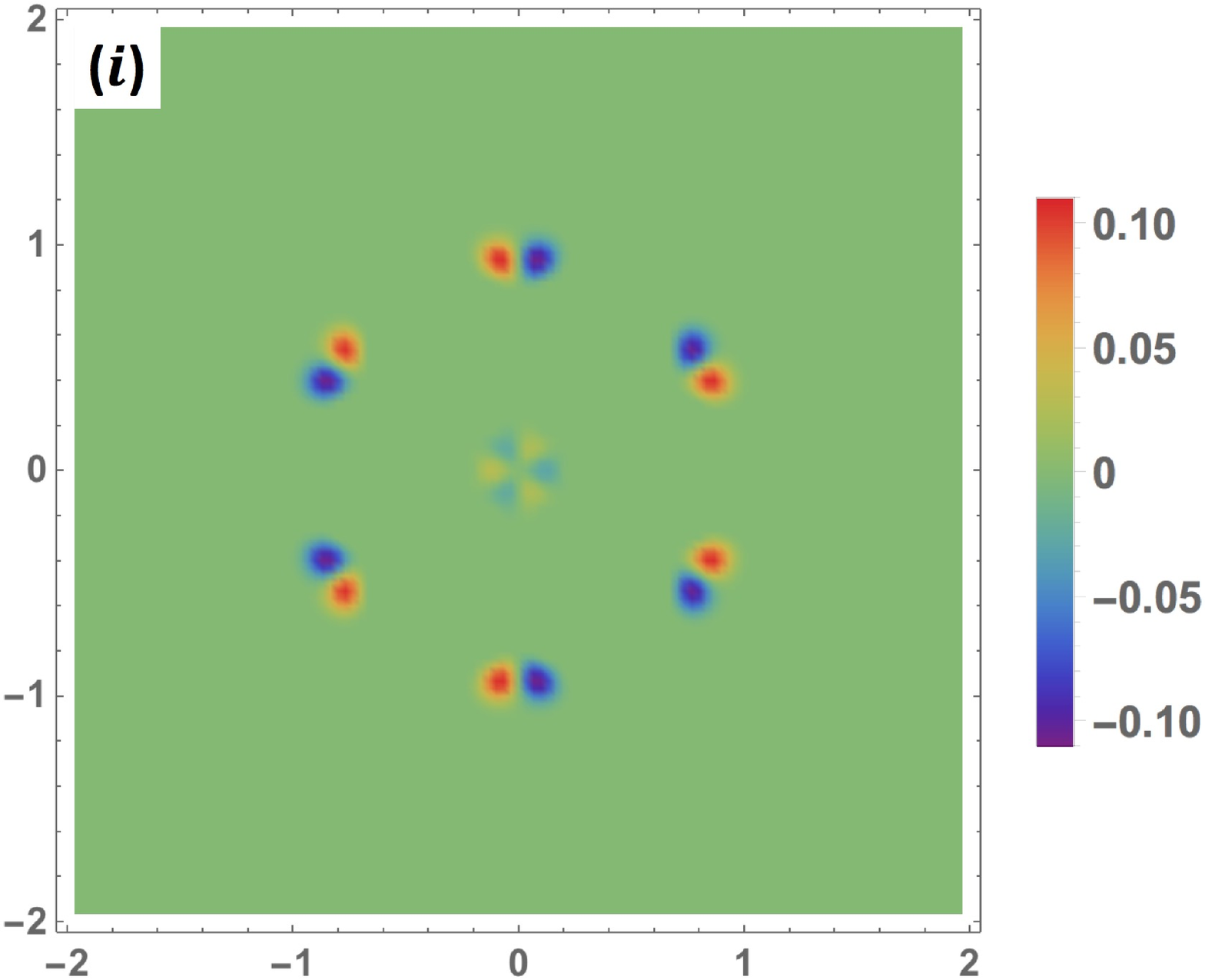,clip=0.45,
width=0.32\linewidth,height=0.26\linewidth,angle=0}
\caption{$\text{Im}\Delta \rho_{sf}^{ij}(\epsilon,\vec{q})$ with
 $ij$ equal to $a)$ $11$, $b)$ $12$, $c)$ 13,
$d)$ 21, $e)$ 22, $f)$ 23, $g)$ 31, $h)$ 32, $i)$ 33.
The parameters are the same as those in
Fig. \ref{fig:Delta03}.
\label{fig:spin_QPI_Im}
}
\end{figure}

In this part, the QPI patterns for the magnetic im- purity are presented, corresponding to $\Delta \rho^{\mu\nu}_{sf}(\omega,\vec{q})$ with $\nu\neq 0$. The magnetic impurity may also trap the Yu- Shiba bound states. The number of such states are finite, and their spatial distributions are only localized around the magnetic impurities. They exhibit sharp resonance peaks in the LDOS in the STM spectra around the impurity. However, the QPI spectra are the Fourier transform of LDOS over a large area, which are mostly related to the scattering states. Hence, the Yu-Shiba states can be neglected in calculating the QPI spectra without affecting any characteristic features. Nevertheless, their possible existence is certainly an interesting question worthwhile for a future investigation.

The magnetic impurity is assumed to be in the classical limit without fluctuations.
Although in principle, a magnetic impurity could also induce
the density response, it is a high order effect not showing up
at the level of the second order Born approximation.
Only $\mu,\nu \neq 0$ are displayed here.
For simplicity, we will use Latin indices to
refer spin directions $1,2,3$.

Fig. \ref{fig:spin_QPI_Re} and \ref{fig:spin_QPI_Im} show the real
and imaginary parts of $\Delta\rho_{sf}^{ij} (\epsilon, \vec{q})$,
respectively.
The magnetic impurity Hamiltonian $H_{\text{imp}}^{j}$ ($j=1,2,3$)
is even under the chiral operation, hence, it only induces scatterings
between Majorana islands with the same chiral index denoted as
scatterings ``$0$'' and ``$2$" in Fig. \ref{fig:surface spectrum}.
For example, these two classes of scatterings are marked in the
QPI patterns of $\Delta \rho_{sf}^{11}(\omega, \vec{q})$ shown in
Fig. \ref{fig:spin_QPI_Re} $(a)$ and Fig.\ref{fig:spin_QPI_Im} $(a)$.

The consequences of the point group symmetry are more complicated.
Let us first consider the magnetic impurity with spin oriented along
the $z$-direction, the impurity Hamiltonian still preserves
the $C_{3}$ symmetry.
Hence, the QPI patterns of $\Delta \rho^{33}(\epsilon,\vec q)$ explicitly
exhibit the $C_3$ symmetry as shown in Fig. \ref{fig:spin_QPI_Re} $i$)
and Fig. \ref{fig:spin_QPI_Im} $i$).
As for $\Delta\rho^{13}(\epsilon,\vec q)$ and $\Delta\rho^{23}(\epsilon,\vec{q})$,
their symmetry properties under the $C_3$ rotation
\bea
\Delta \rho^{i3} (\epsilon,R\vec{q})= R_{ij}\Delta \rho^{j3}
(\epsilon,\vec{q}),
\label{eq:C_3v_in_QPI}
\eea
where $R_{ij}$ refers to the $2\times 2$ rotation matrix of
a $C_3$ rotation.
For example, for the rotation $R(\hat z, \frac{2\pi}{3})$,
$\Delta\rho^{ij}$ satisfies
\bea
\left( \begin{array}{c}
\Delta\rho^{13}(\epsilon,\vec{q})\\
\Delta \rho^{23}(\epsilon,\vec{q})
\end{array}\right)=
\frac{1}{2}\left( \begin{array}{cc}
-1& \sqrt{3}\\
-\sqrt{3}& -1
\end{array}\right)
\left( \begin{array}{c}
\Delta\rho^{13}(\epsilon,\vec{q}^{\prime})\\
\Delta\rho^{23}(\epsilon,\vec{q}^{\prime})
\end{array}\right),
\eea
where $q^\prime= R(\hat{z},\frac{2\pi}{3})\vec{q}$, and
this property has been checked for Fig. \ref{fig:spin_QPI_Re}
($c,f$) and Fig. \ref{fig:spin_QPI_Im} ($c,f$).
On the other hand, the vertical reflection symmetry with respect
to $xz$-plane is broken, nevertheless, it can be restored by
combining with TR operation.
$S^z$ and $S^x$ are odd for this operation, while $S^y$ is even,
hence,
\bea
\Delta\rho^{i3}(\epsilon,\vec q)=\pm\Delta \rho^{i3}(\epsilon,\vec q^\prime),
\eea
where $+$ applies for $i=x,z$ and $-$ applies for $i=y$;
$\vec q^\prime$ is the image of $\vec q$ after the reflection.
Similar transformations can be derived for other planes
equivalent to the $xz$-plane by the $C_3$ rotations.
It is easy to check that all of Fig. \ref{fig:spin_QPI_Re} ($c,f,i$)
and \ref{fig:spin_QPI_Im} ($c,f,i$) satisfy these properties.

Now we consider the case of the impurity spin orientation along
the $x$-direction.
Then the $C_3$ rotation symmetry is no long kept.
The symmetry of the combined reflection followed by TR operation is
still valid, nevertheless, the reflection plane can only be the
$xz$-plane.
Hence, we have
\bea
\Delta\rho^{i1}(\epsilon,\vec q)=\pm\Delta \rho^{i1}(\epsilon,\vec q^\prime),
\eea
where $+$ applies for $i=x,z$, and $-$ applies for $i=y$.
Again this can be checked by examining
Fig. \ref{fig:spin_QPI_Re} ($a,d,g$)
and Fig. \ref{fig:spin_QPI_Im} ($a,d,g$).
At last, we examine the case of the impurity spin orientation along
the $y$-direction.
Again the $C_3$ rotation symmetry is lost, while the refection
symmetry with respect to the $xz$-plane is maintained.
\bea
\Delta\rho^{i2}(\epsilon,\vec q)=\mp\Delta \rho^{i2}(\epsilon,\vec q^\prime),
\eea
where $-$ applies for $i=x,z$, and $+$ applies for $i=y$.
Clearly this symmetry is respected in Fig. \ref{fig:spin_QPI_Re} ($b,e,h$)
and Fig. \ref{fig:spin_QPI_Im} ($b,e,h$).

\section{Summary}
\label{sec:summary}
In summary, the non-centrosymmetric effective spin-$\frac{3}{2}$
systems with cubic group symmetries are discussed.
The emphasis is put on the $T_d$ group, which is relevant to the YPtBi material.
The double degeneracy along $[0\,0\,1]$ and equivalent directions for systems with $T_d$ symmetry is shown to be protected by the little group $SD_{16}$.
Majorana surface states are calculated for the proposed mixed
$s$-wave singlet and $p$-wave septet pairing of the $T_d$ case in $(1\,1\,1)$-surface.
Two representative values of the ratio between $s$- and $p$-wave pairing components are taken as examples for calculations.
The Majorana states form flat bands within the regions enclosed by the projections of
the nodal loops in gap functions on the surface Brillouin zone,
but disappear in the overlapping regions.
The results are consistent with the bulk-edge correspondence principle.
The QPI patterns are computed for the surface states with a single impurity
in Born approximation.
Chiral symmetry forbids scatterings between Majorana islands of same (opposite) chiral index for the non-magnetic (magnetic) impurities.
There are richer structures in the QPI patterns under the transformation of $C_{3v}$ group for the case of magnetic impurity than the case of non-magnetic impurity.
Experimental signatures on the QPI patterns can test the
possible mixed $s,p$ pairing for the half-Heusler compound YPtBi.


{\it Acknowledgments.}
W. Y. and C. W. acknowledge the support from
the NSF DMR-1410375 and AFOSR FA9550-14-1-0168.
T. X is supported by National Key R\&D Program of China 2017YFA0302901
and National Natural Science Foundation of China Grant No 11474331.
C. W. also acknowledges the support from the National Natural
Science Foundation of China (11729402).

{\it Note added.
Near the completion of this manuscript, we learned the recent
works on superconductivity on the half-Heusler material
\cite{Savary2017,Boettcher2017,timm2017}.
}


\appendix
\section{Invariants of the cubic groups}
\label{sec:Cubic groups}

\begin{table}
\begin{center}
\begin{tabular}{| c | c | c | c | }
\hline
 $E$ & $1$ & $(x,y,z)$ & \text{identity}  \\
 \hline
 \multirow{3}{*}{$3C_2$}& $2$ & $(x,-y,-z)$ & $R(OX,\pi)$ \\
 \cline{2-4}
& $3$ & $(-x,y,-z)$ & $R(OY,\pi)$ \\
 \cline{2-4}
& $4$ & $(-x,-y,z)$ & $R(OZ,\pi)$ \\
 \hline
 \multirow{6}{*}{$6C_4$}  & $5$ & $(x,z,-y)$ & $R(OX,\frac{\pi}{2})$ \\
 \cline{2-4}
 & $6$ & $(x,-z,y)$ & $R(OX^{\prime},\frac{\pi}{2})$ \\
 \cline{2-4}
 & $7$ & $(-z,y,x)$ & $R(OY,\frac{\pi}{2})$\\
 \cline{2-4}
 & $8$ & $(z,y,-x)$ & $R(OY^{\prime},\frac{\pi}{2})$ \\
 \cline{2-4}
 & $9$ & $(y,-x,z)$ & $R(OZ,\frac{\pi}{2})$ \\
 \cline{2-4}
 & $10$ & $(-y,x,z)$ & $R(OZ^{\prime},\frac{\pi}{2})$ \\
 \hline
 \multirow{6}{*}{$6C_{2}^{'}$} & $11$ & $(y,x,-z)$ & $R([AC],\pi)$ \\
 \cline{2-4}
 & $12$ & $(-y,-x,-z)$ & $R([BD],\pi)$ \\
 \cline{2-4}
 & $13$ &$(z,-y,x)$ & $R([AB],\pi)$ \\
 \cline{2-4}
 & $14$ & $(-z,y,-x)$ & $R([CD],\pi)$\\
 \cline{2-4}
 & $15$ & $(-x,z,y)$ & $R([AD],\pi)$\\
 \cline{2-4}
 & $16$ & $(-x,-z,-y)$ & $R([BC],\pi)$ \\
 \hline
 \multirow{8}{*}{$8C_3$}&  $17$ & $(y,z,x)$ & $R(OA,\frac{2\pi}{3})$ \\
 \cline{2-4}
 & $18$ & $(z,x,y)$ & $R(OA^{\prime},\frac{2\pi}{3})$ \\
 \cline{2-4}
 & $19$ & $(-y,-z,x)$ & $R(OB,\frac{2\pi}{3})$ \\
 \cline{2-4}
 & $20$ & $(z,-x,-y)$ &  $R(OB^{\prime},\frac{2\pi}{3})$\\
 \cline{2-4}
 & $21$ & $(y,-z,-x)$ &  $R(OC,\frac{2\pi}{3})$\\
 \cline{2-4}
 & $22$ & $(-z,x,-y)$ &  $R(OC^{\prime},\frac{2\pi}{3})$\\
 \cline{2-4}
 & $23$ & $(-y,z,-x)$ &  $R(OD,\frac{2\pi}{3})$\\
 \cline{2-4}
 & $24$ & $(-z,-x,y)$ &  $R(OD^{\prime},\frac{2\pi}{3})$\\
 \hline
\end{tabular}
\caption{List of $24$ group elements of point group $O$.
In accordance with the notations in Fig. \ref{fig:TdCube}, $OM$ represents the vector pointing from the center of the cube (i.e. the point $O$) to the vertex or the direction $M$, where $M$ is one of $A,\,A^{\prime},\,B,\,B^{\prime},\,C,\,C^{\prime},\,D,\,D^{\prime}$ when it is a vertex of the cube, and is one of $X,\,Y,\,Z,\,X^{\prime},\,Y^{\prime},\,Z^{\prime}$ when it represents a direction.
$X,\,Y,\,Z$ represent the positive directions of the three axes $x,\,y,\,z$, and $X^{\prime},\,Y^{\prime}\,Z^{\prime}$ represent the negative directions of the three axes.
The symbol $[MN]$ represents the line passing through the point that bisects the edge $MN^{\prime}$ and the point that bisects $M^{\prime}N$,
where $M,\,N,\,M^{\prime},\,N^{\prime}$ are all vertices of the cube.
\label{table:O}
}
\end{center}
\end{table}

\begin{table}
\begin{center}
\begin{tabular}{| c | c | c | c | c | c | }
\hline
& $O_h$ & $O$ & $T_d$ & $T_h$ & $T$ \\
\hline
$\{k_i\}_{1\leq i \leq 3}$ & $T_{1u}$ & $T_1$ & $T_2$ & $T_u$ & $T$\\
\hline
$k_x^2+k_y^2+k_z^2$ & $A_{1g}$ & $A_1$ & $A_1$ & $A_g$ & $A$ \\
\hline
$\{(k_x^2+k_y^2-2k_z^2)/\sqrt{2},k_x^2-k_y^2 \}$ & $E_g$ & $E$ & $E$ & $E_g$ & $E$ \\
\hline
$\{k_i k_{i+1}\}_{1\leq i \leq 3}$ & $T_{2g}$ & $T_2$ & $T_2$ & $T_g$ & $T$ \\
\hline
$k_x k_y k_z$ & $A_{2u}$ & $A_2$ & $A_1$ & $A_u$ & $A$ \\
\hline
$\{ k_i^3 \}_{1 \leq i \leq 3}$ & $T_{1u}$ & $T_1$ & $T_2$ & $T_u$ & $T$ \\
\hline
$\{ k_i(k_{i+1}^2+k_{i+2}^2) \}_{1\leq i \leq 3}$ & $T_{1u}$ & $T_1$ & $T_2$ & $T_u$ & $T$\\
\hline
$\{ k_i(k_{i+1}^2-k_{i+2}^2) \}_{1\leq i \leq 3}$ & $T_{2u}$ & $T_2$ & $T_1$ & $T_u$ & $T$\\
\hline
\end{tabular}
\end{center}
\caption{Classifications of the momentum spherical harmonics up to
the cubic order according to irreducible representations of
the five cubic point groups.
The subscripts $i,i+1,i+2$ are defined cyclically in $x,y,z$.
\label{table:momentum classify}
}
\end{table}

\begin{table}
\begin{center}
\begin{tabular}{| c | c | c | c | c | c | }
\hline
& $O_h$ & $O$ & $T_d$ & $T_h$ & $T$ \\
\hline
$\{S_i\}_{1\leq i \leq 3}$ & $T_{1g}$ & $T_1$ & $T_1$ & $T_g$ & $T$\\
\hline
$\{(S_x^2+S_y^2-2S_z^2)/\sqrt{2},S_x^2-S_y^2 \}$ & $E_g$ & $E$ & $E$ & $E_g$ & $E$ \\
\hline
$\{S_i S_{i+1} + S_{i+1} S_{i}\}_{1\leq i \leq 3}$ & $T_{2g}$ & $T_2$ & $T_2$ & $T_g$ & $T$ \\
\hline
Sym($S_x S_y S_z$) & $A_{2g}$ & $A_2$ & $A_2$ & $A_g$ & $A$ \\
\hline
$\{ S_i^3 \}_{1 \leq i \leq 3}$ & $T_{1g}$ & $T_1$ & $T_1$ & $T_g$ & $T$ \\
\hline
$\{S_{i+1} S_i S_{i+1} - S_{i+2} S_i S_{i+2}\}_{1\leq i \leq 3}$ & $T_{2g}$ & $T_2$ & $T_2$ & $T_g$ & $T$\\
\hline
\end{tabular}
\end{center}
\caption{Classifications of spin tensors up to the third rank
according to the irreducible representations of the five cubic groups.
The symbol "Sym" represents symmetrization as a sum of all permutations of the objects inside the symbol.
The subscripts $i,i+1,i+2$ are defined cyclically in $x,y,z$.
\label{table:spin classify}
}
\end{table}

\begin{table}
\begin{center}
\begin{tabular}{| c | c | c | c | c | c |}
\hline
& $O_h$ & $O$ & $T_d$ & $T_h$ & $T$ \\
\hline
$k_x k_y k_z\cdot \text{Sym}(S_xS_yS_z)$ & $A_{1u}$ & $A_1$ & $A_2$ & $A_u$ & $A$\\
\hline
$k_i \cdot S_i$ & $A_{1u}$ & $A_1$ & $A_2$ & $A_u$ & $A$ \\
\hline
$k_i\cdot S_i^3$ & $A_{1u}$ & $A_1$ & $A_2$ & $A_u$ & $A$ \\
\hline
$k_i \cdot (S_{i+1}S_iS_{i+1}-S_{i+2}S_iS_{i+2})$ & $A_{2g}$ & $A_2$ & $A_1$ & $A_u$ & $A$ \\
\hline
$k_i^3 \cdot S_i$ & $A_{1u}$ & $A_1$ & $A_2$ & $A_u$ & $A$\\
\hline
$k_i^3 \cdot S_i^3$ & $A_{1u}$ & $A_1$ & $A_2$ & $A_u$ & $A$ \\
\hline
$k_i^3 \cdot (S_{i+1}S_iS_{i+1}-S_{i+2}S_iS_{i+2})$  & $A_{2g}$ & $A_2$ & $A_1$ & $A_u$ & $A$\\
\hline
$k_i(k_{i+1}^2+k_{i+2}^2) \cdot S_i$ & $A_{1u}$ & $A_1$ & $A_2$ & $A_u$ & $A$ \\
\hline
$k_i(k_{i+1}^2+k_{i+2}^2) \cdot S_i^3$ & $A_{1u}$ & $A_1$ & $A_2$ & $A_u$ & $A$ \\
\hline
$k_i(k_{i+1}^2+k_{i+2}^2) \cdot (S_{i+1}S_iS_{i+1}-S_{i+2}S_iS_{i+2})$ & $A_{2g}$ & $A_2$ & $A_1$ & $A_u$ & $A$ \\
\hline
$k_i(k_{i+1}^2-k_{i+2}^2) \cdot S_i$ & $A_{2u}$ & $A_2$ & $A_1$ & $A_u$ & $A$ \\
\hline
$k_i(k_{i+1}^2-k_{i+2}^2) \cdot S_i^3$ & $A_{2u}$ & $A_2$ & $A_1$ & $A_u$ & $A$ \\
\hline
$k_i(k_{i+1}^2-k_{i+2}^2) \cdot (S_{i+1}S_iS_{i+1}-S_{i+2}S_iS_{i+2})$ & $A_{1u}$ & $A_1$ & $A_2$ & $A_u$ & $A$ \\
\hline
\end{tabular}
\end{center}
\caption{The TR and rotation-invariant combinations of momentum and spin tensors up to the third rank.
The second rank combinations preserving the inversion symmetry are not shown since they have been already included in the Luttinger-Kohn Hamiltonian.
"$\cdot$" represents the inner product as a summation of $i$ from $1$ to $3$.
\label{table:A-combination}
}
\end{table}

In this appendix, we briefly describe the five cubic point groups $O_h,\,O,\,T_d,\,T_h,\,T$, then classify the spherical
harmonics of momentum and spin tensors up to the third rank
according to their irreducible representations.
All cubic-group-symmetry invariants up to the third rank
of momentum and spin tensors are presented.

Let us recall some group theory knowledge.
The $O_h$ group is the symmetry group of the cube, containing $48$ elements,
hence, is the largest one among the five cubic groups.
The other four are its subgroups, which are the symmetry groups of
decorated cubes in different ways.
Fig. \ref{fig:TdCube} shows the case of the $T_d$ group.
In the designated coordinate system, $O$ is at the origin.
Among the eight vertices, $A, B, C, D$ are located at
$(-1,-1,-1), (1,-1,1), (1,1,-1)$ and $(-1,1,1)$, respectively,
and $A^\prime, B^\prime, C^\prime, D^\prime$ are their
inversion symmetric partners, respectively.
We use $\mathcal{I}$ to denote the inversion operation,
and $R(\hat{n},\theta)$ the rotation around the direction
$\hat{n}$ by the angle $\theta$.
The group $O_h$ has $24$ proper elements, which corresponds to
rotations, and their conjugation classes $E$, $3C_2$, $6C_4$,
$6C_2^{'}$ and $8C_3$ are listed in Table \ref{table:O}.
The other $24$ elements are improper operations corresponding to
combinations of rotation and inversion.
Their conjugation classes are $i$, $3\sigma_h$, $6S_4$,
$6\sigma_d$, and $8S_6$ by applying the inversion operation to 
$E$, $3C_2$, $6C_4$, $6C_2^{'}$ and $8C_3$, respectively.
For simplicity, they are not listed.
The other four cubic groups are subgroups of $O_h$
represented by the conjugation classes as
\bea
O&=&\{E,3C_2,6C_4,6C_2^\prime,8C_3\}, \nn \\
T_d&=&\{E,3C_2,8C_3,6S_4,6\sigma_d\}, \nn \\
T_h&=&\{E, 3C_2,8C_3,i,3\sigma_h,8S_6\}, \nn \\
T&=& \{E, 3C_2, 8C_3 \}.
\eea

In Table \ref{table:momentum classify}, we list all the momentum
harmonics up to the third rank.
There are three, six, and ten harmonics for rank-1, 2 and 3,
respectively.
The decompositions to cubic groups are
$\text{rank-}1=T$, $\text{rank-}2=A\oplus E \oplus T$,
and $\text{rank-}3=A\oplus T\oplus T\oplus T$.
For different groups, $A$ and $T$ can be further classified
to $A_{1,2}$, and $T_{1,2}$, respectively.
For groups containing inversion operation, the sub-index $g$ and $u$
mean representations of even and odd parities, respectively.
The spherical spin tensors are presented in Tab. \ref{table:spin classify}.
Since $\sum_i S_i^2=S(S+1)$, there are only 5 and 7 independent
spin tensors at rank-2 and 3, respectively.
Since spin operators are parity even, all the corresponding representations
are of the $g$-type.

Here we list all spin-orbit coupled invariants up to the third order
in momentum for three inversion-breaking cubic groups $O, T_d, T$.
For the point group $O$, the invariants are
\bea
& k_i\cdot S_i, \ \ \, k_i\cdot S_i^3,  \nn\\
& k_xk_yk_z\text{Sym}(S_xS_yS_z), \nn\\
& k_i^3\cdot S_i, \ \ \, k_i^3\cdot S_i^3,\nn\\
& k_i(k_{i+1}^2+k_{i+2}^2) \cdot S_i, \ \ \,
k_i(k_{i+1}^2+k_{i+2}^2) \cdot S_i^3,\nn\\
& k_i(k_{i+1}^2-k_{i+2}^2) \cdot (S_{i+1}S_iS_{i+1}-S_{i+2}S_iS_{i+2}).
\label{eq:O combination}
\eea
The $T_d$ invariants are
\bea
& k_i\cdot (S_{i+1}S_iS_{i+1}-S_{i+2}S_iS_{i+2}),\nn\\
& k_i^3\cdot (S_{i+1}S_iS_{i+1}-S_{i+2}S_iS_{i+2}),\nn\\
& k_i(k_{i+1}^2-k_{i+2}^2)\cdot S_i,  \ \ \, k_i(k_{i+1}^2-k_{i+2}^2)\cdot S_i^3, \nn\\
& k_i(k_{i+1}^2+k_{i+2}^2) \cdot (S_{i+1}S_iS_{i+1}-S_{i+2}S_iS_{i+2}).
\label{eq:Td combination}
\eea
For $T$, all the invariants of $O$ and $T_d$ are allowed.
In these expressions, the dot "$\cdot$" represents an inner product
between the two $3$-vectors, in which "$i$" runs over $x,y,z$ and
a summation $\sum_{i=x,y,x}$ is taken.
Inversion symmetry is explicitly broken by all the terms,
hence the double degeneracy in the heavy hole and light
hole bands is in general absent.

The TR symmetry requires that the homogeneity of the combinations be even.
The inversion-preserving combinations have already been included in the Luttinger-Kohn Hamiltonian Eq. (\ref{eq:Luttinger}).
There are $17$ inversion breaking combinations which are listed in Table \ref{table:A-combination}.
All of these combinations belong to $A$-representations of the cubic groups.
Whether they are $A_1$, $A_2$, or $A_g$, $A_u$ of a particular cubic
group can be obtained from the information in Tab.
\ref{table:momentum classify} and \ref{table:spin classify}
and the multiplication rules of the representations.

\section{The semi-dihedral group $SD_{16}$}
\label{sec:little group}

In this appendix, we discuss the little group structure along the
$[0\,0\,1]$ and its equivalent directions of the lattice system with the $T_d$ symmetry
augmented by spinor representations and TR symmetry.
This little group is isomorphic to the semi-dihedral group
(also called the quasi-dihedral group) $SD_{16}$, where
the subscript ``$16$" represents the order of the group.

With the $T_d$ group, its little group $L_0$ along the $[0\,0\,1]$
direction is
\bea
L_0=\{\mathbbm{1},M_{x^{\prime}},M_{y^{\prime}},R(\vec{z},\pi) \},
\eea
in which $\mathbbm{1}$ is the identity element, and the directions of $x^\prime$ and $y^\prime$ are given in Fig. \ref{fig:TdCube}.
The reflection $M_{x^{\prime}}$ can be decomposed as $M_{x^{\prime}}=\mathcal{I} R(\hat{x^{\prime}},\pi)$.
$L_0$ is isomorphic to $D_2$, the dihedral group of order $4$,
which is Abelian and only has 1D irreducible representations.

For half-spin fermions, $L_0$ needs to be doubled to $L_1$
by adding $\bar{\mathbbm{1}}=R(\vec{n},2\pi)$, then $L_1$
is represented by
\bea
L_1&=&\{\mathbbm{1},M_{x^{\prime}},M_{y^{\prime}},R(\vec{z},\pi),\nn\\
&&\bar{\mathbbm{1}},\bar{\mathbbm{1}}M_{x^{\prime}},
\bar{\mathbbm{1}}M_{y^{\prime}},\bar{\mathbbm{1}}R(\vec{z},\pi) \}.
\eea
$\bar{\mathbbm{1}}$ commutes with every element in the group, which
takes $-1$ for half-integer spin representations.
Since the inversion operator $\mathcal{I}$ commutes with all $O(3)$ elements and acts as identity operator in spin space, we can explicitly check that
\bea
M_{x^{\prime}}^2&=&R(\hat{x^{\prime}},2\pi)=\bar{\mathbbm{1}}, \nn \\
M_{x^{\prime}}M_{y^{\prime}}&=&R(\vec{z},\pi)=-M_{y^{\prime}}M_{x^{\prime}},
\eea
hence $L_1$ is non-Abelian.
We can explicitly work out the multiplication rules of $L_1$, which shows
that it is isomorphic to the quaternion group,
\bea
Q_8=\{\pm 1, \pm i,\pm j,\pm k\},
\eea
through the identifications $\mathbbm{1}=1$, $\bar{\mathbbm{1}}=-1$, $M_{x^{\prime}}=i$, $M_{y^{\prime}}=j$, and $R(\vec{z},\pi)=k$.
The multiplication rules of $Q_8$ are given by
\bea
&i^2=j^2=k^2=-1,\nn\\
&ij=-ji=k,jk=-kj=i,ki=-ik=j.
\label{eq:Q8_multiplication}
\eea
$Q_8$ and hence $L_1$ has one 2D irreducible representations
up to isomorphism in which $\mathbbm{1}$ and $\bar{\mathbbm{1}}$
take the value of $1$ and $-1$, respectively.
and four 1D irreducible representations in which $\mathbbm{1}$
and $\bar{\mathbbm{1}}$ are identical.

Now we extend $T_d$ to its magnetic group, and identify its
little group along the $[0\, 0\, 1]$-direction.
The anti-unitary operator $\mathcal{S}$ defined in Eq. \ref{eq:S} as
$\mathcal{S}=R(\hat{z},\frac{\pi}{2})\mathcal{T}^{\prime}$
is an element of this little group, where
$\mathcal{T}^{\prime}=\mathcal{I}\mathcal{T}$.
$\mathcal{T}^{\prime}$  leaves the momentum direction
$[0\,0\,1]$-direction unchanged, and
$\mathcal{T}^{\prime2}=1$ or $-1$ depending on whether the spin is
integer or half-odd integer.
In the case of $\mathcal{T}^{\prime2}=1$, the magnetic little group
is denoted as $L_2$
\bea
L_2&=&\{\mathbbm{1},M_{x^{\prime}},M_{y^{\prime}},R(\vec{z},\pi),\nn\\
&&\mathcal{S},  M_{x^{\prime}}\mathcal{S},M_{y^{\prime}}\mathcal{S},R(\vec{z},\pi)\mathcal{S}\}.
\eea
Defining $r=R(\hat{z},\frac{\pi}{2})\mathcal{T}^{\prime}$, $s=M_{x^{\prime}}$,
$L_2$ can be rewritten as
\bea
L_2=\{1, r, r^2, r^3 , s, sr,sr^2,sr^3 \},
\eea
which is isomorphic to $D_4$, the dihedral group of order $4$, with
the relations of $r^4=s^2=1$, $srs^{-1}=r^{-1}$.

Finally we consider the case of $\mathcal{T}^{\prime2}=-1$.
We define $r^\prime$ as the spinor version of $r$ and
$s^{\prime}=r^\prime M_{x^{\prime}}$.
Then the magnetic little group $L_3$ for the $[0\, 0\, 1]$-direction
can be represented in terms of $r^\prime$ and $s^{\prime}$ as
\bea
L_3&=&\{\mathbbm{1}, r^\prime,r^{\prime 2},r^{\prime 3},\nn\\
&&r^{\prime 4},r^{\prime 5},r^{\prime 6},r^{\prime 7},\nn\\
&&s^{\prime},s^{\prime}r^\prime,s^{\prime}r^{\prime 2},s^{\prime}r^{\prime 3},\nn\\
&&s^{\prime}r^{\prime 4},s^{\prime}r^{\prime 5},s^{\prime}r^{\prime 6},s^{\prime}r^{\prime 7}
\}.
\eea
This is in fact isomorphic to the semi-dihedral group $SD_{16}$ of order $16$ defined in terms of generators and relations as
\bea
SD_{16}=\langle r^\prime,s^\prime | r^{\prime 8}=s^{\prime 2}=1,s^\prime r^\prime s^{\prime -1}=r^{\prime 3}\rangle.
\eea
Here we show that $s^{\prime}r^\prime s^{\prime-1}=r^3$ as follows,
\bea
s^{\prime}r^\prime s^{\prime-1}&=&M_{x^{\prime}} r^\prime M_{x^{\prime}}^{-1} \mathcal{T}^{\prime}\nn\\
&=& R(\hat{x^{\prime}},\pi) R(\hat{z},\frac{\pi}{2}) R(\hat{x^{\prime}},\pi)^{-1} \mathcal{T}^{\prime}\nn\\
&=& R(\hat{z},\frac{7}{2}\pi) \mathcal{T}^{\prime}= \mathbbm{\bar{1}}R(\hat{z},\frac{\pi}{2}\pi) \mathcal{T}^{\prime}\nn\\
&=&R(\hat{z},\frac{3}{2}\pi) \mathcal{T}^{\prime3}=r^{\prime 3},
\eea
in which in the second last line $R(\hat{z},2\pi)=\mathbbm{\bar{1}}$ is used.
We further note that $SD_{16}$ has both $Q_8$ and $D_4$ as subgroups.
The $Q_8$ subgroup is generated by $\{r^{\prime 2},r^\prime s^{\prime}\}$, while the $D_4$ subgroup is generated by $\{r^{\prime 2},s^{\prime}\}$.

\section{The anti-unitary operator $\mathcal{S}$ with $\mathcal{S}^4=-1$}
\label{sec:antiunitary}
In this appendix,
we will give the explicit form of the operator $\mathcal{S}$,
and its action on the two doubly degenerate subspaces along $[0\,0\,1]$ directions.

Since $\mathcal{I}$ acts as identity operator in the spin space,
 the anti-unitary operation $\mathcal{S}$ is given by
\bea
\mathcal{S}=e^{-i S_z (-\pi/2)} R\cdot K,
\eea
in which $K$ is the complex conjugate operation.
Then $\mathcal{S}$ is computed as $\mathcal{S}=M K$, where
\bea
M=\left(\begin{array}{cccc}
0 & 0 & 0 & e^{i\frac{3\pi}{4}}\\
0 & 0 & -e^{i\frac{\pi}{4}} & 0\\
0 & e^{-i \frac{\pi}{4}} &0 & 0\\
-e^{-i \frac{3\pi}{4}} & 0 & 0  & 0\\
\end{array}
\right).
\eea
It is straightforward to verify that $\mathcal{S}^2=\text{diag}\{i,-i,i,-i\}$,
and $\mathcal{S}^4=-1$.

Up to an overall factor, the Hamiltonian Eq. (\ref{eq:H0}) along
the $[0\, 0\, 1]$-direction is
\bea
H_z=S_z^2+\delta (S_x S_z S_x-S_y S_z S_y),
\label{eq:100direction}
\eea
in which $|\delta|<<1$ for the case of small inversion breaking strength.
Since the Hamiltonian Eq. (\ref{eq:100direction}) changes the $S_z$-eigenvalue by $0$ or $2$, the helicity $\frac{3}{2}$ component will mix with the
helicity $-\frac{1}{2}$ component, and similarly, the helicity $\frac{1}{2}$-component mixes with  the helicity $-\frac{3}{2}$ component.

As proved in Sect. \ref{sect:degenercy}, the spectra of
Eq. \ref{eq:100direction} are doubly degenerate.
The two eigenvectors $v_{1,2}$ corresponding to $E_v=\frac{1}{4} (5+ 2 \sqrt{4 + 3 \delta^2})$ are
\bea
v_1&=&\frac{1}{\mathcal{N}}\Big (1, 0, \frac{2}{ \sqrt{3} \delta} (\sqrt{1+\frac{3}{4} \delta^2}-1), 0\Big)^T,\nn\\
v_2&=&\frac{1}{\mathcal{N}}\Big (0,-\frac{2}{ \sqrt{3} \delta} (\sqrt{1+\frac{3}{4} \delta^2}-1),0,1\Big )^T,
\eea
in which $\mathcal{N}$ is the normalization factor.
The two eigenvectors $w_{1,2}$ of $E_w=\frac{1}{4} (5 - 2 \sqrt{4 + 3 \delta^2})$ are
\bea
w_1&=&\frac{1}{\mathcal{N}}\Big( 0, 1, 0, \frac{2}{ \sqrt{3} \delta} (\sqrt{1+\frac{3}{4} \delta^2}-1)\Big)^T,\nn\\
w_2&=&\frac{1}{\mathcal{N}}\Big(-\frac{2}{ \sqrt{3} \delta} (\sqrt{1+\frac{3}{4} \delta^2}-1),0,1,0\Big)^T.
\eea

The anti-unitary operation $\mathcal{S}$ is diagonal-blocked.
Its off-diagonal elements between the two subspaces spanned
by $v_{1,2}$ and $w_{1,2}$ are zero.
In the subspace spanned by $v_{1,2}$ and that by $w_{1,2}$,
it has matrix structure as
\bea
\mathcal{S}=\left(\begin{array}{cc}
0 & e^{ i\pm\frac{\pi}{4}}\\
-e^{-i\mp \frac{\pi}{4}} & 0
\end{array}\right)K,
\eea
in which the upper sign is for subspace $v_{1,2}$ and lower sign for $w_{1,2}$.

\section{Topological index for nodal-line superconductors}
\label{sec:topo}

In this appendix, we first review the definition of the path-dependent topological number for the TR invariant nodal topological
superconductors \cite{Schnyder2012}, then apply it to the spin-$3/2$
case of the current interest.
The pairing strengths of $s$- and $p$-wave components are
parametrized as $\Delta_s=C_s \Delta_0$ and $\Delta_p=C_p \Delta_0$.

In the presence of TR symmetry, the combined operation $C=P_h T$,
dubbed as chiral operator, anti-commutes with the B-deG Hamiltonian.
As a result, $H_{\vec{k}}$ can be transformed into a block-off-diagonal
form as
\bea
W H_{\vec {k}} W^\dagger=
\left( \begin{array}{cc}
0& D^\dagger_{\vec k}\\
D_{\vec k}& 0
\end{array}
\right).
\eea
The matrix $D_{\vec{k}}$ can be decomposed into $D_{\vec{k}}=U_{\vec{k}} \Lambda_{\vec{k}} V_{\vec{k}}$ via singular-value decomposition,
in which $\Lambda_{\vec{k}}$ is a diagonal matrix with non-negative eigenvalues,
and $U_{\vec{k}},V_{\vec{k}}$ are unitary matrices.
Consider a closed path $\mathcal{L}$ in momentum space.
If the gap does not vanish along the path $\mathcal{L}$, then $\Lambda_{\vec{k}}$ on $L$ can be deformed into the
identity matrix, and then $D_{\vec{k}}$ becomes a unitary
matrix denoted as $Q_{\vec{k}}$.
The topological number of the path $\mathcal{L}$ is defined
by the formula \cite{Schnyder2012},
\bea
N_{\mathcal{L}}=\frac{1}{2\pi i} \int_L dk_l \, \text{Tr}[Q_{\vec{k}}^{\dagger} \partial_{k_l} Q_{\vec{k}}].
\label{eq:N_L}
\eea

Next we carry out the calculation of the topological number in Eq. (\ref{eq:N_L}) for the Hamiltonian Eq. (\ref{eq:Hamiltonian}) with
the $T_d$ symmetry.
The chiral operator $C=\tau_1 \otimes R$ is diagonalized by
the following matrix $W$,
\bea
W=\frac{1}{\sqrt{2}} \left( \begin{array}{cc}
I_4 & iR\\
-iR & I_4
\end{array}\right),
\eea
in which $I_4$ is the $4\times 4$ identity matrix.
The Hamiltonian $H_{\vec{k}}$ can be brought into a block off-diagonal form as
\bea
WH_{\vec{k}}W^{\dagger}=\left( \begin{array}{cc}
0&D_{\vec{k}}\\
D_{\vec{k}}^{\dagger}&0
\end{array}\right),
\eea
in which
\bea
D_{\vec{k}}=H_L(\vec{k})+\delta A(\vec{k})-\mu+i \Delta_0 (C_s+C_p A(\vec{k})).
\eea
Treating the inversion breaking term $\frac{\delta}{k_f} A(\vec{k})$
at the level of the first order degenerate perturbation theory,
the band energies $\epsilon^{(1/2)}_{\pm}(\vec{k})$ and $\epsilon^{(3/2)}_{\pm}(\vec{k})$ of
the spin-split light hole and heavy hole bands are given, respectively,  by
\bea
\epsilon^{1/2}_{\pm}(\vec{k})= (2\lambda_2+\lambda_1) k^2 \pm \delta |\Lambda^{(1/2)}(\vec{k})|-\mu, \nn\\
\epsilon^{3/2}_{\pm}(\vec{k})= (2\lambda_2-\lambda_1) k^2 \pm \delta |\Lambda^{(3/2)}(\vec{k})|-\mu.
\eea
Up to the first order in $\delta/|\mu|$ and $\Delta_0 /|\mu|$, the matrix $D_{\vec{k}}$ is diagonalized via a unitary transformation
$U_{\vec{k}}$ as
\bea
D_{\vec{k}}=U_{\vec{k}}
\text{diag}(\epsilon^{(\alpha)}_{\nu} (\vec{k})+i \Delta^{(\alpha)}_{\nu}(\vec{k}) )
U_{\vec{k}}^{\dagger},
\eea
in which $\alpha=\frac{3}{2}, \,\frac{1}{2}$, $\nu=\pm$,
and $\Delta^{(\alpha)}_{\nu} (\vec{k})=\Delta_0 (C_s+\nu C_p |\vec{\Lambda}^{(\alpha)}(\vec{k})|)$.
Then $Q_{\vec{k}}$ is derived as
\bea
Q_{\vec{k}}=U_{\vec{k}}\text{diag}(e^{i \theta^{(\alpha)}_{\nu} (\vec{k})})U^{\dagger}_{\vec{k}},
\eea
in which $\tan \theta^{(\alpha)}_{\nu} (\vec{k})=\Delta^{(\alpha)}_{\nu}(\vec{k}) / \epsilon^{(\alpha)}_{\nu}(\vec{k}) $.
Plugging $Q_{\vec k}$ in Eq. (\ref{eq:N_L}), we arrive at the topological
index $N_L$ as
\bea
N_{\mathcal{L}}=\frac{1}{2\pi} \sum_{\alpha,\nu} \int_{\mathcal{L}} dk_{l} \, \partial_{k_l} \theta^{(\alpha)}_{\nu}(\vec{k}_l).
\label{eq:topo reduced}
\eea

The bands of helicity $\pm \frac{1}{2}$ lie above the Fermi energy
at the energy of the order of $|\mu|$, hence, they will not give
rise to non-trivial contribution to $N_L$.
For the bands of helicity $\pm 3/2$, the formula for $N_L$
can be simplified to \cite{Schnyder2012}
\bea
N_{\mathcal{L}}=-\frac{1}{2} \sum_{\nu} \sum_{\vec{k}_F} \text{sgn} (\partial_{\vec{k}_l} \epsilon^{(3/2)}_{\nu}(\vec{k}_F) ) \cdot \text{sgn} (\Delta^{(3/2)}_{\nu}(\vec{k}_{F})),\nn\\
\eea
where $\vec{k}_F$'s are the wavevectors at which the path
$\mathcal{L}$ crosses the Fermi surfaces.
The equations $\epsilon^{(3/2)}_{\pm }(\vec{k})=0$ determines the smaller
(larger) Fermi surface.
These two Fermi surfaces touch along $[0\, 0\, 1]$ directions
protected by the little group $SD_{16}$ as analyzed in Sect. \ref{sect:degenercy}.
Assuming $\Delta_s,\Delta_p>0$, then the gap function $\Delta^{(3/2)}_+(\vec {k}_F)$ on the smaller Fermi surface
are positive definite.
A closed path $\mathcal{L}$ always crosses the Fermi surface even times,
and the sign of $(\partial_{\vec{k}_l} \epsilon^{(3/2)}_{\nu}(\vec{k_F}) )$ %
for crossing the Fermi surface from the inner to outer direction
is opposite to that from the opposite directions,
hence, the smaller Fermi surface does not contribute
the $N_{\mathcal{L}}$ as well.
Then the formula is simplified to
\bea
N_{\mathcal{L}}=-\frac{1}{2} \sum_{\vec{k}_F} \text{sgn} (\partial_{\vec{k}_l} \epsilon^{(3/2)}_{-}(\vec{k_F}) ) \cdot \text{sgn} (\Delta^{(3/2)}_{-}(\vec{k}_{F})),\nn\\
\label{eq:topo final}
\eea
in which only the larger Fermi surface contributes.

For the Hamiltonian Eq. (\ref{eq:Hamiltonian}) with $T_d$ symmetry,
the sign of the pairing $\Delta^{(3/2)}_{-}(\vec{k}_{F})$ inside the
nodal loop is opposite to that outside the loop.
Hence from Eq. (\ref{eq:topo final}) the topological number for a
closed path that encloses the nodal loop once is $\pm 1$, where
the sign of $N_{\mathcal{L}}$ depends on the direction that the path is traversed.

The physical meaning of Eq. (\ref{eq:topo final}) can be related to the sign structure of the gap functions along the incident and reflected wavevectors \cite{Schnyder2012}.
Let $\vec{k}_{\parallel}$ be an in-plane wavevector within the surface of interest.
The infinite vertical line passing through $\vec{k}_{\parallel}$ crosses the larger Fermi surface at two wavevectors with vertical components $k_{\perp1}$ and $k_{\perp2}$.
Now we can enclose the infinite line with a semi-infinite circle and consider the topological number of the combined path $\mathcal{L}_0$.
From Eq. (\ref{eq:topo final}), $N_{\mathcal{L}_0}$ is related to the sign difference between $\Delta^{(3/2)}_{-}(k_{\perp 1})$ and $\Delta^{(3/2)}_{-}(k_{\perp 2})$.
On the other hand, $N_{\mathcal{L}_0}$ does not change by continuously deforming the path as long as the nodal circles are not touched.
If after such a deformation no nodal loop is enclosed, the pairings
of the incident and reflected wavevectors are of the same sign,
corresponding to a topologically trivial situation.
If two nodal loops are enclosed and
the topological numbers from the two loops cancel, which
is a topologically trivial situation again.
In contrast, if only one nodal loop is enclosed, the topological
number $N_{\mathcal{L}_0}=\pm 1$, for which the Majorana zero modes
appear.

\section{The method for solving surface states}
\label{sec:solve Majorana}

In this appendix, the equation determining surface state
is derived in the limit $\Delta_0\ll \delta \ll |\mu|$.
We first obtain the eight pairs of $\{\vec{k}_l,\Phi_l\}$ from Eq. (\ref{eq:boundary1}) as functions of $E_s$,
then plug them in $\det(\{\Phi_l\}_{1\leq l \leq 8})=0$ to solve for $E_s$.
The pairing strengths and the surface energy are parametrized as $\Delta_s=C_s \Delta_0$, $\Delta_p=C_p \Delta_0$ and $E_s=\epsilon \Delta_0$.

Due to the translation symmetry in $xy$-plane, $(k_x,k_y)$ remain good 
quantum numbers.
The momentum $k_z$ solving Eq. (\ref{eq:boundary1}) can be expanded 
as $k_z=\pm k_{0z}+\zeta \delta -i \xi \Delta_0$,
in which $k_{0z}$ is the magnitude of the $z$'th component of the Fermi wavevector determined by the Luttinger-Kohn Hamiltonian,
$\zeta\delta$ originates from the splitting of Fermi surfaces due to the inversion breaking term in the band structure,
and $-i \xi \Delta_0$ is from the superconducting pairing.
Denote $k_{0z}^{(3/2)}$ and $k_{0z}^{(1/2)}$ to be the above mentioned 
$k_{0z}$ for the heavy and light hole bands, respectively.
Since the Fermi energy crosses the heavy hole bands,
$k^{(3/2)}_{0z}$ is real,
while $k^{(1/2)}_{0z}$ is purely imaginary.
The expressions of $k^{(3/2)}_{0z}$ and $k^{(1/2)}_{0z}$ are
\bea
k^{(3/2)}_{0z}&=&\sqrt{|\mu|/(2\lambda_2-\lambda_1)-k_x^2-k_y^2}, \nn\\
k^{(1/2)}_{0z}&=&-i \sqrt{|\mu|/(2\lambda_2+\lambda_1)+k_x^2+k_y^2}.
\eea
We also denote $k^{(3/2)}_{0}$ and $k^{(1/2)}_{0}$ as
\bea
k^{(3/2)}_{0}&=&\sqrt{|\mu|/(2\lambda_2-\lambda_1)}, \nn\\
k^{(1/2)}_{0}&=&-i \sqrt{|\mu|/(2\lambda_2+\lambda_1)}.
\eea

We will discuss the heavy and light hole bands separately,
and first consider the heavy hole bands.
The superscript ``$3/2$" on momentum will be dropped for simplicity.
Denote $\hat{a}$ to be the unit vector normal to the surface. 
Let $\vec{k}_{0\eta}=(k_x,k_y,\eta k_{0z})$ ($\eta=\pm 1$), and define $\mathcal{U}^{(3/2)}(\vec{k}_{0\eta})$ as
\begin{flalign}
&\mathcal{U}^{(3/2)}(\vec{k}_{0\eta})=\nn\\
&\left(\begin{array}{cc}
U(\hat{a}) & 0 \\
0 & U(\hat{a})^{T,-1}
\end{array}\right)
\left(\begin{array}{cc}
U(\hat{k}_{0\eta}) & 0 \\
0 & U(\hat{k}_{0\eta})^{T,-1}
\end{array}\right),
\end{flalign}
in which $U(\hat{a})=e^{-i S_z \phi_a} e^{-i S_y \theta_a}$ and
$U(\vec{k}_{0\eta})=e^{-i S_z \phi_{\eta}} e^{-i S_y \theta_{\eta}}$,
where $\theta_a$ and $\phi_a$ are the polar and azimuthal angles of $\hat{a}$,
and $\theta_{\eta}$ and $\phi_{\eta}$ are those of the vector $\vec{k}_{0\eta}$.
$\mathcal{U}^{(3/2)}(\vec{k}_{0\eta})$ corresponds to the helicity basis at momentum $\vec{k}_{0\eta}$.
Plug $k_z=\eta k_{0z}+\zeta \delta -i \xi \Delta_0$ ($\eta=\pm 1$) into the Hamiltonian in Eq. (\ref{eq:Hamiltonian}),
perform the transformation $\mathcal{U}^{(3/2)}(\vec{k}_{0\eta})^{-1} H(\vec{k}) \mathcal{U}^{(3/2)}(\vec{k}_{0\eta})$, and project into the heavy hole bands.
Then by keeping the leading order terms in the expansion over $\delta$ and $\Delta_0$, the $H_L$ and $H_A$ terms in Eq. (\ref{eq:Hamiltonian}) become
\bea
H^{\prime}_L-\mu&=&(\zeta \delta-i \xi \Delta_0) \big(2(\lambda_1+\frac{5}{2} \lambda_2) \eta k_{0z} \nn\\
&&-2 \lambda_2 k_0 P_{3/2}\{S_z,U(\vec{k}_{0\eta})^{-1} S_z U(\vec{k}_{0\eta}) \} P_{3/2}\big ), \nn \\
H^{\prime}_A&= &\frac{\delta}{k_f} \,P_{3/2} U(\vec{k}_{0\eta})^{-1} U(\hat{a})^{-1} A(R_a \vec{k}_{0\eta})\nn\\
&&\cdot U(\hat{a}) U(\vec{k}_{0\eta}) P_{3/2}, \nn\\
\label{eq:transformed}
\eea
in which $P_{3/2}$ is the projection operator to the heavy hole bands;
the superscript of prime denotes the terms after the transformation and the projection; $\{,\}$ represents the anti-commutator of two matrices;
and $k_0$ is $k^{(3/2)}_0$ with the superscript ``$3/2$" omitted.
$H^{\prime}_L$ and $H^{\prime}_A$ are $2\times 2$ matrices after the projection to the heavy hole bands.
$H^{\prime}_A$ is traceless, hence can be expanded in terms of Pauli matrices as
\bea
H^{\prime}_A=\delta\, \vec{\Lambda}(\vec{k}_{0\eta}) \cdot \vec{\sigma},
\eea
in which $\vec{\Lambda} (\vec{k}_{0\eta})$ is a three-component vector.
Let $D(\vec{k}_{0\eta})$ be the transformation that diagonalizes $H^{\prime}_A$.
The eigenvalues of $H^{\prime}_A$ are $\pm \delta |\vec{\Lambda}(\vec{k}_{0\eta})|$, which leads to Fermi surface splitting.
The correction of the Fermi wave vector due to the splitting is given by $\zeta \delta$,
and there are two values of $\zeta$ corresponding to the two eigenvalues of $H^{\prime}_A$.
In the following, for simplicity, we will term the basis after the transformation of $D(\vec{k}_{0\eta})$ as band structure basis.

Next we consider the effect of the superconducting pairing.
Since $\Delta_0\ll \delta$, as long as $H^{\prime}_A$ does not vanish,
we can project the superconducting pairing onto the eigen-bases of
the band Hamiltonian, and the corrections from the mixing between 
different spin-split bands are of higher orders in $\Delta_0/\delta$.
Let $P_{\nu}$ ($\nu=\pm 1$) be the projection operator to one of the two band structure basis with eigenvalue $\nu\delta \sqrt{\vec{\Lambda}(\vec{k}_{0\eta})^2}$ of $H^{\prime}_A$.
Plug $k_z=\eta k_{0z}+\zeta \delta -i \xi \Delta_0$ into Eq. (\ref{eq:boundary1}),
and set $k_z$ to be $\eta k_{0z}$ in the pairing Hamiltonian with corrections of high orders.
Then Eq. (\ref{eq:boundary1}) becomes a two-component eigen-equation, as
\bea
\left(\begin{array}{cc}
i\xi \gamma(\vec{k}_{0\eta}) -\epsilon  & (-)^{\frac{1-\nu}{2}} \chi_{\nu} (\vec{k}_{0\eta}) \\
(-)^{\frac{1-\nu}{2}}{\chi}_{\nu}(\vec{k}_{0\eta})^* & -i\xi \gamma(\vec{k}_0) -\epsilon
\end{array} \right) \Phi^{\prime}=0,
\label{eq:2eigen}
\eea
in which
\bea
\gamma(\vec{k}_{0\eta})&=&-2 (\lambda_1+\frac{5}{2} \lambda_2)\eta k_{0z} +\nn\\
&&2 \lambda_2 k_0 P_{\nu} P_{3/2} \{S_z,U(\vec{k}_{0\eta})^{-1} S_z U(\vec{k}_{0\eta})\} P_{3/2} P_{\nu},\nn\\
\chi_{\nu}(\vec{k}_{0\eta})& =&P_{\nu} D(\vec{k}_{0\eta})^{-1} P_{3/2} U(\vec{k}_{0\eta})^{-1} U(\hat{a})^{-1}K(R_a\vec{k}_{0\eta}) \nn\\
 &&\cdot U(\hat{a}) U(\vec{k}_{0\eta}) P_{3/2} D(\vec{k}_{0\eta}) P_{\nu}.
\eea
The $\gamma(\vec{k}_{0\eta})$ term corresponds to the $O(\Delta_0)$ correction to the diagonal block in Eq. (\ref{eq:boundary1}) from $-i\xi \Delta_0$ in $k_z=\eta k_{0z}+\zeta \delta -i \xi \Delta_0$,
and the $\chi_{\nu}(\vec{k}_{0\eta})$ term is the projection to band structure basis of the superconducting pairing.
The solutions of $\xi$ and $\Phi^{\prime}$ are given by
\bea
\xi^{(3/2)}(\vec{k}_{0\eta};\nu)&=&\sqrt{\frac{|\chi_{\nu}(k_{0\eta})|^2-\epsilon^2}{\gamma(\vec{k}_{0\eta})^2}}, \nn\\
\Phi^{\prime(3/2)}(\vec{k}_{0\eta};\nu)&=&
\left(\begin{array}{c}
-(-)^{\frac{1-\nu}{2}} \chi_{\nu}(\vec{k}_{0\eta})\\
i \xi(\vec{k}_{0\eta},\nu) \gamma(\vec{k}_{0\eta}) -\epsilon
\end{array}\right),
\eea
in which $\xi$ is chosen to be positive to match the boundary condition at $z\rightarrow -\infty$.
The eigenvector $\Phi$ can be obtained from $\Phi^{\prime}$ by performing the transformations $D(\vec{k}_{0\eta})$, $U(\vec{k}_{0\eta})$ and $U(\hat{a})$ back in sequence.

Now we turn to the light hole bands.
Again the superscript ``$1/2$" will be dropped in the following expressions for simplicity.
The momentum in $z$-direction is in general $k_z=k_{0z}+\zeta \delta -i \xi \Delta_0$, where $k_{0z}=-i \sqrt{|\mu|/(2\lambda_2+\lambda_1)+k_x^2+k_y^2}$.
$\text{Im}(k_{0z})$ is chosen to be negative so as to match the boundary condition at $z\rightarrow -\infty$.
Let $\vec{k}_0=(k_x,k_y,k_{0z})$, and define
\begin{flalign}
&U^{(1/2)}(\vec{k_0})=
\left(\begin{array}{cc}
U(\hat{a}) & 0 \\
0 & U(\hat{a})^{T,-1}
\end{array}\right)
\left(\begin{array}{cc}
U(\hat{k}_0) & 0 \\
0 & U(\hat{k}_0)^{T,-1}
\end{array}\right),
\end{flalign}
in which $U(\vec{k}_0)=e^{-i S_z \phi} e^{-i S_y \theta}$,
where $\phi=\arctan(k_y/k_x)$ and $\theta=\arccos(k_{0z}/k_0)$.
Unlike the heavy hole bands, here $\theta$ is purely imaginary since $|k_{0z}|>|k_0|$.
Plug $k_z= k_{0z}+\zeta \delta -i \xi \Delta_0$ into the Hamiltonian in Eq. (\ref{eq:Hamiltonian}),
perform the transformation $\mathcal{U}^{(1/2)}(\vec{k}_{0})^{-1} H(\vec{k}) \mathcal{U}^{(1/2)}(\vec{k}_{0})$,
and project into the light hole bands.
Then by keeping the leading order terms in the expansion over $\delta$ and $\Delta_0$,
the $H_L$ and $H_A$ terms become
\bea
H^{\prime}_L-\mu&=&(\zeta \delta-i \xi \Delta_0) \big (2(\lambda_1+\frac{5}{2} \lambda_2) k_{0z} \nn\\
&& -2 \lambda_2 k_0 P_{1/2}\{S_z,U(\vec{k}_0)^{-1} S_z U(\vec{k}_0) \} P_{1/2}\big ), \nn \\
H^{\prime}_A&=& \frac{\delta}{k_f} \,P_{1/2} U(\vec{k}_0)^{-1} U(\hat{a})^{-1} A(R_a \vec{k}_0)\nn\\
&&\cdot U(\hat{a}) U(\vec{k}_0) P_{1/2}, \nn\\
\label{eq:transformed}
\eea
in which $P_{1/2}$ is the projection operator to the helicity $\pm 1/2$ bands,
and the supercript of prime denotes the terms after the transformation and the projection.
The eigenvalues of $H^{\prime}_A$  are $\pm \sqrt{\vec{\Lambda}^2(\vec{k}_0)}$.
$H^{\prime}_A$ introduces the correction of $\zeta \delta$ into the Fermi wave vectors.
Let $D(\vec{k}_0)$ be the transformation that diagonalizes $H^{\prime}_A$. 
It defines the band structure basis for the case of the light hole bands.

Let $P_{\nu}$ ($\nu=\pm 1$) be the projection operator to one of two band structure basis with eigenvalue $\nu \sqrt{\vec{\Lambda}^2(\vec{k}_0)}$ of $H^{\prime}_A$.
For the treatment of the superconducting pairing,
again by assuming $\Delta_0\ll\delta$,
the projection to the band structure basis can be performed,
and the eigen-equation determining $\xi$ and $\Phi$ is
\bea
\left(\begin{array}{cc}
i\xi \gamma(\vec{k}_0) -\epsilon  & -(-)^{\frac{1-\nu}{2}} \chi_{\nu} (\vec{k}_0) \\
-(-)^{\frac{1-\nu}{2}}\tilde{\chi}_{\nu}(\vec{k}_0) & -i\xi \gamma(\vec{k}_0) -\epsilon
\end{array} \right) \Phi^{\prime}=0,
\eea
in which
\bea
\gamma(\vec{k}_0)&=&-2(\lambda_1+\frac{5}{2}\lambda_2) k_{0z}\nn\\
&&+2 \lambda_2 k_0 P_{\nu} P_{1/2} \{S_z,U(\vec{k}_0)^{-1}S_zU(\vec{k}_0)\} P_{1/2} P_{\nu},\nn\\
\chi_{\nu}(\vec{k}_0)&=&P_{\nu} D^{-1}(\vec{k}_0) P_{1/2} U(\vec{k}_0)^{-1} U(\hat{a})^{-1} K(R_a\vec{k}_0)\nn\\
&&  \cdot U(\hat{a}) U(\vec{k}_0) P_{1/2} D(\vec{k}_0) P_{\nu},\nn\\
\tilde{\chi}_{\nu}(\vec{k}_0)&=&P_{\nu} D^{\dagger}(\vec{k}_0) P_{1/2} U(\vec{k}_0)^{\dagger} U(\hat{a})^{-1} K(R_a\vec{k}^*_0)\nn\\
&&  \cdot U(\hat{a}) U(\vec{k}_0)^{\dagger,-1} P_{1/2} D(\vec{k}_0)^{\dagger,-1} P_{\nu}.\nn\\
\eea
Then $\xi$ and $\Phi^{\prime}$ are solved as
\bea
\xi^{(1/2)}(\vec{k}_0;\nu,\iota)&=&\iota \sqrt{\frac{\chi_{\nu}(\vec{k}_0) \tilde{\chi}^*_{\nu} (\vec{k}_0)-\epsilon^2 }{\gamma(\vec{k}_0)^2}},\nn\\
\Phi^{'(1/2)}(\vec{k}_0;\nu,\iota)&=&\left(\begin{array}{c}
(-)^{(1-\nu)/2} \chi_{\nu} (\vec{k}_0)\\
i \xi(\vec{k}_0;\nu,\iota) \gamma(\vec{k}_0)-\epsilon
\end{array}\right),
\label{eq:phi prime lh}
\eea
in which $\iota=\pm 1$.
The eigenvector $\Phi$ can be obtained from $\Phi^{\prime}$ by performing the transformations $D(\vec{k}_0)$, $U(\vec{k}_0)$ and $U(\hat{a})$ back in sequence.

Plugging these expressions into the boundary condition at $z=0$, we obtain the equation determining the energy of the surface states,
\bea
\det \big(\{\Phi^{(1/2)}(\vec{k}^{(1/2)}_0;\nu,\iota) \}_{\nu,\iota=\pm},\{\Phi^{(3/2)}(\vec{k}^{(1/2)}_{0\eta};\nu) \}_{\eta,\nu=\pm}\big  )\nn\\
=0,\nn\\
\label{eq:surface eqn}
\eea
in which
\begin{flalign}
&\Phi^{(1/2)}(\vec{k}^{(1/2)}_0;\nu,\iota)=U^{(1/2)}(\vec{k}_0) \bar{D}^{(1/2)}(\vec{k}_0)  \Phi^{\prime(1/2)}(\vec{k}_0;\nu,\iota), \nn\\
&\Phi^{(3/2)}(\vec{k}^{(3/2)}_{0\eta};\nu)=U^{(3/2)}(\vec{k}_{0\eta}) \bar{D}^{(3/2)}(\vec{k}_{0\eta})  \Phi^{\prime(3/2)}(\vec{k}_{0\eta};\nu),\nn\\
\label{eq:surfacestate}
\end{flalign}
where $\bar{D}=\text{diag} (D,D^{T,-1})$ is the extension of the matrix $D$ to the particle-hole space.
In Eq. (\ref{eq:surfacestate}), all of the terms are in the 8-dimensional space.
For those originally defined not to be 8-dimensional, we need to appropriately embed them into the 8-dimensional space.
Eq. (\ref{eq:surfacestate}) is the general equation for solving surface state energy, without restriction on the form of the inversion breaking term, nor on the form of the pairing Hamiltonian.

We further note that for the present case in which the Fermi energy only crosses the heavy hole bands,
the $8 \times 8$ matrix in Eq. (\ref{eq:surfacestate}) can be reduced to a $4 \times 4$ one when solving Majorana zero modes.
The boundary condition requires that the eight vectors $\Phi^{(\alpha)}_i$ ($\alpha=3/2, 1/2$, $1\leq i \leq 4$) are linearly dependent.
At zero energy, the four vectors $\Phi^{(1/2)\prime}_i$ in the light hole space in Eq. (\ref{eq:phi prime lh}) are linearly independent.
Hence they must also be so at least in a neighborhood of $\epsilon=0$.
The vectors $\Phi^{(1/2)}_i$ are obtained from $\Phi^{(1/2)\prime}_i$ by the same transformation $D^{(1/2)}$.
This means that $\Phi^{(1/2)}_i$ are also linearly independent in a neighborhood of $\epsilon=0$.
Thus the boundary condition can be simplified to $\det \{ \mathcal{P}  \Phi^{(3/2)}_i \}_{1\leq i \leq 4}=0$,
where $\mathcal{P}$ is the projection operator into the linear subspace orthogonal to the space spanned by $\{ \Phi^{(1/2)}_i \}_{1\leq i \leq 4}$.

Although the original boundary condition matrix in Eq. (\ref{eq:surface eqn}) can be reduced to the heavy hole space,
the light hole space cannot be neglected since they touch with the heavy hole bands at the $\Gamma$ point.
If the two sets of bands are separated by a band gap $E_g$ much greater than the value of the chemical potential,
then the light hole bands are inert up to leading order of $|\mu|/E_g$.
In the current situation, they enter into the reduced $4\times 4$ boundary condition matrix,
which is a reflection of the spin-$3/2$ nature of the system.

\section{Symmetry properties in QPI patterns}
\label{sec:ph_QPI}

We discuss in this section the implications of particle-hole, TR, chiral and $C_{3v}$ symmetries on the QPI patterns for $(1\,1\,1)$-surface.

\subsection{Particle-hole symmetry}

In this part, we discuss the consequence of particle-hole symmetry in QPI patterns.

First consider the non-magnetic impurity.
Let $P_{sf}$ be the projection operator to the subspace of the surface states.
In Born approximation and only taking into account the contribution from Majorana zero modes,
$\Delta\rho^{(00)}(\omega,\vec{r})$ can be expressed as
\bea
\Delta\rho^{(00)}(\omega,\vec{r})&=&-\frac{1}{2\pi} \text{ImTr} \big[  (1+\tau_3) \bra{\vec{r}} P_{sf}\frac{1}{\omega-H+i\epsilon}\nn\\
&&P_{sf}H_{\text{imp}}P_{sf}\frac{1}{\omega-H+i\epsilon}P_{sf} \ket{\vec{r}}\big ], \nn\\
&=&-\frac{1}{2\pi} \text{Im} \big [(\frac{1}{\omega+i\epsilon})^2 \text{Tr} \big( (1+\tau_3) \nn\\
&&\bra{\vec{r}}P_{sf}V_{\text{imp}}P_{sf} \ket{\vec{r}} \big)\big ],
\eea
in which $V_{\text{imp}}$ is a scalar potential.
Particle-hole symmetry leads to
\bea
\bra{\vec{r}}P_{sf}H_{\text{imp}}P_{sf} \ket{\vec{r}}=-\tau_1\bra{\vec{r}}P_{sf}V_{\text{imp}}P_{sf} \ket{\vec{r}}^{*}\tau_1.
\label{eq:ph_1}
\eea
Since $\bra{\vec{r}}P_{sf}H_{\text{imp}}P_{sf} \ket{\vec{r}}$ is hermitian, $\text{Tr}\bra{\vec{r}}P_{sf}H_{\text{imp}}P_{sf} \ket{\vec{r}}$ vanishes due to Eq. (\ref{eq:ph_1}).
Thus
\bea
\Delta\rho^{(00)}(\omega,\vec{r})=-\frac{1}{2\pi} \text{Im} (\frac{1}{\omega+i\epsilon})^2 \text{Tr}\big[  \tau_3  \bra{\vec{r}}P_{sf}H_{\text{imp}}P_{sf} \ket{\vec{r}}  \big].\nn\\
\eea

Next consider the magnetic impurity.
In the same approximations, we have
\bea
\Delta \rho^{ij}(\omega,\vec{r})&=& -\frac{1}{2\pi} \text{ImTr} \big[ (\frac{1}{\omega+i\epsilon})^2 (1+\tau_3)\nn\\
&& \Sigma^{i} \bra{\vec{r}} P_{sf} \Sigma^{j} V_{\text{imp}} P_{sf} \ket{\vec{r}}\big].
\eea
Particle-hole symmetry implies that
\bea
\bra{\vec{r}}P_{sf}\Sigma^j H_{\text{imp}}P_{sf} \ket{\vec{r}}=-\tau_1\bra{\vec{r}}P_{sf}\Sigma^j H_{\text{imp}}P_{sf} \ket{\vec{r}}^{*}\tau_1.
\eea
Combining with $\tau_1 \Sigma^i \tau_1=-\Sigma^{i*}$, we obtain
$\text{Tr}(\tau_3 \Sigma^i \bra{\vec{r}}P_{sf}\Sigma^j V_{\text{imp}}P_{sf} \ket{\vec{r}})=0$,
and
\bea
\Delta \rho^{ij}(\omega,\vec{r})&=& -\frac{1}{2\pi} \text{Im}\big[ (\frac{1}{\omega+i\epsilon})^2 \nn\\
&&\text{Tr} \big(  \Sigma^{i} \bra{\vec{r}} P_{sf} \Sigma^{j} V_{\text{imp}} P_{sf} \ket{\vec{r}}\big)\big].
\eea

\subsection{Time reversal symmetry}
\label{sec:QPI_TR}

In addition to the suppression of scatterings between TR related Majorana islands,
TR symmetry also requires that $\Delta \rho^{\mu 0}=0$, and $\Delta \rho^{ 0\nu}=0$ in Born approximation,
where $\mu,\nu \neq 0$.
We show these properties in this part.

It can be proved that in Born approximation, TR symmetry leads to
\begin{flalign}
&\Delta \rho^{\mu\nu}(\omega,\vec{r})\nn\\
&=-\frac{1}{2\pi} \text{ImTr} \bra{\vec{r}} (1+\tau_3) (T^{-1}\Sigma^{\mu}T) \frac{1}{\omega-H+i\epsilon} \nn\\
&\cdot V_{\text{imp}}(T^{-1} \Sigma^{\nu}T) \frac{1}{\omega-H+i\epsilon} \ket{\vec{r}}.\nn\\
\end{flalign}
Since $\Sigma^{0}$ is invariant under TR operation and $\Sigma^{i}$ ($i=1,2,3$) changes sign under TR operation,
it is clear that $\Delta \rho^{\mu\nu}$ vanishes when $\mu=0$, $\nu\neq 0$ or $\mu\neq 0$, $\nu= 0$.

\subsection{Chiral symmetry}

TR operation reverses the sign of $\Sigma^{\nu}$ when $\nu=1,2,3$, and keeps it invariant when $\nu=0$.
Particle-hole operation reverses the sign of $\Sigma_{\nu}$ for all $\nu=0,1,2,3$.
Chiral operation is the composition of TR and particle-hole operations.
Hence the selection rule of chiral symmetry is that
$H^{\nu}_{\text{imp}}$ couples Majorana islands with opposite chiral indices when $\nu=0$,
and those with the same chiral index when $\nu=1,2,3$.

\subsection{$C_{3v}$ symmetry}
\label{sec:C_3v}

The little group of the $(1\,1\,1)$-surface within the $T_d$ group is $C_{3v}$.
In this part, we analyze the consequence of $C_{3v}$ symmetry on the QPI patterns in Born approximation.
Since $\Delta\rho_{sf}^{00}(\omega,\vec{q})$ is invariant under $C_{3v}$,
and $\Delta\rho_{sf}^{\mu\nu}(\omega,\vec{q})=0$ if one of $\{\mu,\nu\}$ is zero,
here we consider $\mu,\nu\neq 0$.

In Born approximation
\bea
\Delta \rho^{\mu\nu}(\omega,\vec{r}) &=& -\frac{1}{2\pi} \text{Im} \text{Tr} [(1+\tau_3) \Sigma^{\mu} \bra{\vec{r}} \frac{1}{\omega-H+i\epsilon} \nn\\
&&V_{\text{imp}} (\vec{r}) \Sigma^{\nu}\frac{1}{\omega-H+i\epsilon} \ket{r} ].
\eea
Since $C_{3v}$ is the symmetry group of the system, for $\mathcal{C} \in C_{3v}$, $\Delta \rho^{\mu\nu}(\omega,\vec{r})$ can also be written as
\begin{flalign}
&\Delta \rho^{\mu\nu}(\omega,\vec{r}) \nn\\
&= -\frac{1}{2\pi} \text{Im} \text{Tr} [(1+\tau_3) \mathcal{C} \Sigma^{\mu} \mathcal{C}^{-1} \bra{ C \vec{r}}  \frac{1}{\omega-H+i\epsilon} \nn\\
&\cdot V_{\text{imp}} (\vec{r})\mathcal{C}  \Sigma^{\nu}\mathcal{C}^{-1}\frac{1}{\omega-H+i\epsilon} \ket{C\vec{r}} ],\nn\\
\end{flalign}
in which $C$ is the corresponding $3\times 3$ rotation matrix.
Using $\mathcal{C}\Sigma^{\mu}\mathcal{C}^{-1}=\Sigma^{\alpha} C_{\alpha \mu}$,
$\Delta \rho^{\mu\nu}(\omega,\vec{r})$ satisfies
\bea
\Delta \rho^{\mu\nu}(\omega,\vec{r})=\Delta \rho^{\alpha\beta}(\omega,C\vec{r})C_{\alpha\mu}C_{\beta_{\nu}},
\eea
for any $\mathcal{C}\in C_{3v}$.

For fixed $\omega,\vec{q}$, define $\Delta\rho_{sf}(\omega,\vec{q})$ as the $3\times 3$ matrix whose $\mu\nu$ element is $\Delta\rho_{sf}^{\mu\nu}(\omega,\vec{q})$.
The above analysis shows that
\bea
\Delta\rho_{sf}(\omega,C\vec{q})=C\Delta\rho_{sf}(\omega,\vec{q}) C^{T},
\eea
for any $\mathcal{C}\in C_{3v}$.
This is the relation that $\Delta\rho_{sf}(\omega,\vec{q})$ must satisfy due to the $C_{3v}$ symmetry.



\end{document}